\documentclass{ws-ijmpa}

\def\lsim{\raise0.3ex\hbox{$\;<$\kern-0.75em\raise-1.1ex\hbox{$\sim\;$}}}
\def\gsim{\raise0.3ex\hbox{$\;>$\kern-0.75em\raise-1.1ex\hbox{$\sim\;$}}}
\def\bmat{\left(\begin{array}}
\def\emat{\end{array}\right)}
\def    \be            {\begin{equation}}
\def    \ee            {\end{equation}}
\def    \bea           {\begin{eqnarray}}
\def    \eea           {\end{eqnarray}}
\def    \nn            {\nonumber}

\begin{document}
\markboth{Carlos Mu\~noz}
{Dark Matter Detection 
in the Light of Recent Experimental Results}

%
\catchline{}{}{}{}{}
%


\title{DARK MATTER DETECTION\\ 
IN THE LIGHT OF RECENT EXPERIMENTAL RESULTS
}

\author{\footnotesize CARLOS MU\~NOZ
}

\address{Departamento de F\'{\i}sica
Te\'orica C-XI
and 
Instituto de F\'{\i}sica Te\'orica C-XVI,\\
Universidad Aut\'onoma de Madrid,
Cantoblanco, 28049 Madrid, Spain. 
}



\maketitle


\begin{abstract}
The existence of dark matter was suggested,
using simple gravitational arguments,
seventy years ago.
Although we are now convinced that most of the mass in the
Universe is indeed some non-luminous matter, we still do not know its
composition. The problem of the dark matter in the Universe
is reviewed here.
Particle candidates for dark matter are discussed with particular
emphasis on Weakly Interacting Massive Particles (WIMPs).
Experiments searching for these relic particles,
carried out by many groups around the world, 
are also reviewed, paying special attention to their direct detection
by observing the elastic scattering on target nuclei through nuclear
recoils.
Finally, we concentrate on the theoretical models
predicting WIMPs, and in particular on supersymmetric extensions
of the standard model,
where the leading candidate for WIMP, the neutralino, is present.
There, we compute the cross section for the direct detection of 
neutralinos, and compare it with the sensitivity of detectors.
We mainly discuss supergravity, superstring and M-theory
scenarios.



\end{abstract}

\section{Introduction}

One of the most evasive and fascinating enigmas in physics is 
the problem of the dark matter in the Universe.
Substantial evidences exist suggesting that at least 90\%
of the mass of the Universe is due to some non-luminous matter, the so-called
`dark matter'.
However, although the existence of dark matter was
suggested 70 years ago, we still do not know its composition.

In 1933, the astronomer Fritz Zwicky \cite{Zwicky}
provided evidence that the mass of the luminous matter (stars) 
in the Coma cluster, which consists of about 1000 galaxies, 
was much smaller than its total mass implied by the 
motion of cluster member galaxies.
However, only in the 1970s the existence of dark matter began to be
considered seriously. Its presence in spiral galaxies was
the most plausible explanation for  
the anomalous rotation curves of these galaxies,
as we will discuss in detail in Section~2. 
In summary, 
the measured rotation velocity of isolated stars or
gas clouds in the outer parts of galaxies was not
as one should expect from the gravitational attraction due to 
the luminous matter. This led astronomers to assume that
there was dark matter in and around galaxies. 
We will see that nowadays there is overwhelming observational evidence
for the presence of dark matter. It not only clusters  
with stellar matter forming the galactic halos, but also
exists as a background density over the entire Universe.
Thus the problem is no longer to explain the rotation curves
but to decipher the nature of the dark matter.
In any case, 
it is fair to say that some authors still suggest
that dark matter is not necessary to explain the rotation
curves. They try to avoid the introduction of this
hypothesis modifying the
Newton's law at galactic scales. 
Although these attempts are not very convincing,
we will briefly discuss them

As we will explain in detail in Section~3, the search of the dark
matter
provides
a potentially important
interaction between particle physics and cosmology, since
only elementary particles are reliable candidates for the dark matter
in the Universe.
In particular we will see that 
baryonic objects, such as e.g. gas, brown dwarfs, etc.,
can be components of the dark matter, but more candidates are needed.
The reason being that they cannot be 
present in the right amount to explain the observed matter density
of the Universe, 
$0.1\lsim \Omega h^2\lsim 0.3$. 
Fortunately, particle physics, and mainly
extensions of the standard model 
offer candidates for dark matter.
Indeed, detecting non-baryonic dark matter
in the Universe might be a signal for new physics beyond the Standard
Model.
It is fair to say that many of these candidates are quite exotic
and most of them are ephemeral.
However,
we will also see that very interesting and plausible candidates for dark
matter are Weakly Interacting Massive Particles (WIMPs), since
long-lived or stable WIMPs can be left over from the
Big Bang
in sufficient number to account for a significant
fraction of relic 
matter density.
As suggested 
in 1985
by
Goodman and Witten \cite{Witten}, and also 
by 
Wasserman \cite{Wasserman}, 
this raises the hope of detecting relic WIMPs directly.

WIMPs cluster gravitationally with ordinary stars in
galactic halos, and therefore they must be present in our own
galaxy, the Milky Way. As a consequence there will be a flux of
these dark matter particles on the Earth, and experiments
can be carried out to detect this flux 
by observing its elastic scattering on target nuclei through nuclear
recoils.
As we will mention, inelastic scattering is also possible but 
the event rates seem to be suppressed.
Thus in Section~4 we will review the current and projected experiments
for detecting WIMPs directly in underground laboratories.
In fact, 
a dark matter experiment of this type,
the DAMA collaboration \cite{experimento1,masdama},
has reported data recently favouring
the existence of a WIMP signal.
It was claimed that 
the preferred range of the WIMP-nucleon cross section
is $\approx$
$10^{-6}-10^{-5}$ pb 
for a WIMP mass of about 100 GeV.
Although this could be a great discovery, it is fair to say that
this experiment is controversial, mainly because 
other collaborations reported negative search results.
In particular, 
CDMS \cite{experimento2}, and more recently EDELWEISS \cite{edelweiss} 
and ZEPLIN I \cite{zeplin1}, 
claim to have excluded important regions of the DAMA 
parameter space. 
In any case, the whole region allowed by DAMA 
will be covered by these and
other experiments in the near future. 
Besides, there are projected experiments which will be able to measure
a cross section as small as $10^{-9}$ pb.
In addition, there are also promising methods for the indirect
detection of WIMPs
by looking for evidence of their annihilation, as we will briefly discuss.
Thus it
seems plausible that the dark matter will be found in the near future.
Given this situation, 
it is crucial to analyze in detail the
present theoretical models predicting WIMPs.

In Section~5 we will review the situation concerning this point.
In particular, 
we will concentrate on the leading candidate for WIMP,
the 
lightest supersymmetric particle (LSP).
As is well known, supersymmetry (SUSY) \cite{susy,susy2}  is 
a new type of symmetry, since it relates bosons and fermions,
and this ensures 
the stability of the hierarchy between the weak and the Planck
scales. This is the most relevant theoretical argument in its favour.
In SUSY models the so-called R-parity is often imposed in order to avoid
fast proton decay or lepton number violation. 
This yields important phenomenological implications.
SUSY particles are produced or destroyed only in pairs
and therefore the LSP
is absolutely stable, implying that it might  
constitute a possible candidate for dark matter. 
So SUSY, 
whose original motivation has nothing to do
with the dark matter problem, 
might solve it.

The first discussion of SUSY dark matter was by Goldberg
\cite{Goldberg} 
in 1983. He considered the SUSY fermionic partner of the photon, the photino,
as the LSP and pointed out the strong constraints on its mass
from its relic abundance. Soon followed two works considering also this
possibility \cite{Srednicki,Krauss}, but it was in Ref.~\refcite{Ellis}
were the analysis of the most general case was carried out.
In particular, one has to take into account that in SUSY
the photino mixes with the fermionic partners of the $Z^0$
and the two neutral Higgs bosons. 
Therefore one has  
four particles called neutralinos,
$\tilde{\chi}^0_i~(i=1,2,3,4)$.
Obviously, the lightest neutralino, $\tilde{\chi}^0_1$, 
will be the 
dark matter candidate. The fact that the LSP is chosen to be
an electrically neutral particle (also with no
strong interactions) is welcome since otherwise it 
would bind to nuclei and would be excluded as a candidate
for dark matter from unsuccessful searches for exotic heavy 
isotopes.
It is remarkable that in large regions
of the parameter space of the simplest SUSY extension of the
standard model, the so-called Minimal Supersymmetric Standard 
Model (MSSM), the 
LSP is indeed the lightest neutralino.

It is worth noticing here that 
several accelerator experiments are in preparation 
in order to detect SUSY particles. For example,
LHC at CERN will probably start operations in 2007 producing energies
about 1 TeV. 
Concerning the LSP, one will be able 
to detect events with SUSY particle decays which
produce lots of missing energy.
However, let us remark that, even if neutralinos are discovered in this
way,
only their direct detection in underground experiments
due to their presence in our galactic halo
will confirm that they are the sought-after dark matter of the
Universe. 
As a matter of fact, given the present and projected experiments
in underground laboratories
in order to detect WIMPs, there may be a competition 
between these experiments and those in accelerators 
in the hunt for the first SUSY particle.

Thus we will review 
how big the cross section for the direct detection of neutralinos 
can be, in the generic context of the MSSM.
In fact, the cross section for the 
elastic scattering of
$\tilde{\chi}^0_1$
on nucleons
has been examined exhaustively in the 
literature since many years ago \cite{kami}.
Obviously, this computation is crucial to analyze the compatibility
of the neutralino as a dark matter candidate, with the sensitivity
of detectors. In the light of the recent experimental results
mentioned above, many theoretical works studying possible values of the
cross section in different scenarios
have appeared. Here
we will concentrate on the most recent results in 
supergravity (SUGRA),
superstring and M-Theory scenarios \cite{Gondolo}$^-$\cite{softrelation}.

Let us recall that in the framework of SUGRA \cite{susy}
the scalar and gaugino masses and the bilinear and
trilinear couplings are generated once SUSY is softly broken
through gravitational interactions, and in addition radiative electroweak
symmetry breaking is imposed. 
First, we will analyse this framework in the usual context
of a Grand Unified Theory (GUT) with scale
$M_{GUT} \approx 2\times 10^{16}$ GeV and universal soft 
terms \cite{Bottino,phases,arna2,masfases,Falk,masfases2,Bott,Ellis2,Arnowitt,focus,Bed,Drees,Gomez,large,Nojiri,tuning,darkcairo,Drees2,EllisOlive,Arnowitt3,focusito,nosopro,Lahanas,Dutta,Elliswmap,Lahanas2,Wnath,Vergados,darkufb,Farrill,softrelation},
the so-called minimal supergravity (mSUGRA) scenario.
Second, we will discuss 
how the results are modified
when the GUT condition
is relaxed. In particular, we will consider the case of an 
intermediate scale
\cite{muas,Bailin,nosotros,darkcairo,nosopro,darkufb}, 
a possibility inspired by experimental observations
and also by some superstring constructions. 
Then, 
a more general situation in the context of SUGRA, 
non-universal 
scalar \cite{Bottino,arna2,Arnowitt,Santoso,Drees,Nojiri,darkcairo,Arnowitt3,nosopro,Dutta,Rosz,darkufb,Farrill}
and 
gaugino \cite{Nath2,darkcairo,nosopro,Orloff,Dutta,Birkedal,Roy2,darkufb}
masses, will
be studied.
On the other hand, 
a phenomenological SUSY model, with parameters defined
directly at the electroweak scale, the so-called
effMSSM, has also been studied in the 
literature \cite{Gondolo,muchos,Bed,Bottinoeff,Austri}. 
We will discuss this approach briefly.

Let us remark that in these analyses to reproduce the correct
phenomenology is crucial. In this sense it is important to check
that the parameter space of the different scenarios 
fulfills the most recent experimental and astrophysical constraints.
Concerning the former, 
the lower bound on the Higgs mass \cite{Higgs},
the $b\to s\gamma$ branching ratio \cite{bsgamma}, and the
muon \cite{marci} $g-2$ will be taken into account.
The astrophysical bounds on the dark matter density
mentioned above
will also be imposed on the 
theoretical computation \cite{kami} of the relic neutralino density,
assuming thermal production.
In addition, 
the constraints that the absence of dangerous charge
and colour breaking minima imposes on the parameter space \cite{discussion}
will also be taken into account.

Finally, since 
generically the low-energy limit of
superstring theory \cite{superstring} 
is $4$-dimensional 
SUGRA, the neutralino is also a candidate for dark matter
in superstring constructions. Taking into account
that the soft terms can in principle be computed in these
constructions, we will review the associated $\tilde{\chi}^0_1$-nucleon
cross section. In particular we will discuss this first in the
context of the (weakly-coupled) heterotic 
superstring \cite{Shafi2,Drees}.
Then, we will consider
D-brane configurations from the type I 
superstring
\cite{Khalil,Nath2,Arnowitt2,Bailin,nosotros,darkcairo,nosopro},
which are explicit scenarios where intermediate scales,
and also non-universality of soft terms, may occur.

On the other hand, $4$-dimensional SUGRA is also the low-energy limit of
$11$-dimensional M-Theory \cite{sch}. The latter has been
proposed as the fundamental
theory which contains the five $10$-dimensional superstring theories,
and, in particular, compactified on an orbifold is indeed the
strong coupling limit of the $E_8\times E_8$ heterotic
superstring theory. 
Therefore, the same review for the cross section 
as above will be carried out
for this strongly-coupled heterotic superstring
obtained from M-Theory \cite{bailinlove2,krani,cerde}, 
the so-called heterotic
M-Theory.
The conclusions are left for Section~6.

\section{The dark matter problem}



To compute the rotation velocity of stars or hydrogen clouds
located far away from galactic centres 
is easy. One only needs to extrapolate
the Newton's law, which works outstandingly well for nearby
astronomical phenomena, to galactic distances. For example,  
for an average distance $r$ of a planet from the center 
of the Sun,
Newton's law implies that 
$v^2(r)/{r} = {G M(r)}/{r^2}$,
where $v(r)$ is the average orbital velocity of the planet,
$G$ 
is the Newton's 
constant and $M(r)$ is the total mass inside 
the orbit.
Therefore one obtains
%
\begin{equation}
v(r)= \sqrt{\frac{G\ M(r)}{r}}
\ .
\label{Newton}
\end{equation}
Clearly, $v(r)$ decreases with increasing radius
since $M(r)$ is constant and given by the solar mass. 

\begin{figure}[t]
\begin{center}
\begin{tabular}{c}   
\epsfig{file= 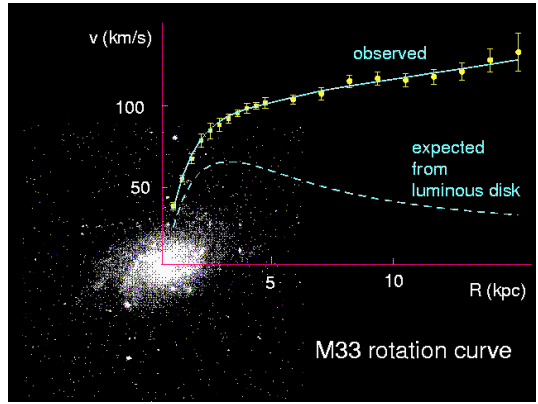, width=7cm}\\
\end{tabular}
\end{center}
\caption{Observed rotation curve of the nearby dwarf spiral galaxy M33,
superimposed on its optical 
image.
}
\label{flat}
\end{figure}

In the case of a galaxy, if its
mass distribution can be approximated as
spherical or ellipsoidal, Eq.~(\ref{Newton}) can also be used
as an estimate.
Thus
if the mass of the galaxy 
is concentrated in its visible part, one would
expect $v(r)=\sqrt{G M_{vis}/r} \propto 1/\sqrt{r}$ 
for distances far beyond the visible radius.
Instead, astronomers, by means of the Doppler effect, observe that
the velocity rises towards a constant value  
$v_c \approx$ 100 to 200 km\ s$^{-1}$.
An example of this can be
seen in Fig.~\ref{flat} (from Ref.~\refcite{Roy}), 
where the rotation curve of M33, 
one of the about 45 galaxies which form our small cluster, the Local
Group,
is shown. For comparison, the expected velocity from luminous disk
is also shown. 
This phenomenon has already been observed for about a thousand 
spiral galaxies \cite{Persic,Sofue,Roncadelli}, 
and in particular also for
our galaxy, the Milky Way.
Although this observation is more problematic in
galaxies other than spirals, such as ellipticals,
dwarf irregulars, dwarf spheroidals, lenticulars, etc.,
they also produce similar results \cite{Salucci,Battaner,Roncadelli}.

The most common explanation for these flat rotation curves
is to assume that disk galaxies are immersed in extended dark matter
halos.
Thus for large distances $M(r)/r$ 
is generically constant because
the mass interior to $r$ increases
linearly with $r$, $M_{tot} (r)=G^{-1} v_c^2 r$.
In fact, a self-gravitating ball of ideal gas at an
uniform temperature of $kT=\frac{1}{2}m_{\mbox{\tiny DM}}v_c^2$,
where $m_{\mbox{\tiny DM}}$ is the mass of one dark matter particle,
would have this mass profile \cite{binney}.
In addition, dark matter is
necessary to explain structure formation, as deduced from the
cosmic microwave background measurements. 
Clumps of neutral particles arose first through gravitational
attraction
and later, when the neutral atoms were formed, these were
gravitationally
attracted by the dark matter to form the galaxies.


The above analysis of rotation curves
implies that 90\% of the mass in galaxies
is dark.
Whereas current observations of luminous matter in galaxies determine
$\Omega_{\mbox{\tiny lum}} \lsim 0.01$, 
analyses of rotation curves imply in fact 
$\Omega \approx 0.1$.
Let us recall that
$\Omega = \rho/\rho_c$ is 
the present-day mass density averaged over the Universe,
$\rho$, in units of the critical density, $\rho_c=10^{-5} h^2$ GeV\ cm$^{-3}$,
with $h\sim 0.7$.
In fact, the previous value,
$\Omega \approx 0.1$,
is really a 
lower bound, since 
almost all rotation curves remain flat out to the largest values 
of $r$ where one 
can still find objects orbiting galaxies. We do not really
know how much further the
dark matter halos of these galaxies extend (see e.g. Fig.~\ref{flat}).
Thus we can conclude that galactic rotation curves imply 
$\Omega \gsim 0.1$.

Although
the analysis of dark matter in cluster of galaxies becomes more
involved than in galaxies, different techniques have been used
to compute \cite{Freedman,Roncadelli} the value of $\Omega$. 
For example, the conventional method of studying the motion of
cluster member galaxies seems to point at
values of $\Omega$ larger than those obtained in galactic 
scales. The high temperature of the gas detected in clusters
through its X-ray emission also points at large amounts
of dark matter. 
Finally, the more reliable method of
studying the gravitational lensing confirms the previous conclusions.
Here a cluster acts as a lens which distorts
the light emitted
by quasars and field galaxies in its background, due to the gravitational
bending of light. All these analyses favour a value
$\Omega\approx 0.2-0.3$.
Moreover, measurements of velocity flows at scales 
larger than that of clusters
favour \cite{Freedman} also very large amounts of dark 
matter:
$\Omega > 0.3$. Thus 
the following astrophysical bounds are commonly used
in the literature:
%
\begin{equation}
0.1\lsim\Omega_{\mbox{\tiny DM}} h^2\lsim 0.3
\ .
\label{omega}
\end{equation}
As a matter of fact, the lower part of this range seems
to be
preferred by comparing models with a set of cosmological 
observations \cite{Silk,wmap}.
In particular, the (cold) dark matter range
\begin{equation}
0.094\lsim\Omega_{\mbox{\tiny DM}} h^2\lsim 0.129
\ ,
\label{wmaprange}
\end{equation}
can be deduced from the recent data obtained by the
WMAP satellite \cite{wmap}.


Clearly, any sensible candidate for dark matter
must be able to reproduce the observations summarized
by the above equations, i.e. it must be 
present in the right amount to explain the observed matter density
of the Universe.
This is the issue that we will review in the next section.

\vspace{0.2cm}

Before finishing this section, 
it is fair to say that a small number of authors suggest
that dark matter is not really necessary to explain rotation
curves \cite{Battaner}. 
Basically their approach consists of modifying the
Newton's law at galactic scales. 
This has been done first in the pure classical 
theory \cite{Finzi,Milgrom,Sanders}. For example, in
Ref.~\refcite{Milgrom} the authors assume that the behaviour
of the 
potential at galactic distances is $V\sim \ln r$.
In the context of the relativistic theory, several 
authors 
have studied the modification induced to the
scalar potential by higher-derivative gravity \cite{Mannheim,Kenmoku}, 
or by a
cosmological constant \cite{Kokorelis}.
These modifications have also been studied in other contexts, such as
five dimensional gravity inspired in strings \cite{who} (D-branes),
and geodetic brane gravity \cite{Davidson}.
However, 
these attempts are not only rather ad hoc
in general (the authors impose specific values 
for the free parameters of the
theory in order to reproduce {\it some} of the rotation curves
that have been observed) 
but also insufficient to account for the necessity
of dark matter in scales larger than galactic ones \cite{Mond}  
(values of the parameters necessary to reproduce galactic rotation
curves cannot reproduce the observations at larger scales).
Recently, the authors of Ref.~\refcite{sloan},
using the Sloan Digital Sky Survey, have shown the first observational
evidence that the halo density decline as $1/r^3$, as predicted
by cold dark matter cosmological models. Alternative theories
of gravity are in contradiction with this result.

\section{Dark matter candidates}

Here we will review possible candidates for dark matter.
Since ordinary matter is baryonic, 
the most straightforward possibility is to assume also
this composition for the evasive dark matter. The contribution
from gas is not enough, 
so astrophysical bodies collectively known as MAssive Compact Halo 
Objects (MACHOs) are the main candidates \cite{Sadoulet}.
This is the case of brown and white dwarfs, Jupiter-like objects, 
neutron stars,
stellar black hole remnants. 
However, the scenario of Big-Bang nucleosynthesis, which
explains the origin of the elements after the Big Bang, 
taking into account measured abundances of helium, deuterium and
lithium sets a limit to the 
number of baryons that can exist in the Universe, 
namely 
$\Omega_{\mbox{\tiny baryon}}h^2\lsim 0.05$.
This density is clearly small to account for the whole dark matter
in the Universe (see bounds (\ref{omega})).
The conclusion is that baryonic objects are likely
components of the dark matter 
but more candidates are needed.
This result is also confirmed by observations of MACHOs in our galactic 
halo
through their gravitational lensing effect 
on the light emitted by stars. Their contribution to the 
dark matter density is small.
Thus non-baryonic matter is required in the 
Universe. 

Particle physics provides this type of candidate for dark matter.
In principle, the three most promising are 
`axions', `neutrinos' and `neutralinos' with 
masses of the order of $10^{-5}$ eV, 30 eV and 100 GeV, 
respectively.
As a matter of fact, neutrinos are the only candidates which
are known to exist. However, although the other particles 
are not
present in the standard model, they are crucial to
solve important theoretical problems of this model. 
Thus they are well 
motivated by extensions of the standard model.


Before analyzing these three (standard) candidates, 
it is fair to say that many others have also been proposed.
Although some of them are quite exotic, we will also discuss
them in some detail below.

\subsection{Standard candidates}

\subsubsection{Neutrinos}


In recent years, observation of solar and
atmospheric 
neutrinos have indicated that one flavour might change to 
another \cite{Torrente}.
As is well known, these neutrino oscillations 
can only happen if neutrinos have mass. The best evidence 
for neutrino masses comes
from the SuperKamiokande experiment in Japan
concerning atmospheric neutrino oscillations. The results of this experiment 
indicate a mass
difference of order 0.05 eV
between the muon neutrino and the tau 
neutrino. If there
is a hierarchy among the neutrino masses 
(as it is actually the case not only for quarks 
but also for  
electron-like leptons),
then such a small mass difference implies that the neutrino masses
themselves lie well below 1 eV. This
is not cosmologically significant as we will show below,
since a light ($m_{\nu}\lsim 100$ eV) neutrino
has a cosmological density \cite{Cowsik} 
$\Omega_{\nu}\approx m_{\nu}/30\ \mathrm{eV}$.  
On the other hand, there could be near mass degeneracy among the neutrino
families. In this case, if
neutrino masses $m_{\nu}\approx 30$ eV, they 
could still contribute significantly to the
non-baryonic dark matter in the Universe.

These neutrinos left over from the Big Bang
were in thermal equilibrium in the early Universe
and decoupled when they were moving with relativistic velocities.
They fill the Universe in 
enormous quantities and their current number density
is similar to the one of photons
(by entropy conservation in the adiabatic expansion
of the Universe). In particular,  
$n_\nu=\frac{3}{11}\ n_\gamma$.
Moreover, the number density of photons 
is very accurately obtained from the cosmic microwave 
background measurements. The present temperature
$T\approx 2.725\ K$ implies
$n_{\gamma}\approx 410.5$ cm$^{-3}$. 
Thus one can compute the neutrino mass
density $\rho_\nu = m_{tot}\ n_\nu$, where 
$m_{tot}$
is basically the total mass due to all flavours of neutrino.
Hence,
\begin{equation}
\Omega_\nu \approx \frac{m_{tot}}{30\ \mathrm{eV}}\ .
\label{neutrinos}
\end{equation}
Clearly, neutrinos with $m_{\nu}\lsim 1$ eV cannot solve the
dark matter problem, but a neutrino with
$m_{\nu}\approx 30$ eV would give
$\Omega_{\nu}\approx 1$
solving it.
Unfortunately, due to the small energies involved,
detection of these cosmological neutrinos in the laboratory is
not possible. 



However, even if $m_{\nu}\approx 30$ eV, there is 
now significant evidence against neutrinos as the bulk of the
dark matter. 
Neutrinos belong to the so-called `hot' dark matter because
they were moving with relativistic velocities at the time the galaxies
started to form. But hot dark matter cannot reproduce correctly the
observed structure in the Universe.
A Universe dominated by neutrinos would form large structures first,
and the small structures later by fragmentation of the larger
objects. 
Such a Universe would produced a `top-down' cosmology, in which the
galaxies
form last and quite recently. 
This time scale seems incompatible with our
present ideas of galaxy evolution. 

This lead to fade away 
the initial enthusiasm for a
neutrino-dominated Universe. Hence, many cosmologists now favour an alternative
model, one in which the particles dominating the Universe are `cold'
(non-relativistic) rather than hot.
This is the case of the axions and neutralinos which we will study below.

\subsubsection{Axions}

Axions are spin 0 particles with zero charge
associated with the spontaneous breaking of the
global $U(1)$ Peccei-Quinn symmetry, which was introduced
to dynamically solve the strong CP problem \cite{PQ}.
Let us recall that this important problem arises because
QCD
includes in its Lagrangian the term 
$\frac{\theta}{16\pi^2} tr (F_{\mu\nu} {\tilde F}^{\mu\nu})$
violating CP, and therefore
there are important experimental bounds against it.
In particular, the stringent upper limit
on neutron dipole electric moment implies the bound $\theta<10^{-9}$.
This is the problem. Why is this value
so small, when a strong interaction parameter would be expected
to be ${\cal O}(1)$?

Although axions are 
massless at the classical level they pick up a small mass
by non-perturbative effects. 
The mass of the axion, $m_a$, and its 
couplings to ordinary matter, $g_a$, are proportional to $1/f_a$,
where $f_a$ is the 
(dimensionful) axion decay constant which is related to 
the scale of the symmetry breaking. 
In particular, the coupling of an axion with two fermions of
mass $m_f$, is given by $g_a\sim m_f/f_a$. Likewise, 
$m_a \sim \Lambda^2_{\mbox{\tiny QCD}}/f_a$, i.e. 
\begin{equation}
m_a 
\sim 10^{-5}~\mathrm{eV} \times \frac{10^{12}~\mathrm{GeV}}{f_a}\ .
\end{equation}

A lower bound
on $f_a$ can be obtained from the requirement that
axion emission does not over-cool stars. 
The supernova SN1987 put the strongest bound,
$f_a \gsim 10^9$ GeV. 
On the other hand, 
since coherent oscillations of the axion 
around the minimum of its potential may give an important contribution
to the energy density of the Universe, the requirement
$\Omega\lsim 1$ 
puts a lower bound on the axion mass implying $f_a \lsim 10^{12}$ GeV.
The combination of both constraints, astrophysical and cosmological,
give rise to the following window for the value of the axion constant
\begin{equation}
10^9~\mathrm{GeV}\lsim f_a \lsim 10^{12}~\mathrm{GeV}\ .
\end{equation}

The lower bound 
implies an extremely small coupling of the axion to 
ordinary matter and therefore a very large lifetime, larger than the
age of the Universe by many orders of magnitude. 
As a consequence, the axion is a candidate for dark 
matter \cite{axions}.
Axions would have been produced copiously in the Big Bang, they
were never in thermal equilibrium and are always nonrelativistic
(i.e. they are cold dark matter).
In addition the upper bound implies that $m_a\sim 10^{-5}$ eV
if the axion is to be a significant component of the dark matter.

Since the axion can couple to two photons via fermion vacuum loops,
a possible way to detect it is through conversion to photon
in external magnetic field. Unfortunately, due to the small couplings
to matter
discussed above, we will not be able to produce axions in the
laboratory. On the other hand, relic axions are very abundant
(as we will discuss in Section~4, the density of dark matter particles
around the Earth is about 0.3 GeV\ cm$^{-3}$, since $m_a\sim 10^{-5}$
eV
there will be about $10^{13}$ axions per cubic centimeter)
and several experiments are trying already to detect axions or
are in project.
For example, an experiment at Lawrence Livermore National Laboratory 
(California, US) has reported in 1998 its first results
excluding a particular kind of axion of mass $2.9\times 10^{-6}$ eV
to 
$3.3\times 10^{-6}$ eV as the
dark matter in the halo of our galaxy.

\vspace{0.2cm}

Let us finally mention that links to web pages
of axion detection experiments
can be found in the web page
http://cdms.physics.ucsb.edu/others/others.html.

\subsubsection{WIMPs}

As mentioned in the Introduction, WIMPs 
are very interesting candidates for dark matter
in the Universe. They were in thermal equilibrium with the standard
model particles in the early Universe, and decoupled when
they were non-relativistic.
The process was the following.
When the temperature $T$ of the Universe was larger than the mass of the
WIMP, 
the number density of WIMPs and photons was roughly the same, 
$n\propto T^3$,
and
the WIMP was annihilating with its own antiparticle into lighter
particles and vice versa.
However, after the temperature dropped below the mass of the
WIMP,
$m$,
its number density dropped exponentially,
$n\propto e^{-m/T}$,
because only a small fraction of the light particles mentioned above
had sufficient kinetic energy to create WIMPs.
As a consequence, the WIMP annihilation rate 
$\Gamma = 
\langle\sigma_{ann} v \rangle n$
dropped below the expansion
rate of the Universe, $\Gamma \lsim H$. At this point  
WIMPs came away, they could not annihilate, and
their density is the same since then (freeze-out 
typically occurs at $T_F\approx m/20$).  
This can be obtained using the Boltzmann equation, which
describes the time evolution of the number density $n(t)$ of WIMPs
\begin{equation}
\frac{d n}{d t} + 3 H n = - \langle \sigma_{ann} v \rangle \left[(n)^2 - 
(n^{eq})^2 \right]\ ,
\end{equation}
where $H$ is the Hubble expansion rate, 
$\sigma_{ann}$ is the total cross section for annihilation of a
pair
of WIMPs into
standard model particles, $v$ is the relative velocity between the two WIMPs, 
$\langle...\rangle$ denotes
thermal averaging, and
$n^{eq}$ is 
the number density
of WIMPs in thermal equilibrium. 
One can discuss qualitatively the solution using
the freeze-out condition
$\Gamma = 
{\langle\sigma_{ann} v \rangle}_F n=H$.
Then, the current WIMP 
$\Omega h^2 =(\rho/\rho_{c}) h^2$, 
turns out to be
%
\begin{equation}
\Omega  h^2
\simeq \frac{3\times 10^{-27}\ \mathrm{cm^3\ s^{-1}}}
{\langle\sigma_{ann}\ v \rangle}\ , 
\label{relicdensity}
\end{equation}
where 
the number in the numerator
is obtained using the value of the 
temperature of the cosmic background radiation,
the Newton's constant, etc. 
As expected from the above discussion about the early Universe, 
the relic WIMP density decreases with
increasing annihilation cross 
section\footnote{Let us remark that the theoretical computation of the
relic density 
depends on assumptions about the evolution of the early
Universe, and therefore cosmological scenarios 
different from the thermal production of neutralinos discussed here 
would give
rise to different results \cite{relic}.}.

Now we can understand easily why WIMPs are so good candidates for dark
matter.
If a new particle with weak interactions exists in Nature,
its cross section will be 
$\sigma\simeq \alpha^2/m_{\mbox{\tiny weak}}^2$,
where $\alpha\simeq {\cal O}(10^{-2})$ 
is the weak coupling and $m_{\mbox{\tiny weak}}\simeq {\cal O}(100$ GeV)
is
a mass of the order of the W gauge boson.
Thus 
one obtains
$\sigma\approx 10^{-9}$ GeV$^{-2}$ $\approx 1$ pb
(recall that in natural units
1 GeV$^{-2}=0.389\times 10^{-27}$ cm$^2$ $=0.389\times 10^9$ pb).
Since at the freeze-out temperature $T_F$ the velocity $v$ is a
significant fraction of the speed of light ($v^2\approx c^2/20$),
one obtains 
$<\sigma_{ann}\ v>\approx 10^{-26}$ cm$^3$\ s$^{-1}$.
Remarkably, this number 
is close to the value that we need in Eq.~(\ref{relicdensity})
in order to obtain the observed density of the Universe.
This is a possible hint that new physics at the weak scale
provides us with a reliable solution to the dark matter problem,
and also a qualitative improvement with respect to the axion dark matter
case, where a small mass for the axion 
about $10^{-5}$ eV has to be postulated. 

\begin{figure}[t]
\begin{center}
\epsfig{file= 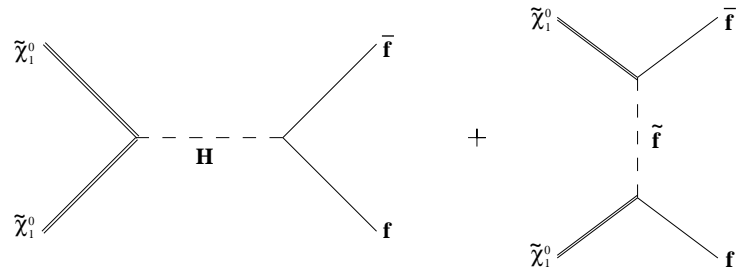,height=3.5cm,angle=0}
\end{center}
\caption{Feynman diagrams contributing to early-Universe 
neutralino ($\tilde{\chi}^0_1$) annihilation into 
fermions through neutral Higgses ($H\equiv H,h,A$) 
and squarks and sleptons ($\tilde f$).}
\label{annihilation}
\end{figure}

SUSY is a theory
that introduces new physics precisely at the weak scale, and that,
as discussed in the Introduction,
predict a new particle, the neutralino, which could be stable.
These are the reasons to consider the neutralino as a very serious
candidate for
the
sought-after dark matter.   
Concerning the annihilation cross section \cite{kami} 
contributing to the
density of the Universe 
in Eq.~(\ref{relicdensity}), there are numerous final states into which the 
neutralino can annihilate. The most important of these are the two body final 
states which occur at the tree level. Specially these are 
fermion-antifermion pairs \cite{Ellis,Gri,kturner,Lalak,drno}
($f\bar f$ where $f$ are the standard model quarks and leptons), 
as those shown in
Fig.~\ref{annihilation}. Others are weak gauge bosons 
pairs \cite{olis,kturner,drno} $W^+W^-$, $Z^0Z^0$, and those containing Higgs 
bosons \cite{olis,kturner,jojo,drno} such as
$W^+H^-$, $W^-H^+$, $Z^0A$, $Z^0H$, $Z^0h$, $H^+H^-$
and all six combinations of $A,h,H$.
Different subtleties of the analyses has been discussed in 
Refs.~\refcite{Seckel,gondolgel}.
The annihilation cross section
is of the form
\begin{equation}
\sigma_{ann}
\simeq N_{ann} m_{\tilde{\chi}_1^0}^2 |{\cal A}_{ann}|^2\ ,
\label{sigmann}
\end{equation}
where ${\cal A}_{ann}$ is the amplitude which depends on the
dynamics of the collision and $N_{ann}$ is the number of
annihilation channels. Generically ${\cal A}_{ann}\sim 1/{\cal M}^2$,
where ${\cal M}$ is the mass of the particles mediating the interaction.
Thus
$\sigma_{ann}\sim 1/m_{\tilde{\chi}_1^0}^2$, and therefore
in order to satisfy the upper bound in Eq.~(\ref{omega}), 
moderate values of the LSP mass are necessary.
Several exceptions to this rule will be discussed in 
Subsection~5.2.1. Let us mention here one of them which is 
particularly interesting.

In principle to use the above discussed neutralino annihilation cross section
is sufficient, since the effects of heavier particles are 
suppressed by the Boltzmann factor.
However, the next to lightest supersymmetric particle (NLSP)
may lie near in mass to 
the LSP so that both particles freeze out
of equilibrium at approximately the same temperature.
Thus the NLSP should be included in principle in the 
reaction network, since 
coannihilation channels NLSP-LSP (and also channels NLSP-NLSP) might be 
now relevant \cite{Seckel}.
In fact, this is so only
when 
$(m_{\mbox{\tiny NLSP}}-m_{\tilde{\chi}_1^0})/m_{\tilde{\chi}_1^0}\lsim 0.2$,
since the NLSP number density is suppressed by 
$e^{-(m_{\mbox{\tiny NLSP}}-m_{\tilde{\chi}_1^0})/T_F}$
relative to the neutralino number density,
where recall that
$T_F\approx m_{\tilde{\chi}_1^0}/20$.
We will see in Section~5 that 
special regions of the parameter space of the MSSM
fulfil this condition, and therefore including coannihilations
is important.
Coannihilation  
channels in SUSY,
in particular 
$\tilde{\chi}_1^0$-stau, 
$\tilde{\chi}_1^0$-chargino and
$\tilde{\chi}_1^0$-stop,
have been exhaustively studied in the 
literature \cite{Mizuta,Yamaguchi,stau,char,stop,jeong,Profumo}.

Taking into account all the above comments concerning
SUSY scenarios, we will see in detail in 
Section~5 that 
significant regions of the parameter space of the MSSM
produce values of the cross section 
in the interesting range mentioned below Eq.~(\ref{relicdensity}).


On the other hand, since neutralinos, or WIMPs 
in general, interact with ordinary matter with
roughly weak strength, their presence in galactic scales, and in particular
in our galaxy, raises
the hope of detecting relic WIMPs directly in a detector
by observing their
scattering on target nuclei through nuclear recoils.  
This will be the subject of the next section, but
before analyzing it let us briefly discuss here other
candidates for dark matter.

\subsection{Other candidates}

The first proposed candidate for dark matter in the context of WIMPs 
was in fact a {\it heavy} (Dirac or Majorana) 
fourth-generation
{\it neutrino} \cite{fourth} with the usual standard model
couplings. 
It is not easy to construct theoretical 
models with such a stable heavy particle, but in any case
LEP limits from the $Z^0$ width imply
$m_{\nu}\gsim m_{Z^0}/2$, and this is not compatible
with a correct relic abundance. The relic abundance can 
be studied as in Subsection~3.1.3 with 
the annihilation of the heavy neutrinos 
into light fermions through the
exchange of $Z^0$. The final result is that
neutrinos with the above lower bound
give rise to a too small relic abundance \cite{fourth,malam}.
In addition, direct \cite{Gedetectors} and  indirect \cite{kamioka} detection
experiments rule out 
$m_{\nu}\lsim 1$ TeV.
Attempts to consider still massive fourth-generation neutrinos
as dark matter candidates, can be found in Ref.~\refcite{volovik}.

In addition to the neutralino there are other potential SUSY candidates
for dark matter. In principle one of them might be the 
{\it sneutrino} \cite{iba} of the MSSM.
One finds \cite{fom} that the sneutrino relic density is in the region
$0.1\lsim \Omega_{\tilde{\nu}}h^2\lsim 1.0$
for 550 GeV$\lsim m_{\tilde{\nu}}\lsim$ 2300 GeV.
This is consistent with the LEP limits on
$Z^0\to$ invisible neutral particles suggesting
$m_{\tilde{\nu}}\gsim m_{Z^0}/2$, as in the case of the heavy neutrino above.
However, sneutrino-nucleus interaction is similar to the
heavy neutrino-nucleus interaction, and therefore
direct detection experiments \cite{Gedetectors} impose similar limits 
on the sneutrino mass as above.
Such a heavy sneutrino cannot be the LSP in SUSY models.

The supersymmetric partner of the graviton, the {\it gravitino},
has also been proposed as a candidate for dark matter \cite{gravitino}.
In the absence of inflation, it could give rise to the
correct relic abundance if its mass is of order keV.
This is unlikely in specific theoretical models. For example,
in gravity mediated SUSY breaking the masses of the superpartners
are of order the gravitino mass, and therefore this should
be of order 1 TeV.
Gravitinos as dark matter would be undetectable since their
interactions with ordinary matter are extremely weak.

Another candidate for dark matter 
that can arise in SUSY is the {\it axino} \cite{axino}.
Models combining SUSY and the Peccei-Quinn solution to the strong
CP problem necessarily contain this particle.
Unlike the neutralino and gravitino, the axino mass is
generically not of order the SUSY-breaking scale and can be much
smaller. For a recent work with masses in the range
of tens of MeV to several GeV as corresponding to cold axino relics
see e.g. Ref.~\refcite{axikim}.

The possibility that dark matter could be made of 
additional {\it scalars} in the standard model
has also been analyzed \cite{Burgess}. Scalar candidates
have also been studied in a prototypical theory space `Little
Higgs Model' \cite{little}, and in
theories with $N=2$ extended supersymmetry and/or extra space
dimensions \cite{Fayet}.

More (exotic) candidates can be found in the literature.
Let us mention some of them.
Strongly Interacting Massive Particles, {\it SIMPs},
have been proposed \cite{raros,tech,simps}.
For example, bound state of ordinary quarks and gluons
with a heavy stable quark, scalar quark, gluino, or colored
technibaryon.

Mirror matter is predicted to exist if Nature exhibits an exact
unbroken mirror symmetry, and then could be a candidate for dark 
matter \cite{Foot}.

Dark matter made of CHArged Massive Particles, {\it CHAMPs}, has also
been proposed \cite{champs}. Apparently,
if their masses are larger than 20 TeV these particles
will very rarely be trapped in ordinary matter, and are safe from
bounds on exotic heavy isotopes.

A potential conflict between
collisionless dark matter scenarios and observations
motivated the authors of Ref.~\refcite{sidm}
to propose Self Interacting Dark Matter, {\it SIDM},
i.e. dark matter with a large scattering cross section but negligible
annihilation.
A concrete realization of this idea, using a scalar gauge singlet, can
be
found in Ref..~\refcite{orfeu}.
On the other hand, the authors of Ref.~\refcite{jean}
solved the problem using a mechanism of non-thermal production
of WIMPs.

Instead of using WIMPs with typical masses of order a hundred GeV,
the authors of Ref.~\refcite{wimpzillas}
studied scenarios with nonthermal WIMPs in the range
$10^{12}$ to $10^{16}$ GeV. They called these objects
{\it WIMPZILLAS}.

The lightest Kaluza-Klein particle, {\it LKP}, in models 
with TeV extra dimensions has been studied as a viable dark
matter candidate \cite{KK}. 
It is actually a typical WIMP (the most studied possibility is
a Kaluza-Klein 'photon'), with a mass in the range 400-1200 GeV.

Superweakly-interacting massive particles, 
{\it superWIMPs}, were also proposed in Ref.~\refcite{superwimps}.
They naturally inherit the desired relic density from late
decays of metastable WIMPs. Well-motivated examples
are weak-scale gravitinos in supergravity and Kaluza-Klein gravitons
from extra dimensions.

In the context of the heterotic string $E_8\times E_8$, 
where there is a natural hidden sector associated with the second $E_8$,
it was mentioned in Ref.~\refcite{heterotic} 
that this sector which only interacts with ordinary
matter through gravitational interactions
could be a {\it shadow world}. The astrophysical and cosmological 
implications of this comment were analyzed in detail in 
Ref.~\refcite{shadow}

Still, in the context of strings, other possibilities were also
analyzed. For example, in Ref.~\refcite{cryptons} the authors
proposed {\it Cryptons}, which are stable bound states of matter
in the hidden sector. In Ref.~\refcite{sidmstrings}
glueballs of the {\it hidden-sector} non-Abelian gauge group were
proposed as candidates for SIDM.
{\it Brane-world} dark matter was studied in Ref.~\refcite{brane}.

Let us finally mention that the list of dark matter candidates discussed
here is large but by no means complete. 
The reason being that
different candidates have been
proposed during many years, and therefore 
it is almost impossible not to forget some of them.
In fact, this list is still increasing (see e.g. the works
in Ref.~\refcite{more}).

\section{Dark matter detection}

Given the discussion in the previous section, one can say that
there are good particle candidates for dark matter. 
As a matter of fact, we saw that 
WIMPs and axions are particularly interesting. 
Since the former can be left over from the Big Bang in sufficient
number to account for the relic matter density in a natural way,
we will concentrate on them.
Thus we will review in this section the current and projected
experiments for detecting WIMPs 
(e.g. the detection of axions was briefly discussed in Section~3.1.2).
In particular we will analyze the two possible ways of detecting them:
direct and indirect detection.
Of course, the most clear one is the direct detection 
by observing their scattering on the material in a detector,
and we will study it in detail.
Finally, we will discuss briefly the indirect detection
by looking for evidence of their annihilation.


\subsection{Direct detection}

As discussed in the Introduction, 
if neutralinos, or WIMPs in 
general, are the bulk of the dark matter,
they will form not only a background density in the Universe, but also
will cluster gravitationally with ordinary stars in the galactic halos.
In particular they will be present in our own galaxy, the Milky Way.
This raises the hope of detecting relic WIMPs directly, by experiments
on the Earth. 
In particular,
through scattering with the material
in a detector.
In fact general studies of the possibility of dark matter
detection
began around 1982.
Since the detection will be on the Earth we need to know
the properties of our galaxy in order to be sure that
such a detection is feasible \footnote{Direct 
detection of extragalactic WIMPs has also been analyzed
recently \cite{gondo}.
Although of much lower flux than typical galactic halo WIMPs,
it seems that they have a number of interesting features for their 
detectability.}.

%
\begin{figure}[t]
\begin{center}
\begin{tabular}{c}
\hspace{-2.0cm}\epsfig{file= 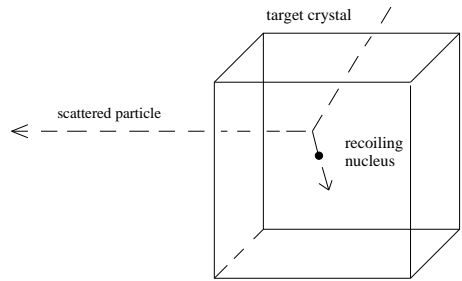,width=6cm,
}\\
\end{tabular}
\end{center} 
\caption{Elastic scattering of a dark matter particle with an
atomic nucleus in a detector.
} 
\label{scattering}
\end{figure}
As a matter of fact, rotation curves are much better known for external
galaxies than for ours, due to the position of the Earth inside the galaxy.
In any case, analyses have been carried out with the conclusion
that indeed the Milky Way contains large amounts of dark 
matter \cite{ourgalaxy}.
Besides, 
some observational evidence seems to point at
a roughly spherical distribution of dark matter in the galaxy.
At the position of the Sun, around 8.5 kpc away from the galactic center,
the mean density of elementary particles trapped in the gravitational 
potential well of the galaxy is expected to be 
$
\rho_0
\approx 5\times 10^{-25}$ gr\ cm$^{-3} \simeq 0.3$ GeV\ cm$^{-3}$. 
For WIMPs with masses about 100 GeV this means a number
density 
$
n_0
\approx 3\times 10^{-3}$ cm$^{-3}$. 
In addition,
their velocity will be similar to the one of the Sun since they
move in the same gravitational potential well,
$v_0\approx 220$ km\ s$^{-1}$, implying a flux of dark matter particles
$
J_0
=n_0\ v_0\approx 10^{5}$ cm$^{-2}$ s$^{-1}$
incident on the 
Earth\footnote{Clumps of dark matter crossing the Earth 
with a density
larger by a factor $10^8$, and
with a periodicity of 30-100 Myrs, 
have been proposed to explain double mass extinctions of paleontology.
First, the direct passage of a dark matter clump may lead to 
a preliminary extinction \cite{Collar} by causing lethal carcinogenesis
in organisms \cite{Zioutas}. Later, 
the accumulation of dark matter in the
center of the Earth and the subsequent annihilation would produce
a large amount of heat with the consequent ejection of superplumes,
followed by massive volcanism leading to the second burst of
more severe extinction \cite{Abbas}.
For another exotic mechanism 
solving mass extinctions, and related to dark matter, see
Ref.~\refcite{Sila} where mirror matter is used.}.
Although this number is apparently large, the fact that WIMPs
interact weakly with matter makes their detection very 
difficult. Most of them will pass through matter without prevention.
In any case, as suggested 
in 1985 \cite{Witten,Wasserman}, 
direct experimental detection of WIMPs is in principle possible.
The two types of direct detection experiments will be discussed below.

\vspace{0.2cm}

\noindent {\it (i) Elastic scattering}

\vspace{0.2cm}

\noindent The detection of WIMPs 
through elastic scattering with nuclei
in a detector is shown schematically in Fig.~\ref{scattering}.
As we can see 
the nucleus recoils as a whole. 
A very rough estimate of the rate $R$ in a detector is the following.
A particle with a mass of order 100 GeV and electroweak interactions
will have a cross section 
$\sigma\approx 1$ pb,
as discussed in Subsection~3.1.3.
Thus for a material with nuclei composed of about 100 nucleons,
i.e. $M_{N}\sim 100$ GeV $= 177\times 10^{-27}$ kg, one obtains
$R\sim J_0\ \sigma/M_N \approx 10$ 
events\ kg$^{-1}$\ yr$^{-1}$.
This means that every day a few WIMPs, the precise number
depending on the number of kilograms of material, 
will hit an atomic nucleus in a detector.

\begin{figure}[t]
\begin{center}
\epsfig{file= 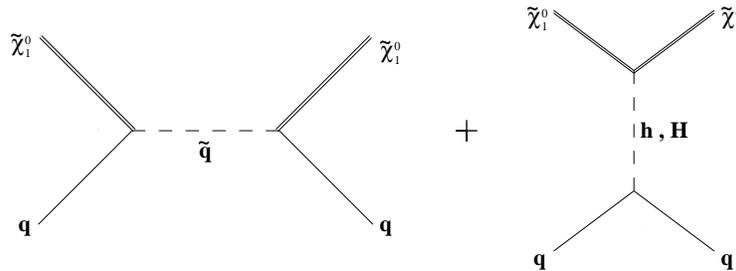,
height=3.5cm,angle=0}
\end{center}
\caption{Feynman diagrams contributing to neutralino-nucleon cross 
section through 
squark ($\tilde q$) exchange and 
CP-even light ($h$) and heavy
($H$) neutral Higgs 
exchange.}
\label{Feynman}
\end{figure}

Of course the above computation is just an estimate and one should
take into account in the exact computation the interactions of WIMPs
with quarks and gluons, the translation of these
into interactions with nucleons, and finally the translation of
the latter into interactions with
nuclei. 
In the case of neutralinos as WIMPs, diagrams
contributing to neutralino-quark cross section are shown
in 
Fig.~\ref{Feynman}. 
The relevant (scalar) $\tilde{\chi}^0_1$-nucleus 
cross section is of the form
\begin{equation}
\sigma_{scat}
\simeq  M_r^2 |{\cal A}_{scat}|^2\ ,
\label{sigmscat}
\end{equation}
where 
$M_{r}=M_{N} m_{\tilde{\chi}_1^0}/(M_{N}+m_{\tilde{\chi}_1^0})$
is the reduced mass with $M_{N}$ the mass of the nucleus,
and ${\cal A}_{scat}$ is the amplitude which depends on the
dynamics of the collision. 
In particular, 
the quarks masses $m_q$, the 
hadronic matrix elements
$f^{(p)}_{T_q}$, the proton mass $m_p$,
and the masses of the particles mediating the interaction,
such as $m_{\tilde q}$, $m_H$, $m_h$, enter in ${\cal A}_{scat}$.
We will check in Section~5 that significant 
regions of the parameter space of the MSSM
produce values of the neutralino-nucleus cross section 
$\sigma_{scat}\simeq 1$ pb, and therefore giving rise to
a reasonable number of events.
As a matter of fact, in the experimental results that one
finds in the literature the authors prefer to give the 
WIMP-nucleon cross section. As will be discussed in
Section~5, this is about eight orders of magnitude smaller
than the WIMP-nucleus cross section, and therefore a typical
value is $\approx 10^{-8}$ pb. In the next Sections, when talking
about scattering cross section, we will always consider this one.

Let us finally remark that
the diagrams for neutralino annihilation (see Fig.~\ref{annihilation}) 
are related to these
by crossing symmetry. Thus,
provided that the main annihilation channel is into fermions,
the  amplitudes of annihilation and scattering with nucleons are
similar, and this leads 
for the amplitudes with the nucleus to the relation 
%
$|{\cal A}_{scat}|^2\simeq c^2 A^2 |{\cal A}_{ann}|^2$,  
where $A$ is the atomic weight and $c^2$ is a 
constant (we can deduce from Section~5 that $c\simeq f^{(p)}_{T_s}
m_p/m_s={\cal O}(1)$).
From Eqs.~(\ref{sigmann})
and (\ref{sigmscat}) it is obvious that
${\sigma_{scat}}/{\sigma_{ann}} \simeq  const.   $
%
However, if the neutralinos are heavy they have other annihilation
channels,
such as Higgs bosons or vector boson pairs, and therefore the 
crossing argument does not apply.

\vspace{0.5cm}

\noindent {\it (ii) Inelastic scattering}

\vspace{0.5cm}

\noindent There are two possible ways of detecting WIMPs directly
through inelastic scattering.
These can be with nuclei in a detector \cite{excited} or
with orbital electrons \cite{electrons}.
The former produces an
excited nuclear state, and has the double signature of recoil energy
followed $\sim 1$ ns later by the emission of a decay photon.
However, the event rates seem to be suppressed and
problems of natural radioactivity and expense disfavour the materials
that might be used for the detector\footnote{For a recent analysis
with a different conclusion, see Ref.~\refcite{quentin}.}.
A similar conclusion concerning the event rates was obtained for the
case
of scattering of WIMPs from orbital electrons. These could
leave an excited state emitting a photon by the deexcitation.

\subsubsection{Experiments around the world}

More than 20 experiments for the direct detection of
dark matter are running or in preparation around the world.
Given the above results,
all of them use the elastic-scattering technique.
For example, Germanium is a very pure material and has been
used for many years for detecting dark matter in this way.
In this type of experiments, 
in order to detect the nuclear recoil energy, they
measure the ionization produced by collision with electrons.
In fact, $^{76}$Ge ionization detectors has been applied to WIMP searches
since 1987 \cite{Gedetectors}. In 2000 the situation was the following.
The best combination of data from these experiments, together with
the last data from the Heidelberg-Moscow 
\cite{HeidelbergMoscow} 
and IGEX experiments 
\cite{IGEX} 
located at
the Gran Sasso (L'Aquila, Italy) and Canfranc (Huesca, Spain) 
Underground Laboratories, respectively, 
were able to exclude
a WIMP-nucleon cross section larger than about $10^{-5}$ pb 
for masses of WIMPs $\sim 100$ GeV, due to the negative search result.
Although this was a very interesting bound, it was still well above the
expected value $\sim 10^{-8}$ pb.

Let us remark that it is 
convenient to carry the experiments out in the deep underground.
For a 
slow moving ($\sim$ 300 km\ s$^{-1}$) and heavy
($\sim 100-1000$ GeV) particle forming the dark matter halo, 
the kinetic energy is very small, around
100 keV, and in fact
the largest recoil energy transfer to a nucleus in the detector
will only be a few keV.   
Since cosmic rays with energies $\sim$ keV-MeV 
bombard the surface of the Earth, the experiments must have an
extremely good background discrimination.
In particular, neutrons coming from collisions between
cosmic-ray muons and nuclei produce nuclear recoils similar to
those expected from WIMPs  at a rate
$\sim$ 10$^{3}$ events kg$^{-1}$ day$^{-1}$.
Thus detectors
located 
in the deep underground,
reduce the background 
by orders of magnitude 
in 
comparison with the sea level intensity.

In fact, this is still not enough since  
the detector has to be protected also 
against the natural radioactivity from the
surroundings (e.g. the rocks) and the materials of the detector itself.
This produces again neutrons but also X rays, gamma rays and beta rays giving
rise to electron recoils.
The latter may be a problem for detectors based only on
ionization or scintillation light since nuclear recoils with energies
of a few keV are much less efficient in ionizing or
giving light than electrons of the same energy.
Various protections aim to reduce these backgrounds. 
In particular, low radioactive materials, such as e.g.
high-purity copper or aged lead, are used for the shielding.
In addition, high-purity materials for the detector are also used.

\begin{figure}[t]
\begin{center}
\begin{tabular}{c}
\epsfig{file= 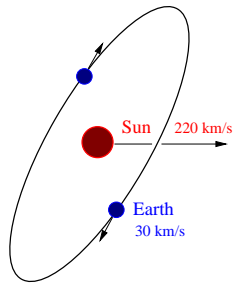,width=3cm
}\\
\end{tabular}
\end{center} 
\caption{
Earth's motion around the Sun.}
\label{sun} 
\end{figure}

Summarizing, with this type of experiments
the WIMP nuclear recoil 
signal will appear as an excess of recoil events above 
the expected 
background rate. However, 
it would be very interesting
to also look for some additional 
feature of the WIMP signal that positively identifies it as galactic in origin.
In this sense
a different method for discriminating a dark matter signal from
background is the so-called
annual modulation signature \cite{anual}$^,$\footnote{Another novel
possibility to search for characteristic signatures has recently been
proposed, 
namely by observing the electrons which follow the ionization of the
atom during the WIMP-nucleus collision \cite{verga3}.}. 
As it is shown schematically 
in Fig.~\ref{sun}, as the Sun orbits the galaxy with velocity
$v_0\approx 220$ km\ s$^{-1}$, the Earth orbits the Sun with
velocity $\approx 30$ km\ s$^{-1}$ and with the orbital plane
inclined at an angle of $60^{\circ}$ to the galactic plane.
Thus e.g. in June the
Earth's rotation velocity adds to the Sun's velocity through the
halo with a maximum around June 2, 
whereas in December the two velocities are in opposite
directions.
When this is taken into account the Earth velocity is given by
\begin{equation}
v_E=v_0 \left\{1.05 + 0.07\cos\left[\frac{2\pi(t -
      t_m)}{1\ year}\right]\right\}\ ,
\label{Earthvelocity}
\end{equation}
where $t_m$= June 2$\pm$ 1.3 days.
This fluctuation produces a rate variation 
between the two extreme conditions. The variation is
so small $\approx 7\%$ that the experiment
can only work if large number of events are found, implying that large
mass apparata are necessary.

\begin{figure}[t]
\centering
\vspace{0.1cm}
\includegraphics[height=10cm]{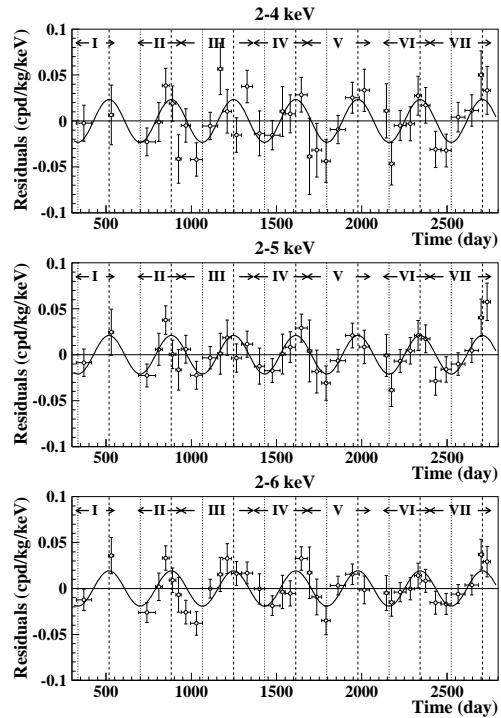}

\caption{
Model independent residual rate for events in the (2--4), (2--5) 
and (2--6) keV energy intervals as a function of the time elapsed since
January 1-st of the first year of data taking. The
experimental points present the errors as vertical
bars and the associated time bin width as horizontal bars. The 
superimposed curves represent the cosinusoidal functions 
behaviours expected for a WIMP signal 
with a period equal to 1 year and phase at $2^{nd}$ June.
}
\label{rate}
\end{figure}

\begin{figure}[t]
\begin{center}
\epsfig{file=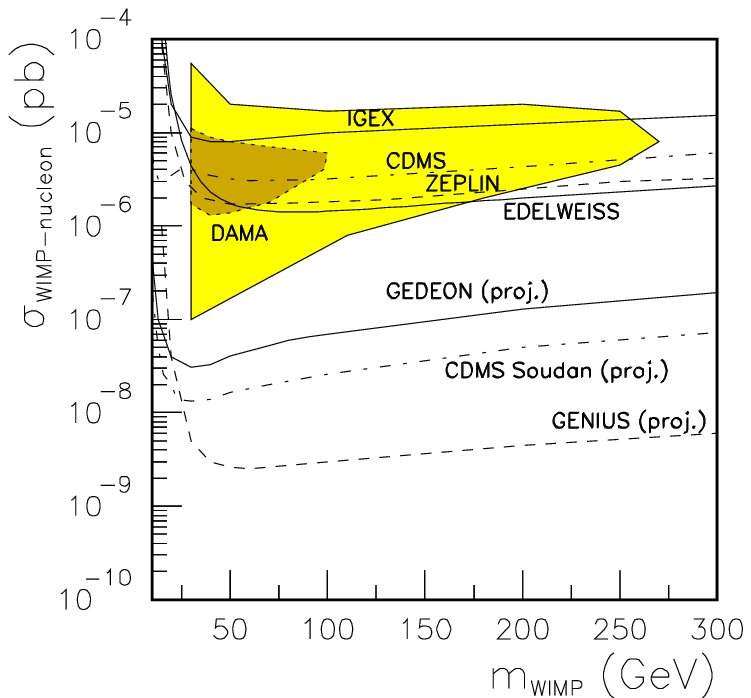,width=7cm}
\end{center}
\caption{
Areas allowed by the different experiments for the direct 
detection of dark matter in the parameter
space 
($\sigma_{\mbox{\tiny WIMP-nucleon}}$, $m_{\mbox{\tiny WIMP}}$), where
$\sigma_{\mbox{\tiny WIMP-nucleon}}$ is the
WIMP-nucleon cross section and
$m_{\mbox{\tiny WIMP}}$ is the WIMP mass.
The light shaded (yellow) area is allowed by the DAMA 
experiment, including uncertainties on the halo model.
The dark shaded (brown) area inside the previous one corresponds to the simple
case of an isothermal sphere halo model.
This case is also the one analyzed by the other experiments: 
The (lower) areas
bounded by solid, dot-dashed, dashed and (again) solid lines are
allowed by IGEX, CDMS, ZEPLIN I 
and EDELWEISS  
current experimental
limits.
The (upper) areas bounded by solid, dot-dashed and dashed lines 
will be analyzed by the 
projected GEDEON, CDMS Soudan and GENIUS experiments.
\label{limits}}
\end{figure}

\begin{figure}[t]
\centering
\includegraphics[height=4.7cm]{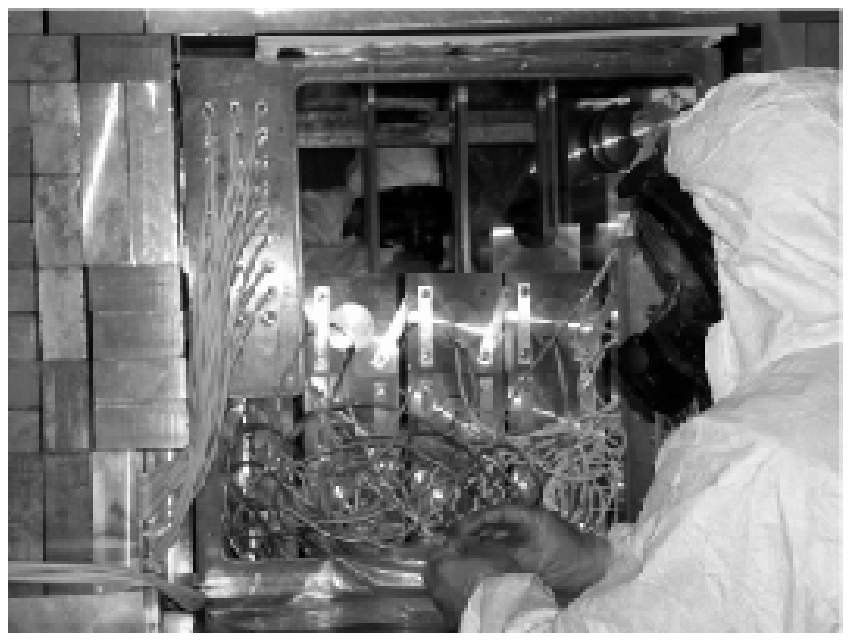}
\includegraphics[height=4.7cm]{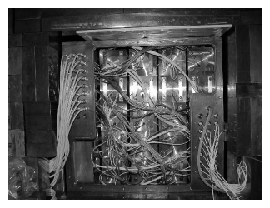}
\caption{Left picture: during the LIBRA detectors installation
in HP Nitrogen atmosphere.
Right picture: view at end of the detectors installation.
All the used materials have been deeply selected for radiopurity 
(see for example the cables with teflon envelop).}
\label{fg:fig11}
\end{figure}

The DArk MAtter (DAMA) 
experiment 
\cite{experimento1,masdama} 
investigates the annual modulation
of this signature rather than discriminating
signal events against background. It 
consists of 
about 100 kg of material in a small room 
at the Gran Sasso National Laboratory located 
beside the Gran Sasso Tunnel on the highway connecting Teramo to Rome
(see Fig.~\ref{fg:fig11} for a similar experiment DAMA/LIBRA).
The 
maximum thickness of the rock overburden is 1400 m.
In the experiment they use nine 9.70 kg NaI crystal scintillators
which measure the ionization produced by the nuclear recoil
through the emission of photons.
Remarkably, they found that 
their detectors flashed more times in June than in December.
The data collected \cite{experimento1} 
over four yearly cycles, DAMA/NaI-1,2,3,4,
until the second half of August 1999,
strongly favour  
the presence of a yearly modulation of the rate of the
events.
Moreover, very recently, more data have been reported \cite{masdama}
confirming this result.
In particular
the DAMA/NaI-5 data have been collected 
from August 1999 to end of July 2000.
Afterwards, the DAMA/NaI-6 data have been collected from
November 2000 to end of July 2001, while
the DAMA/NaI-7 data have been collected from August 2001 to July 2002.
The analysis of the data of the seven annual cycles 
offers an immediate evidence of the presence of 
an annual modulation of the rate of the events
in the lowest energy region  
as shown in Fig.~\ref{rate} (from Ref.~\refcite{masdama}),
where the time behaviours of the (2--4), (2--5) and (2--6) keV
residual rates are
depicted.
This signal is compatible \cite{experimento1}
with WIMP masses up to 100 GeV
and WIMP-nucleon cross sections in the interval
$10^{-6}-10^{-5}$ pb,
as shown with a dark shaded area in Fig.~\ref{limits},
where the data \cite{nai0} from DAMA/NaI-0 have also been taken into
account.
The lower bound of 30 GeV in the figure 
is purely theoretical, and inspired by the
lower bound on the lightest neutralino of SUSY models, as derived
from the LEP data using GUT conditions.

It is worth remarking that this result has been obtained assuming the
simple isothermal sphere halo model with dark-matter density
$\rho_0 = 0.3$ GeV\ cm$^{-3}$ 
and WIMP velocity
$v_0= 220$ km\ s$^{-1}$ 
(to be precise one assumes a Maxwell-Boltzmann local
velocity distribution 
$f(v)\propto e^{-v^2/v_{0}^2}$ producing a velocity dispersion
$\bar v = <v^2>^{1/2}=(3/2)^{1/2}v_0\approx 270$ km s$^{-1}$).
However,
when uncertainties on the halo model
are taking into account,
the signal is consistent with a larger region of the
parameter space.
In particular, in Refs.~\refcite{experimento1} and \refcite{halo} 
modifications in the velocity distribution function
for different galactic halo models were considered, using
in addition the allowed ranges for $v_0$ and $\rho_0$ in each model
(see Ref.~\refcite{otros} for other 
studies of uncertainties on the halo model). 
The final result of the analyses is shown in Fig.~\ref{limits} 
with a light shaded area. One sees that 
the signal is compatible with larger values of the parameters 
(for a different opinion see Ref.~\refcite{different}),
i.e. WIMP masses up to 270 GeV,
and WIMP-nucleon cross sections in the interval
$10^{-7}- 6\times 10^{-5}$ pb.
This result corresponds at a rate of about 1 event per kg per day.
In fact, as discussed in 
Ref.~\refcite{halo} (see also Ref.~\refcite{masdama}),
when co-rotation of the galactic halo is also considered,
the mass range extends further to $500-900$ GeV,
for cross sections in the interval
few$\times 10^{-6}- 2\times 10^{-5}$ pb.


Although the DAMA group is confident about the data, 
since they claim
to have ruled out 
systematic effects which could fake
the signature \cite{fake,pictures}, as e.g. temperature changes,
it is worth remarking that 
the above values for the cross section are generically above the
expected weak-interaction value, 
and therefore they are not easy to obtain 
in 
SUSY models with
neutralino dark matter, as we will discuss in Section~5.
But in fact, 
the DAMA result is controversial, mainly because 
the negative search result obtained by other recent experiments.
The first of these was
the Cryogenic Dark Matter Search
(CDMS) experiment \cite{experimento2} in the US.
This was located just 10 metres below ground at Stanford University
in California,
and therefore it must discriminate 
WIMPs signals against interactions of background particles.
Two detection techniques are used for
this discrimination, both the ionization and the temperature rise
produced during a recoil are measured.
The latter can be observed since the recoiling nucleus is stopped
within $10^{-7}-10^{-6}$ cm ($\sim 10^{-14}$ s) releasing
a spherical wave of phonons traveling at $\sim 5\times 10^{5}$ cm
s$^{-1}$,
and subsequently converted to a thermal distribution. 
These two techniques allow to discriminate electron recoils 
caused by interactions of background particles from
WIMP-induced nuclear recoils.
The ratio of deposited energies heat/ionization would be $\sim 2-3$
for the former and larger than 10 for the latter.
However, although neutrons are moderated by a 
25-cm thickness of polyethylene between the outer lead shield and
cryostat, an unknown fraction of them still survives.
Two data sets were used in this analysis: one consisting of 33 live days
taken with a 100-g Si detector between April and July 1998,
and another consisting of 96 live days taken with three 165-g Ge detectors
between November 1998 and September 1999.
Although four nuclear recoils are observed in the Si data set, they
cannot
be due to WIMPs, they are due in fact to the unvetoed neutron
background.
On the other hand, in the Ge detector thirteen nuclear recoils are 
observed in the 10.6 kg per day exposure between
10 and 100 keV, which is a similar rate to that expected
from the WIMP signal claimed by DAMA.
However, 
the CDMS group concludes that
they are also due to neutrons.
These data exclude (see also Ref.~\refcite{moredata} for more recent data) 
much of the region allowed by the DAMA annual
modulation signal\footnote{For some attempts to show that
both experiments, DAMA and CDMS, might not be in conflict, see
Ref.~\refcite{conflict}.}, as shown in 
Fig.~\ref{limits} \footnote{
As mentioned in the caption of Fig.~\ref{limits}, unlike DAMA
for CDMS and for the other experiments analyses
taking into account the uncertainties in the galactic halo are not shown 
and we only see the effect of the
standard halo model on their results. Including
those uncertainties, the light shaded area favoured by DAMA 
and not excluded
by the null searches would be in principle smaller than the one
shown here (for an analysis of this see Ref.~\refcite{KKrauss}).}.

In addition, a small part of the region excluded by CDMS
has also been excluded by IGEX \cite{IGEX2}, as shown in 
Fig.~\ref{limits}. To obtain this result, 40 days of data
from the IGEX detector were analyzed.
Let us recall that this experiment is located
at the Canfranc Tunnel Astroparticle Laboratory
in the
international railway tunnel of Somport at Canfranc (Spanish pyrenees),
under a 860 m rock overburden,
and that consists of 
2 kg Germanium ionization detector.
A region similar to the one excluded by IGEX has also been
excluded by 
the Heidelberg Dark Matter Search (HDMS) experiment 
\cite{HDMS}.
This operates two ionization Ge detectors in
a unique configuration, and was installed
at Gran Sasso in August 2000.
The data used for the analysis were taken from
February 2001 to September 2001.

But more disturbing 
are the recent results from 
EDELWEISS and ZEPLIN I collaborations
excluding even larger regions than CDMS, as shown also in Fig.~\ref{limits}.
The EDELWEISS experiment \cite{edelweiss} is located
at the Frejus Underground Laboratory
in the Fr\'ejus Tunnel
under the French-Italians Alps, under a 1780 m rock overburden.
As CDMS, this experiment also uses a heat-and-ionization
cryogenic Ge detector, although they differ by their
mass, geometry and electrode implantation scheme.
First, this collaboration was using a 320 g detector, 70 mm
in diameter and 20 mm in height, but this was not sufficient
to extend the sensitivity to the values obtained by DAMA. 
Following this result, three new detectors with the same
characteristics as the first one were put in operation
simultaneously.
The data consists of all physics runs recorded over a period
from February to May 2002.

On the other hand, the ZEPLIN project \cite{zeplin1} at the
Boulby salt mine (Yorkshire, UK) 
consists of a series of liquid Xenon detectors
operating some 1100 m underground, where the
nuclear recoil produces both an ionization and a scintillation signal,
potentially capable of providing much greater signal 
discrimination than NaI.
The ZEPLIN I detector
has been running for approximately one year with
a mass of about 4 kg. 
The collaboration 
is concentrating on new developments for future ZEPLIN detectors.


\subsubsection{Future dark matter searches}

Owing to this controversy
between DAMA and the other experiments,
one cannot be sure whether or not the first direct evidence for the existence
of dark matter has already been observed.
Fortunately, the complete DAMA region will be tested by current dark matter
detectors.
This is for example the case of 
the IGEX experiment mentioned above,
with a moderate improvement of the detector
performances \cite{IGEX2}.
In addition, a new experimental project,
GErmanium DEtectors in ONe cryostat (GEDEON), is planned \cite{IGEX3}.
It will use the technology developed for the IGEX experiment
and it would consist of a set of $\sim 1$ kg Germanium crystals,
of a total mass of about 28 kg, placed together in a compact
structure inside one only cryostat. 
GEDEON would be massive enough to search for the
annual modulation effect and explore positively a WIMP-nucleon 
cross section
$\sigma \gsim 3\times 10^{-8}$ pb (see Fig.~\ref{limits}).
Likewise an improved HDMS experiment  
should be able to test the DAMA evidence region within about two 
years \cite{HDMS}.

DAMA and CDMS collaborations also plan to expand their
experiments.
The DAMA collaboration 
dismounted the 100 kg NaI set-up and installed
the new LIBRA set-up. The latter consisting
of about 250 kg of NaI made of 25 detectors, 9.70 kg each one.
This will make the 
experiment more sensitive to the annual modulation signal. 
Some pictures taken during the installation can be seen 
in Fig.~\ref{fg:fig11} (from Ref.~\refcite{pictures}). 
This was completed at the end of 2002.
On the other hand, the CDMS collaboration
is moving its detector 
to the abandoned Soudan mine in Minnesota (approximately 700 metres
below ground), increasing also  
the mass of its 
Ge/Si targets to about 10 kg by 2006.
This experiment, CDMS Soudan, will be able to test a WIMP-nucleon cross section
$\sigma \gsim 10^{-8}$ pb (see Fig.~\ref{limits}).






But this is not the end of the story, since
a new generation of very sensitive 
experiments have been proposed
all over the world. For instance, only in the Gran Sasso laboratory 
there will be five experiments searching for WIMPs.
Apart from the two already discussed DAMA and HDMS, 
there are three other experiments in prospect, 
CRESST, CUORE and GENIUS. 
For example, 
the Cryogenic Rare Event Search using Superconducting Thermometers (CRESST)
experiment \cite{CRESST}
measures simultaneously phonons and
scintillation light distinguishing the nuclear recoils
from the electron recoils cause by background radioactivity.
In contrast to other experiments, CRESST detectors allow the
employment of a large variety of target materials,
such as e.g. sapphire or tungsten. This allows a better sensitivity
for detecting the WIMPs. Although they are already using
sapphire, for the next project, CRESST II,
it is planned to install a mass of about 10 kg, 
consisting of 33 modules of 300 g tungsten crystals.
With 3 years of measurement time
the experiment
will be able to test a WIMP-nucleon cross section slightly smaller than 
the CDMS Soudan discussed above.

\begin{figure}[t]
\centering
\includegraphics*[width=75mm, height=60mm]
{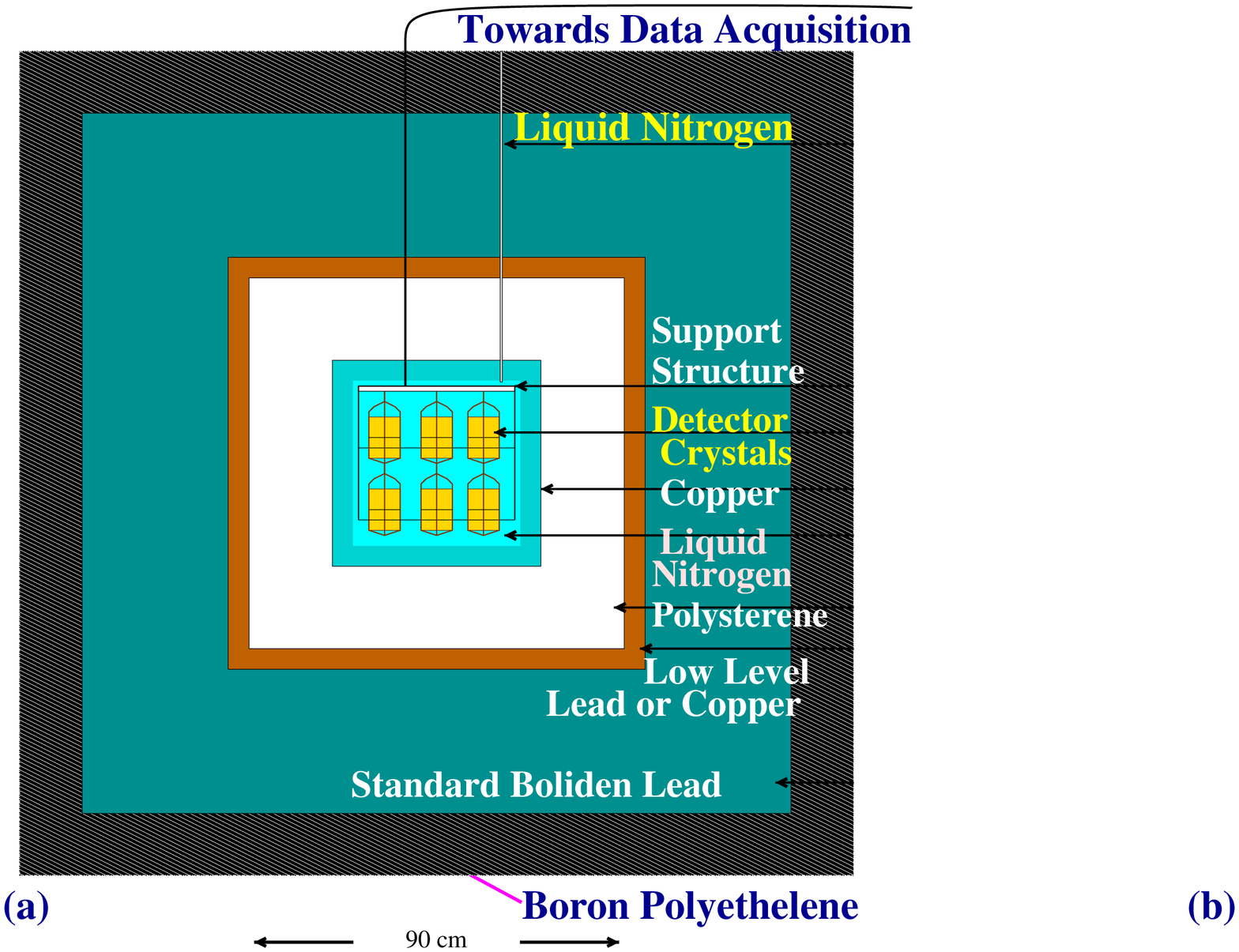}
\hspace{-0.3cm}
\includegraphics*[scale=0.25]{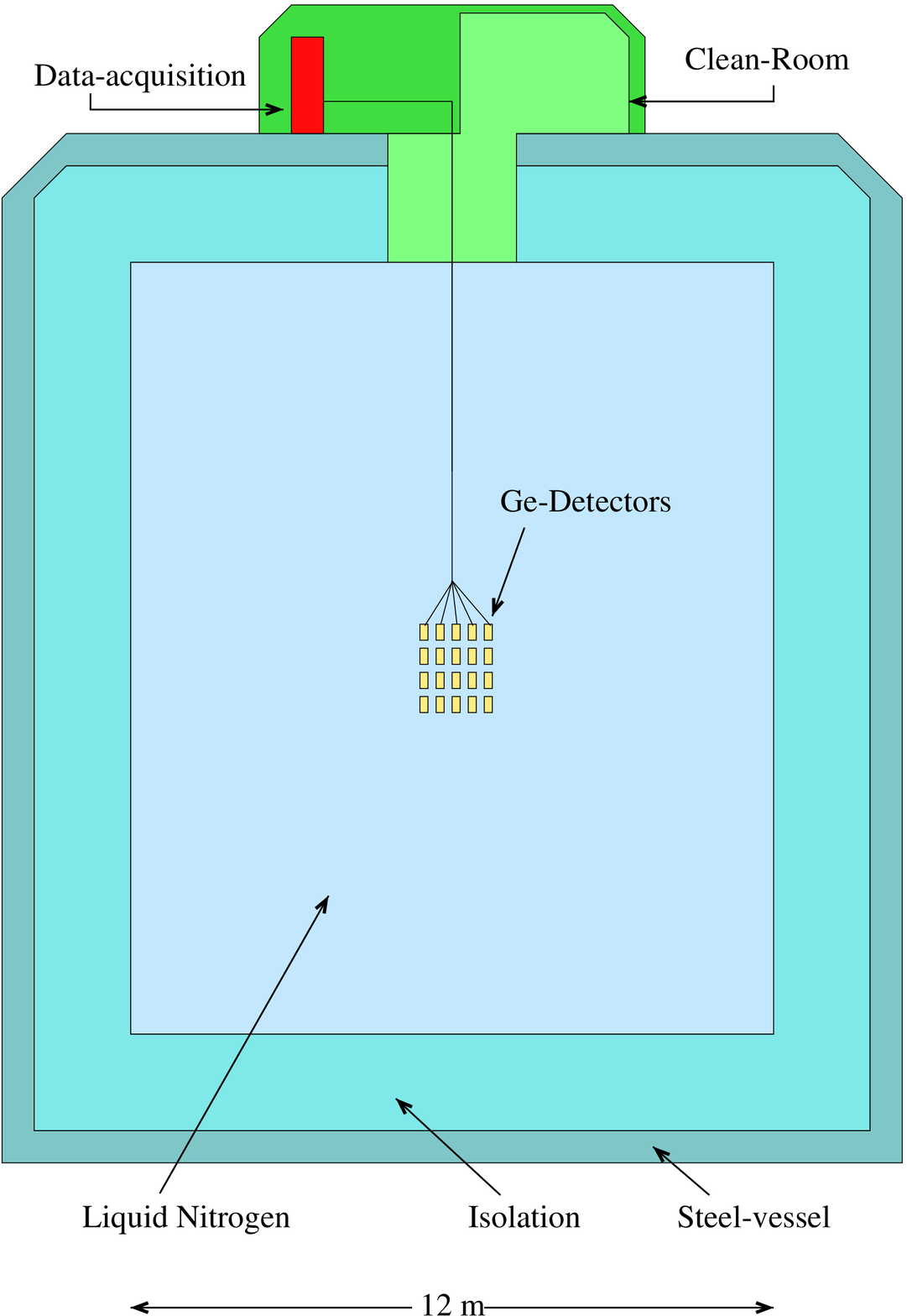}
\caption[]{
       (a): Conceptual design of the Genius TF. Up
   to 14 detectors will be housed in the inner detector chamber, 
   filled with liquid nitrogen. As a first shield 5 cm 
   of zone refined Germanium, or extremely low-level copper
   will be used.
   Behind the 20 cm of polystyrene isolation another 35 cm 
   of low level lead
   and a 15 cm borated polyethylene shield will complete the setup.
   (b): GENIUS - 100\,kg of Ge detectors 
   are suspended in a large liquid nitrogen tank.
\label{genius}}
\end{figure}

\begin{figure}[t]
\centering
\includegraphics*[scale=0.4]
{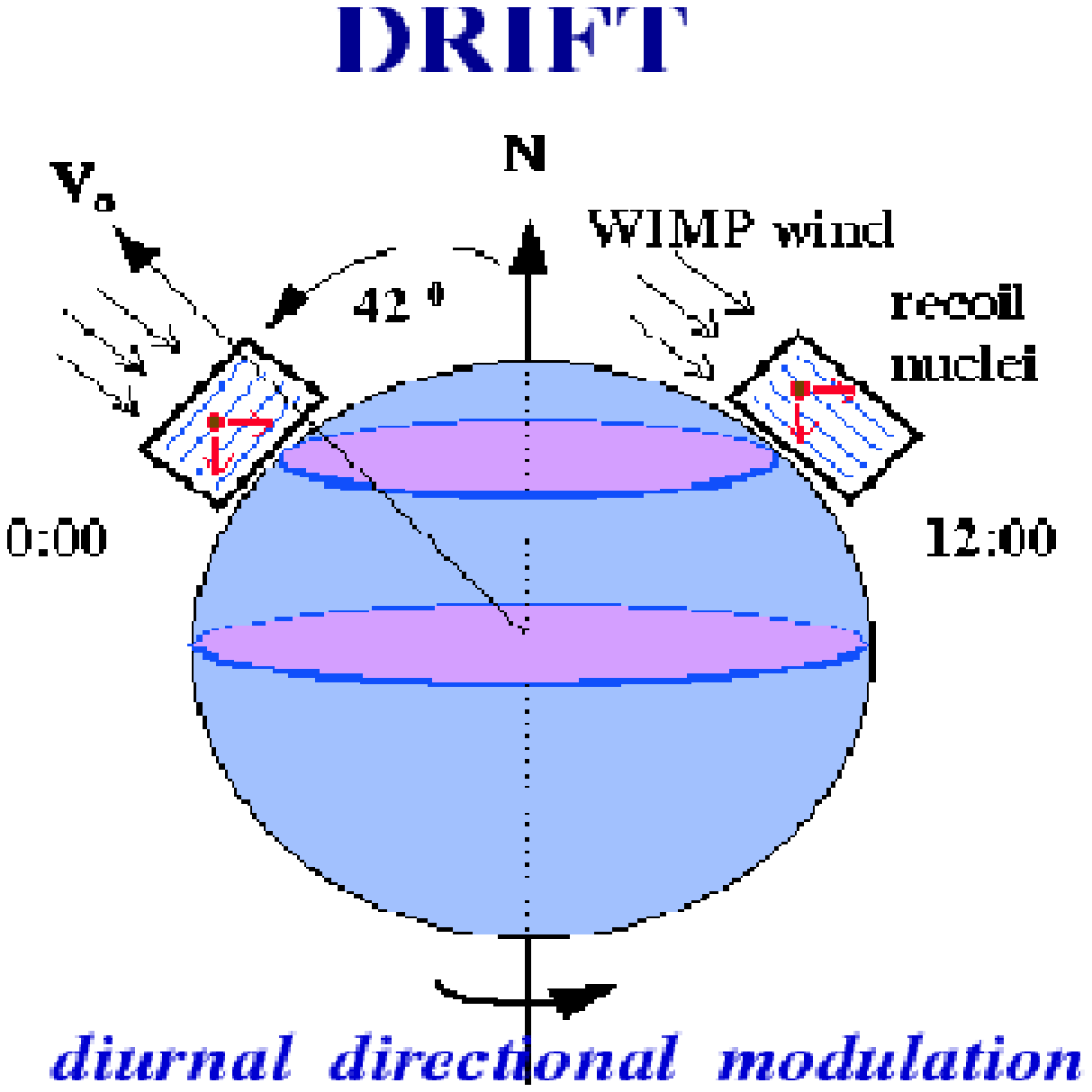}
\hspace{1cm}
\includegraphics*[scale=0.4]{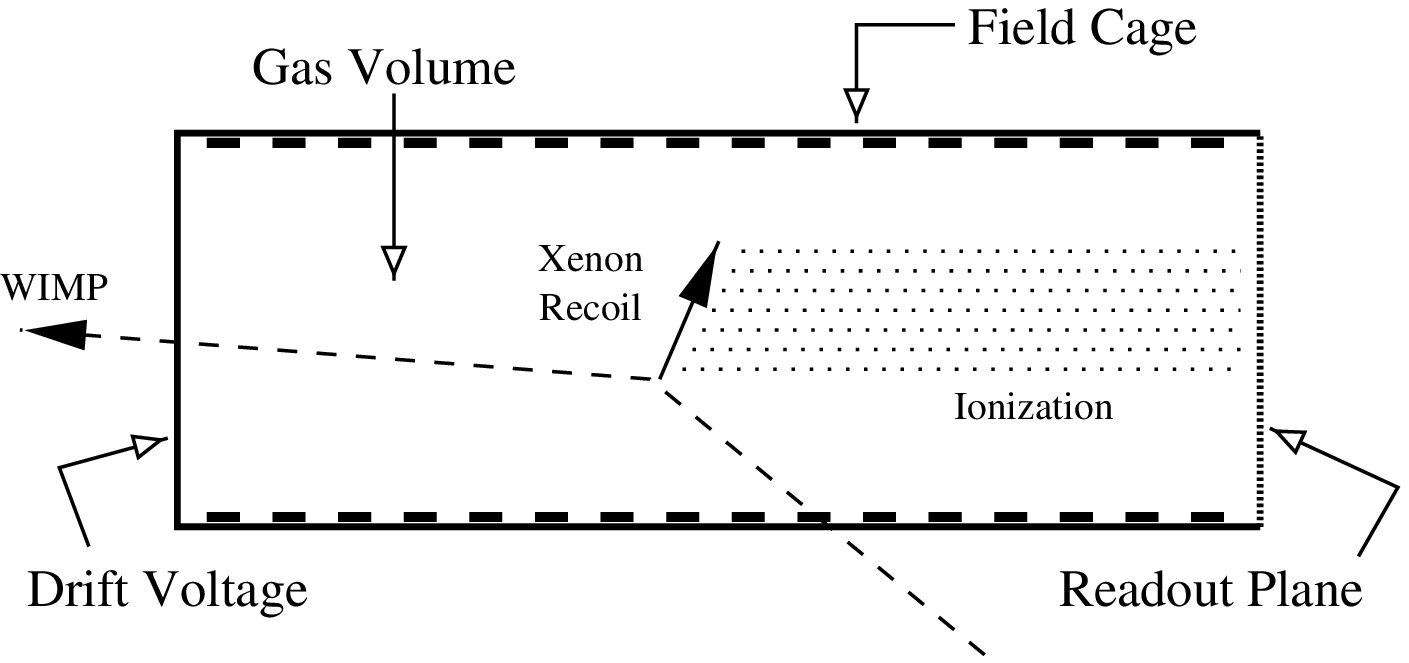}
\caption[]{
       Left picture: DRIF detector sited on the Earth through the WIMP halo.
       Right picture: Schematic of TPC 
\label{driftfigure}}
\end{figure}

The most sensitive 
detector will be the GErmanium in liquid NItrogen Underground Setup
(GENIUS) \cite{HDMS2} shown in Fig.~\ref{genius}b 
(from Ref.~\refcite{figuras}), 
which will be able to test a WIMP-nucleon cross section
as low as $\sigma\approx 10^{-9}$ pb (see Fig.~\ref{limits}). 
Indeed such a sensitivity covers a large range
of the parameter space of SUSY models with
neutralinos as dark matter, as we will see in the next 
section.
The GENIUS project 
is based on the idea to operate an array of 100 kg of Ge crystals directly in 
liquid nitrogen.
The latter,
which is very clean with respect to radiopurity, can act 
simultaneously 
as cooling medium and shield against external activities,
using a sufficiently large tank $\sim$ 12 metres in diameter at least.
It has been shown using Monte Carlo simulations
that with this 
approach the unwanted background is reduced by three to four orders 
of magnitude. 
In order to demonstrate the feasibility of the GENIUS project
a GENIUS Test-Facility (GENIUS-TF), shown in Fig.~\ref{genius}a,
is under installation since March 2001 with
40 kg of detectors \cite{HDMS3}.
It could test the DAMA evidence for dark matter by the
annual modulation signature (similarly to the
GEDEON project mentioned above) within about two years of measurement.

Other experiments are e.g.
the French project MACHe3 at the Joseph Fourier University,
the PICASSO project at the University of
Montreal, or the ORPHEUS project at the University of Bern,

Efforts to build detectors sensitive to the directional
dependence, i.e. recoil away from direction of Earth motion,
has also been carried out. 
This is an extension of the idea of annual modulation, but 
clearly more powerful \cite{Spergel,Vergados2}. 
A detector sited on the Earth can look for the signal 
as shown in the left frame of  
Fig.~\ref{driftfigure} (from Ref.~\refcite{figuras}).
The detector will see the mean recoil direction 
rotate and back again over one day.
Studies indicate that a WIMP signal could be identified with
high confidence from as few as 100 detected WIMP-nucleus
scattering events, compared with thousands for an annual modulation.

Of course, to reconstruct such a three-dimensional direction is not
simple, but a standard Time Projection Chamber (TPC)
where the arrival time of the ionization signal is measured 
may be used for it.
Although the energies of WIMP induced nuclear recoils are smaller
than 100 keV, implying short ranges for the ionization tracks,
this problem can be attacked using a gas such as Xenon or Argon
at low pressure 
to extend
these ranges to a few millimetres. 
The recoil energy and direction can be measured by drifting 
the ionisation created along the recoil track to a suitable 
readout plane, as shown in the right frame of  
Fig.~\ref{driftfigure} (from Ref.~\refcite{drift}). 
The low pressure and therefore the low
target mass is not a problem, since as mentioned above
only a small number of WIMP events are needed.
Another advantage of this technique 
is that its resolution should allow virtually all Compton electrons
to be rejected, since the electron range at a given total ionization
is predicted to be many times longer than for a Xenon recoil.
The Directional Recoil Identification From Tracks (DRIFT) project \cite{drift}
is precisely using this technique to detect dark matter.
For the moment they are using a detector, DRIFT I, 
consisting of 
Carbon Disulfide CS$_2$ with 1 m$^3$ volume. 
It was installed at the Boulby mine in summer 2001.
Although DRIFT-I is currently taking data, the next stage in the DRIFT 
project, DRIFT-II, is being planned. In order to achieve sensitivity to 
scattering cross sections below $10^{-6}$ pb, a scale up in mass is needed. 
DRIFT-II is thus proposed to have increased volume.



\vspace{0.2cm}

Let us finally mention that links to web pages
of the direct detection experiments mentioned in this section, and
others
experiments,
can be found in the
Dark Matter Portal 
http://lpsc.in2p3.fr/tep/fred/dm.html.
The same web page can be used to look for indirect detection experiments.
This is precisely the subject of our next discussion.

\subsection{Indirect detection}

There are promising methods for the indirect detection 
of WIMPs in the halo \cite{Wilczek,recent}. 
For example, WIMPs passing through the Sun and/or Earth 
may be slowed below escape velocity by elastic scattering.
From subsequent scattering they will sink
to the center and accumulate there,
annihilating with another WIMP into quarks and 
leptons (see Fig.~\ref{annihilation}), gauge
bosons, etc. 
Decays of these annihilation products will produce energetic neutrinos
that can be 
detected\footnote{Another proposal for studying the WIMPs accumulated
in the center of the Sun is through helioseismology \cite{silks}.}.
In particular, in underground, underwater and under-ice 
experiments through the `upgoing' muons produced 
by their interactions in the rock, water, and ice respectively.
Although some underground neutrino telescopes,
such as e.g. Kamiokande or MACRO,
have already reported null results,
in the next years underwater (e.g. NESTOR, ANTARES)
and under-ice (e.g. AMANDA) experiments
with sizes of about $10^{3,4}$ m$^2$ 
($10^6$ m$^2$ in the case ICECUBE, a proposed expansion of AMANDA)
will have a much greater sensitivity.

Another way of detecting WIMPs indirectly is through anomalous
cosmic rays produced by their annihilations in the galactic 
halo\footnote{Detection of extragalactic cosmic gamma rays
has also been analyzed recently \cite{extrag}.}.
Whereas positrons produced in this way 
are difficult to distinguish from the standard background
of typical cosmic-ray positrons, there are other products of
annihilation that can be distinguished.
In particular,
the most promising are cosmic-ray antiprotons and cosmic gamma rays.
As discussed above, WIMPs annihilate into quarks, leptons,
gauge bosons, etc. These will hadronize and produce, among other
products, antiprotons.
Unlike the usual cosmic-ray antiprotons, the flux of these
does not decrease dramatically at energies less than a GeV.
Space-based antimatter experiments, such as PAMELA and AMS,
will study this possibility.
On the other hand, two WIMPs can annihilate into gammma rays through
loop diagrams, and this might be 
detected
using
atmospheric Cherenkov telescopes (e.g. CELESTE, MAGIC)
or space-based $\gamma$ ray detectors (e.g. AMS, GLAST).
In particular through the observation of monochromatic
photons at an energy equal to the WIMP mass. No typical
cosmic gamma rays are monochromatic at energies
about 100 GeV. Another contribution of photons is the continuum
emission
from all gamma rays produced by the cascade decays of other 
primarily annihilation products (e.g. from the $\pi^0$
decay originated by the fragmentation of quarks).
Although the signal is less distinctive than the previous one, it has
the
advantage of being unsuppressed by loop effects, and therefore
large flux of photons might be observed\footnote{It is worth noticing that
the EGRET telescope has identified a gamma-ray
source at the galactic center that, apparently,
has no simple explanation with standard processes \cite{egret}.
Likewise, the HEAT balloon experiment has confirmed an excess
of cosmic ray positrons \cite{posi}.}.

Finally, it is worth noticing that in some regions of SUGRA scenarios, 
experiments such as
AMANDA, NESTOR, ANTARES, MAGIC, GLAST, AMS,
might be competitive with direct detection 
experiments
such as CDMS Soudan \cite{Wilczek}.

\section{Theoretical models}

As discussed in the Introduction and Subsection~3.1.3,
the leading candidate for WIMP is the 
neutralino,
a particle predicted by
SUSY extensions of the standard model.
In this section we will review the known SUSY scenarios based
on neutralinos as dark matter candidates, in the context of the
MSSM. In particular, we will discuss
how big the cross section for the direct detection of these neutralinos 
can be.
Obviously, this analysis is crucial in order to know the
possibility of detecting the dark matter 
in current and projected experiments (see Fig.~\ref{limits}).

Since 
in the MSSM
there are four neutralinos, $\tilde{\chi}^0_i~(i=1,2,3,4)$,
the lightest, $\tilde{\chi}^0_1$, will be the dark matter
candidate.
The neutralinos 
are the physical 
superpositions of the fermionic partners of the neutral electroweak 
gauge bosons, 
called bino ($\tilde{B}^0$) and wino ($\tilde{W}_3^0$), and of the
fermionic partners of the  
neutral Higgs bosons, called Higgsinos ($\tilde{H}^0_u$, 
$\tilde{H}_d^0$). 
The neutralino mass matrix 
with the conventions for gaugino and Higgsino masses in the Lagrangian,
${\cal L} =\frac{1}{2}\sum_{a} M_a {\lambda}_a \lambda_a + 
\mu \tilde{H}^0_u \tilde{H}^0_d$ + h.c.,
is given by
\begin{equation}
{
\small 
\bmat{cccc}M_1 & 0 &-M_Z\cos\beta \sin\theta_W &M_Z\sin\beta 
\sin\theta_W \\ 0 & M_2 & M_Z \cos \beta \cos\theta_W & -M_Z \sin \beta 
\cos\theta_W \\ -M_Z \cos \beta \sin\theta_W & M_Z \cos \beta\cos\theta_W&
0& -\mu \\ M_Z \sin \beta \sin\theta_W & -M_Z \sin \beta\cos\theta_W&-\mu 
& 0 \emat,}
\label{mass}
\end{equation}
in the above basis 
($\tilde B^0=-i{\lambda}'$, $\tilde W_3^0=-i{\lambda}^3$, $\tilde H_u^0$, 
$\tilde H_d^0$). 
Here 
$ \tan\beta= \langle H_u^0\rangle/\langle H_d^0\rangle$ 
is the ratio of Higgs vacuum expectation values (VEVs),
$M_1$ and $M_2$ are the soft bino and wino masses respectively, and 
$\mu$ is the Higgsino mass parameter determined
by the minimization of the Higgs effective 
potential
\begin{equation}
\mu^2 = \frac{m_{H_d}^2 - m_{H_u}^2 \tan^2 \beta}{\tan^2 \beta -1 } - 
\frac{1}{2} M_Z^2\ .
\label{electroweak}
\end{equation} 
Thus one 
parameterizes the gaugino and Higgsino content
of the lightest neutralino according to 
\begin{equation}
\tilde{\chi}^0_1 = N_{11} \tilde{B}^0 +N_{12} \tilde{W}_3^0 +
N_{13} \tilde{H}^0_d + N_{14} \tilde{H}^0_u\ .
\label{lneu}
\end{equation}
It is commonly defined that $\tilde{\chi}^0_1$ is mostly gaugino-like 
if $P\equiv \vert N_{11}\vert^2 + \vert N_{12} \vert^2 > 0.9$, Higgsino-like
if $P<0.1$, and mixed otherwise.

The relevant effective Lagrangian describing
the elastic $\tilde{\chi}^0_1$-nucleon scattering in the MSSM 
has been examined exhaustively in the literature \cite{kami,summary}.
It 
is given by 
\begin{equation}
\mathcal{L}_{eff} = 
\alpha_{2i} \bar{\chi} \gamma^{\mu}  \gamma^{5} \chi\ 
\bar{q_i} \gamma_{\mu}  \gamma^{5} q_i
+
\alpha_{3i} \bar{\chi}\chi \bar{q_i} q_i\ .
\label{lagrangiano}
\end{equation}
The Lagrangian is to be summed over the quark generations, and the 
subscript i refers to up-type quarks (i=1) and down-type quarks
(i=2). The couplings $\alpha_{2,3}$  
can be found e.g. in Ref.~\refcite{Ellis2} 
with the sign conventions for Yukawa couplings in the Lagrangian,
${\cal L}=-\lambda_u H_u^0 \bar u_L u_R -\lambda_d H_d^0 \bar d_L d_R 
- \lambda_e H_d^0 \bar e_L e_R $ + h.c.
In particular, the coupling $\alpha_{3i}$ associated to the
scalar (spin-independent) interaction
include contributions from 
squark ($\tilde q$) exchange and 
CP-even light ($h$) and heavy
($H$) neutral Higgs 
exchange, as illustrated in Fig.~\ref{Feynman},
and can be approximated as \cite{Falk}
\begin{eqnarray}
\alpha_{3i} & \simeq &  \frac{1}{2(m^{2}_{1i} -
  m^{2}_{\tilde{\chi}^0_1})} 
\left(
  g'^2 N_{11}^2
\eta_{11} \eta_{12} e_{i} \frac{y_{i}}{2}\right) 
+  \frac{1}{2(m^{2}_{2i} - m^{2}_{\tilde{\chi}^0_1})} 
\left(  g'^2 N_{11}^2 
 \eta_{21} \eta_{22} e_{i} \frac{y_{i}}{2} 
\right)  \nonumber \\
& & \mbox{} + \frac{g g' m_{q_i}}{4 m_{W}   B_{i}} \left[  
\delta_{1i} N_{11}  D_{i} C_{i} 
\left(\frac{1}{m_h^2}- \frac{1}{m_H^2} \right)
 + \delta_{2i} N_{11}
\left(\frac{{D_{i}^{2}}}{m_h^2} +  \frac{{C_{i}^{2}}}{m_H^2} \right)
\label{sie}
 \right]\ ,
\end{eqnarray}
where we are neglecting CP-violating phases\footnote{Analyses of the
effects of the CP phases on the cross section
can be found in Refs.~\refcite{phases,masfases,Falk} 
and \refcite{masfases2}.}.
Here, $\delta_{1i}$ is $N_{13}$ ($N_{14}$), 
$\delta_{2i}$ is $N_{14}$ ($-N_{13}$),
$ B_{i} $ is $ \sin{\beta} $ ($ \cos{\beta} $), $ C_{i} $ is 
$ \sin{\alpha} $ ($ \cos{\alpha} $)
with 
$ \alpha $ is the Higgs mixing angle, and $ D_{i} $ is 
$ \cos{\alpha} $ ($ - \sin{\alpha} $) 
for up (down) type quarks.  
$y_i$ is the hypercharge defined by
$e_i = T_{3i} + y_i/2$.
$m_{q_i}$ are the masses of the quarks, and
the masses $m_1$
and $m_2$ correspond to the two squark mass eigenstates,
and the $\eta$'s are the entries of the matrix diagonalizing
the sfermion squared mass matrix
which can be parameterized by an angle $ \theta_{f} $
\[ \hspace{1cm} \left( \begin{array}{cc}
\cos{\theta_{f}} & \sin{\theta_{f}} \nonumber \\
-\sin{\theta_{f}} & \cos{\theta_{f}}
\end{array} \right) 
\hspace{0.5cm}
 \equiv 
\hspace{0.5cm}
 \left( \begin{array}{cc}
\eta_{11} & \eta_{12} \nonumber \\
\eta_{21} & \eta_{22}
\end{array} \right)\ .  \]
The first two terms in Eq.~(\ref{sie}) arise from interactions of the type
$g' q \tilde{q} \tilde{B}^0\propto g' N_{11} q \tilde{q} \tilde{\chi}^0_1$, 
and the last two from ${\lambda}_q H_{u,d}^0 \bar{q} q$ (note 
that $g m_{q_i}/(m_W B_i)={\sqrt 2} \lambda_{q_i}$),
$g' H_{u,d}^0 \tilde{H}_{u,d}^0 \tilde{B}^0\propto 
g' N_{13,14} N_{11} H_{u,d}^0  \tilde{\chi}^0_1  \tilde{\chi}^0_1$.
Due to the relative size of the 
lightest Higgs mass to the squark masses, and the smallness
of $\eta_{12}$ to $\eta_{11}$, the Higgs exchange term can dominate in
$\alpha_3$ \cite{Falk,mura}. 
Note also that the heavy Higgs can make a significant
contribution to the down-type-quark part of $\alpha_3$.
The reason being that the latter is proportional to
$C_d^2=\cos^2\alpha$
whereas the light Higgs contribution is proportional
to $D_d^2=\sin^2\alpha$, and for a wide range of parameters
one finds $\tan\alpha=O(1/10)$.
In addition, as we will discuss in Subsection~5.2.1,
$m_H$ may decrease significantly for large $\tan\beta$.

The contribution of the scalar interaction to the $\tilde{\chi}^0_1$-nucleus 
cross section, $\sigma_{\tilde{\chi}_1^0-N}$,
is given by \cite{kami,ellissum}
\begin{equation}
\sigma_{3N}
= \frac{4 M_{r}^2}{\pi}[Z f_p + (A-Z) f_n]^2\ ,
\label{crosssection2}
\end{equation}
where the reduced mass
$M_{r}=M_{N} m_{\tilde{\chi}_1^0}/(M_{N}+m_{\tilde{\chi}_1^0})$,
with $M_{N}$ the mass of the nucleus,
and the relevant contributions to $f_p$ are
\begin{equation}
\frac{f_{p}}{m_{p}} = \sum_{q=u,d,s} f_{Tq}^{(p)} 
\frac{\alpha_{3q}}{m_{q}} +
\frac{2}{27} f_{TG}^{(p)} \sum_{c,b,t} \frac{\alpha_{3q}}{m_q}\ ,
\label{f}
\end{equation}
where $m_p$ is the mass of the
proton, $m_q$ the masses of the quarks,
the parameters $f_{Tq}^{(p)}$ 
are defined
by $ \langle p | m_{q} \bar{q} q | p \rangle = m_{p} f_{Tq}^{(p)} $, 
while $ f_{TG}^{(p)} = 1 - \sum_{q=u,d,s} f_{Tq}^{(p)}$.  
$f_n$ has a similar expression.  
Typical numerical values of the hadronic matrix elements $f_{Tq}^{(p,n)}$ 
can be found e.g. in Ref.~\refcite{Ellis2}.
These are:
\begin{eqnarray}
f_{Tu}^{(p)}= 0.020 \pm 0.004\ , \hspace{0.5cm} f_{Td}^{(p)}= 0.026 \pm
0.005\ ,
\hspace{0.5cm} f_{Ts}^{(p)}= 0.118 \pm 0.062\ ,
\nonumber\\
f_{Tu}^{(n)}= 0.014 \pm 0.003\ , \hspace{0.5cm} f_{Td}^{(n)}= 0.036 \pm
0.008\ ,
\hspace{0.5cm} f_{Ts}^{(n)}= 0.118 \pm 0.062\ ,
\label{central}
\end{eqnarray}
and we will use throughout this review the central 
values\footnote{Larger values of the hadronic matrix elements
have also been considered in the
literature. For analyses of the effect induced on the
cross section by the present uncertainties in these values,
see e.g. Refs.~\refcite{Bott} and \refcite{Arnowitt}.
A rough estimate can be obtained from Eq.~(\ref{approx}) below.}.
%
We see that $f_{Ts}^{(n)}=f_{Ts}^{(p)}$ and much larger
than $f_{Tq}$ for $u$ and $d$ quarks, and therefore
$f_p$ and $f_n$ are
basically equal (note that the relevant part of the 
couplings $\alpha_{3q}$ are
proportional to $m_q$, so the fraction
$\alpha_{3q}/m_q$ does not become large for small $m_q$).
Thus we can write\footnote{The contribution to this cross section
from two-nucleon currents from pion exchange in the nucleus
has recently been discussed in the last work of 
Ref.~\refcite{conflict}.}
\begin{equation}
\sigma_{3N}
\simeq \frac{4 M_{r}^2}{\pi}A^2 f_p^2\ .
\label{crosssection22}
\end{equation}

It is worth noticing here
that
the spin-independent scattering adds
coherently giving rise to a cross section proportional
to the squared of the atomic weight, $A$.
However, the axial-vector (spin-dependent) interaction, 
the one proportional to $\alpha_{2i}$ in Eq.~(\ref{lagrangiano}),
which is nonzero only if the nucleus has a non-vanishing spin,
is incoherent. Thus,
for heavy targets,
the scalar cross section is generically larger implying
$\sigma_{\tilde{\chi}_1^0-N}\simeq \sigma_{3N}$.
Recall that this is the case of the experiments discussed in Section~4.
For example, Ge with a mass $\sim 73$ GeV
and NaI with masses $\sim 23$ GeV for Na and $\sim 127$ GeV
for I are used.
In what follows we will concentrate on the scalar
cross section\footnote{Recent analyses
of the spin-dependent part of the cross section can be found
e.g. in Refs.~\refcite{Falk,Ellis2,Santoso,EllisOlive} and \refcite{spin}.
In Ref.~\refcite{Klapd} it was pointed out that for targets
with spin-non-zero nuclei, as e.g. $^{73}$Ge where spin = 9/2,
it might be the spin-dependent interaction the one determining the
lower bound for the direct detection rate, when the cross section
of the scalar interaction drops below about $10^{-12}$ pb.},
and in particular on
the one for protons
%
\begin{equation}
\sigma_{3p}\equiv 
\sigma_{\tilde{\chi}_1^0-p} 
= \frac{4 m_{r}^2}{\pi}f_p^2\ ,
\label{crosssection}
\end{equation}
where
$m_{r}=m_{p} m_{\tilde{\chi}_1^0}/(m_{p}+m_{\tilde{\chi}_1^0})\approx 
m_p$. In fact, this is the cross section shown in the
experimental papers (see Fig.~\ref{limits}), and therefore the one
that we will compute in the different theoretical models. Note that
for a material with heavy nuclei, $A\approx 100$, 
$M_N\approx 100$ GeV $\approx m_{\tilde{\chi}_1^0}$ and therefore
\begin{equation}
\sigma_{\tilde{\chi}_1^0-N}\approx 10^8\ \sigma_{\tilde{\chi}_1^0-p}\ .
\label{crosssection3}
\end{equation}
For $\sigma_{\tilde{\chi}_1^0-p}\approx 10^{-8}$ pb
one recovers the rough estimate of Subsection~3.1.3,
$\sigma_{\tilde{\chi}_1^0-N}\approx 1$ pb,
for a particle with weak interactions.

The predictions for the scalar neutralino-proton
cross section, $\sigma_{\tilde{\chi}_1^0-p}$, 
are usually studied in the framework
of SUGRA, and we will review them below.
We will see that $\sigma_{\tilde{\chi}_1^0-p}\approx 10^{-8}$
pb
can be obtained, but that smaller (or larger)
values are also possible depending on the parameter space 
chosen (e.g. $m$, $M$, $A$, $\tan\beta$ in mSUGRA as we will discuss 
in Subsection~5.2.1), 
since
very different values for $m_h$, $m_H$, $N_{11}$, $N_{13}$, etc.
can arise.  
Note in this sense that, from the above results,
the neutralino-proton cross section
can be approximated as
\begin{eqnarray}
\sigma_{\tilde{\chi}_1^0-p} 
& \approx & 
\frac{1}{4\pi}\left (\frac{g g' f_{Ts}^{(p)} m_{p}^2}{m_{W}\cos\beta}
\right)^2
\nonumber \\
& & \mbox{}
\times
\left[ N_{11} N_{14}  \sin\alpha \cos\alpha 
\left(\frac{1}{m_h^2}- \frac{1}{m_H^2} \right)
+ N_{11} N_{13} 
\left(\frac{\sin^2\alpha}{m_h^2} +  \frac{\cos^2\alpha}{m_H^2} \right)
\right]^2
\label{approx}
\end{eqnarray}
For example, assuming the typical values 
$N_{11}\approx 1$, $N_{13}\approx 0.1$,
$\tan\alpha\approx 1/10$, $\tan\beta\approx 10$, 
and $m_H\approx 100 (500)$ GeV,
one obtains $\sigma_{\tilde{\chi}_1^0-p} \approx 10^{-8} (10^{-11})$ pb.
%
%

Before analyzing in detail the cross section, it is worth remarking 
that in these analyses to reproduce the correct
phenomenology is crucial. 
Thus, in the next Subsection,
the most recent experimental and astrophysical constraints
which can affect this computation will be discussed.

\subsection{Experimental and astrophysical constraints}

We list here the most recent experimental and astrophysical results
which are relevant when computing the neutralino-nucleon
cross section. They give rise to important constraints on the
SUSY parameter space.

\vspace{0.2cm}

\noindent {\it (a) Higgs mass}

\noindent Whereas in the context of the standard model
the negative direct search for the Higgs at the LEP2 collider
implies
a lower bound on its mass of about 114.1 GeV,
the situation in SUSY scenarios is more involved.
In particular, in the framework of mSUGRA (to be discussed in detail below), 
one obtains for the lightest CP-even Higgs
$m_h \gsim 114.1$ GeV when $\tan\beta \lsim 50$,
and $m_h \gsim 91$ GeV when $\tan\beta$ is larger \cite{Heinemeyer}.
Recall in this sense that
$\sigma_{SUSY} (e^+ e^- \to Zh)= \sin^2(\alpha-\beta)\ \sigma_{SM} (e^+ e^- \to Zh)$ with
$\alpha$ the Higgs mixing angle
in the neutral CP-even Higgs sector \cite{sine}.
Thus when the $ZZh$ coupling, $\sin^2(\alpha-\beta)$, which controls the detection
of the lightest MSSM Higgs at LEP, is 
$\sim 1$ one recovers the standard model bound.
However, when
a suppression of $\sin^2(\alpha-\beta)$ is obtained the bound will be smaller.
Such a suppression 
occurs with
$\tan\beta > 50$. 
In this case, in some regions of the parameter space
$m^2_A=m_{H_d}^2+m_{H_u}^2+2\mu^2$ becomes small,
$m_A\lsim 150$ GeV, because
the bottom Yukawa coupling entering in the 
renormalization group equation (RGE) for
$m_{H_d}^2$ is large. And a
small $m_A$ gives rise to $\sin^2(\alpha-\beta)<1$, as can 
be understood in terms of the relations
between the angles $\alpha$, $\beta$ and $m_A$ \cite{porejemplo}. 

In any case, let us remark that generically $\tan\beta$ is constrained to be 
$\tan\beta\lsim 60$, since otherwise several problems arise. 
For example, as mentioned above, the bottom Yukawa coupling is large
and for $\tan\beta> 60$, even for moderate values of $m$ and $M$,
$m_{H_d}^2$ becomes negative.
As a consequence $m^2_A$ 
becomes also negative 
unless a fine-tuning (in the sense that only certain
combinations of $m$ and $M$ are possible) is carried out.
In this review we study only cases with $\tan\beta \leq 50$. Thus for 
mSUGRA we will always have $\sin^2(\alpha-\beta)\sim 1$. 
However, we will also be interested in relaxing 
the mSUGRA framework and therefore 
$\sin^2(\alpha-\beta)$ must be computed in this case for all points of the parameter
space, in order to know which bound for 
the lightest MSSM Higgs must be applied.
For the latter one can use the plot $\sin^2(\alpha-\beta)$ versus $m_h$ shown
in Ref.~\refcite{barate}.



\vspace{0.2cm}

\noindent Let us finally mention that a very convenient program to evaluate
$m_h$ is the so-called {\tt FeynHiggs} \cite{FeynHiggs}
which contains the complete one-loop and dominant two-loop corrections.
Higher-order corrections introduce an 
uncertainty of about 3 GeV in the result.
In addition, there is
a simplified version of the program,  
{\tt FeynHiggsFast}. 
The value of $m_h$ obtained with this version is
approximately 1 GeV below the one obtained using {\tt FeynHiggs}.
The figures shown throughout this review were obtained using
{\tt FeynHiggsFast}, and neglecting the uncertainty due to
higher-order corrections.
Recently, another program to study Higgs phenomenology has been
constructed. See Ref.~\refcite{newHiggs} for details.

\vspace{0.2cm}

\noindent {\it (b) Top mass}

\noindent The central experimental value for the top mass, 
$m_t(pole)=175$ GeV, is used
throughout this review. 
However, let us remark that a modification in this mass by
$\pm 1$ GeV implies, basically, a modification also of $\pm 1$ GeV
in the value of $m_h$ \cite{Heinemeyer}.

\vspace{0.2cm}

\noindent {\it (c) Bottom and tau masses}

\noindent For the bottom mass the input  
$m_b(m_b)=4.25$ GeV is used throughout this review, which,  
following the analysis of Ref.~\refcite{santamaria}
with $\alpha_s(M_Z)=0.1185$,
corresponds to $m_b(M_Z)=2.888$
GeV.
In the evolution of the bottom mass the SUSY
threshold corrections at $M_{SUSY}$ \cite{pierce} are taken into
account.
These are known to be significant, specially for large
values of $\tan\beta$. 
A similar analysis must be carried out for the 
tau mass, using as input
$m_{\tau}(M_Z)=1.7463$ GeV.

\vspace{0.2cm}

\noindent {\it (d) SUSY spectrum}

\noindent The present experimental lower
bounds on SUSY masses coming from LEP and Tevatron must be imposed. 
In particular,
using the low-energy relation from mSUGRA,
$M_1=\frac{5}{3}\tan^2\theta_W M_2$,
one obtains for the lightest chargino mass the bound \cite{chargino}
$m_{\tilde\chi_1^{\pm}}>103$ GeV.
Likewise, one is also able to obtain the following 
bounds for sleptons masses \cite{sleptons}:
$m_{\tilde e}>99$ GeV,
$m_{\tilde\mu}>96$ GeV,
$m_{\tilde\tau}>87$ GeV.
Finally, 
for the masses of the 
sneutrino, the lightest stop, the rest of squarks, and
gluinos, one can use the following bounds: 
$m_{\tilde\nu}>50$ GeV,
$m_{\tilde t_1}>95$ GeV,
$m_{\tilde q}>150$ GeV,
$m_{\tilde g}>190$ GeV.

\vspace{0.2cm}

\noindent {\it (e) $b\to s\gamma$}

\noindent The measurements of $B\to X_s\gamma$ decays 
at 
CLEO \cite{cleo} 
and BELLE \cite{belle}
lead to bounds on the branching ratio 
$b\to s\gamma$. In particular we impose throughout this
review:
$2\times 10^{-4}\leq BR(b\to s\gamma)\leq 4.1\times
10^{-4}$.
Let us mention that a routine to carry out this
evaluation is provided e.g. by
the program {\tt micrOMEGAs} \cite{micromegas}. 
A description of this procedure can be found in 
Ref.~\refcite{routine}. Although the improvements
of Ref.~\refcite{impro} are not included in this routine,
they are not so important for this review since, as discussed
below,
only $\mu>0$ will be considered.

\vspace{0.2cm}

\noindent {\it (f) $g_{\mu}-2$}

\noindent The new measurement of the anomalous
magnetic moment of the muon, $a_\mu=(g_{\mu}-2)/2$, 
in the E821 experiment at the 
BNL \cite{muon} 
deviates by 
$(33.7\pm 11.2) \times 10^{-10}$
from the recent standard model calculation of Ref.~\refcite{news} using
$e^+e^-$ data.
Assuming that the possible 
new physics is due to SUGRA, we will show in the figures
throughout this review the 
constraint
$11.3\times 10^{-10}\leq a_{\mu} (SUGRA)\leq 56.1\times
10^{-10}$ at the
2$\sigma$ level. This excludes the case $\mu<0$\footnote{However, 
it is worth noticing that this result for 
$a_{\mu}$
is in contradiction 
with the one obtained by using 
tau decay data (instead of $e^+e^-$ ones) which only implies a deviation
$(9.4\pm 10.5) \times 10^{-10}$ from the standard model 
calculation \cite{news}.
In any case, concerning the value of $\mu$, it is worth recalling
that constraints coming from the $b\to s\gamma$ process
highly reduce the $\mu<0$ parameter space.}.


\vspace{0.2cm}

\noindent {\it (g) $B_s \to \mu^+ \mu^-$}

\noindent The branching ratio for the $B_s \to \mu^+ \mu^-$ decay
has been experimentally bounded by CDF \cite{Abe} with the result
$BR(B_s \to \mu^+ \mu^-)< 2.6\times 10^{-6}$.
Although in the standard model this branching ratio is very small,
of order $3\times 10^{-9}$, it might be in principle significant in the
SUSY case for large $\tan\beta$, due to Higgs ($A$) mediated 
decay \cite{Bsmu}. This issue has been analyzed recently in the context
of dark matter. It seems that the current experimental constraint
does not eliminate in fact any relevant part of the parameter space of
the MSSM \cite{Wnath}. Analyses of other scenarios can be found
in Refs.~\refcite{Rosz} and \refcite{Farrill}.

\vspace{0.2cm}

\noindent {\it (h) LSP}

\noindent As mentioned in the Introduction, 
the LSP, with mass of order GeV, 
is stable and therefore must be an electrically neutral
(also with no strong interactions) particle,
since otherwise it 
would bind to nuclei and would be 
detectable in the Earth as an exotic heavy isotope with 
abundance $n/n(proton)\sim 10^{-10} (10^{-6})$ in the case
of strong (electromagnetic) interactions \cite{raros}.
This is not consistent with the experimental upper limits \cite{exli}
$n/n(proton)\lsim 10^{-15}$ to $10^{-30}$.
Although the lightest neutralino, $\tilde{\chi}_1^0$, is
the LSP in most of the parameter space of the MSSM,
in some regions
one of the staus, $\tilde{\tau}_1$, can be lighter.
Therefore, following the above discussion,
these regions must be discarded.

\vspace{0.2cm}

\noindent {\it (i) Relic $\tilde{\chi}_1^0$ density}

\noindent 
The preferred astrophysical bounds 
on the dark matter density (see Eq.~(\ref{omega})),
$0.1\lsim \Omega_{\mbox{\tiny DM}}h^2\lsim 0.3$, must be imposed
on the
theoretical computation of the relic $\tilde{\chi}_1^0$ density, once
$\Omega_{\tilde{\chi}_1^0}=\Omega_{\mbox{\tiny DM}}$ is 
assumed\footnote{If neutralinos 
do not constitute all the dark matter in the Universe, and other
particles discussed in Section~3 also contribute, then 
$\Omega_{\tilde{\chi}_1^0}<\Omega_{\mbox{\tiny DM}}$.
This may allow points in the parameter space which would be excluded
by
$\Omega_{\tilde{\chi}_1^0}h^2< 0.1$.
However, notice that in this case the density of neutralinos
in the galaxy would be 
$\rho_{\tilde{\chi}_1^0}= \xi\cdot\ 0.3$ GeV\ cm$^{-3}$
with $\xi <1$, and therefore the 
results (event rates) would imply a larger value of the experimental 
cross section \cite{nueva}.
Since the DAMA area is already difficult to reproduce in SUSY models,
now the situation will be even worse. In 
Fig.~\ref{limits} we should substitute
$\sigma_{\mbox{\tiny WIMP}-nucleon}$ by
$\xi\cdot\ \sigma_{\mbox{\tiny WIMP}-nucleon}$.}.
For the sake of completeness, we also show in the figures below 
the bounds
$0.094\lsim \Omega_{\mbox{\tiny DM}}h^2\lsim 0.129$ 
deduced from the WMAP satellite (see Eq.~\ref{wmaprange}).
As mentioned in footnote $a$, 
different cosmological scenarios give
rise to different results in the computation of the relic density.
We will consider throughout this review the standard mechanism of 
thermal production of neutralinos (see Section 3.1.3).

\vspace{0.2cm}

\noindent Let us finally mention that there are 
programs \cite{programs,micromegas}
to evaluate $\Omega_{\tilde{\chi}_1^0}$.
A very convenient one used in the figures shown
here is {\tt microMEGAs} \cite{micromegas}.
In this program the 
exact tree-level cross sections for all possible annihilation \cite{kami}
and coannihilation \cite{stau,char,stop} 
channels are included in the code through a link
to {\tt CompHEP} \cite{comphep}, and accurate thermal average of them is used.
Also, poles and thresholds are properly handled and one-loop
QCD corrected Higgs decay widths \cite{width}
are used. The SUSY corrections included in the latest version of the
code \cite{width} are not implemented yet by {\tt micrOMEGAs}. Fortunately,
in our case, their effect is much smaller than that of the QCD
corrections.
Good agreement between {\tt micrOMEGAs} and other independent
computations of $\Omega_{\tilde{\chi}_1^0}$ including
$\tilde{\chi}^0_1-\tilde{\tau}_1$ coannihilations can be found in 
Ref.~\refcite{compare}.

\vspace{0.2cm}

\noindent {\it (j) UFB}

\noindent The constraints that
arise from imposing the absence of charge and colour breaking 
minima can also be considered \cite{darkufb}.
As is well known, the presence of scalar fields with colour and
electric
charge in SUSY theories induces the possible existence of 
dangerous charge and colour breaking minima, which would make
the standard vacuum unstable \cite{discussion}.
The presence of these instabilities may imply either that the
corresponding model is inconsistent or that it requires non-trivial
cosmology to justify that the Universe eventually fell in the
phenomenologically realistic (but local) minimum \cite{reviewccb}.
There are two types of constraints:
the ones arising from directions in the field-space along
which the (tree-level) potential can become unbounded from below (UFB),
and those arising from the existence of charge and color
breaking (CCB) minima in the potential deeper than the
standard minimum. By far, the most restrictive are the
UFB bounds, and therefore these are the ones used 
in the figures shown here.
There are three UFB directions, labelled as UFB-1, UFB-2, UFB-3
in Ref.~\refcite{clm1}. It is worth mentioning here that in general the
unboundedness is only true
at tree-level since radiative corrections eventually raise the potential for
large enough values of the fields, but still these minima can be deeper than
the realistic one (i.e. the SUSY standard-model vacuum) and thus dangerous.
The UFB-3 direction, which involves
the scalar fields
$\{H_u,\nu_{L_i},e_{L_j},e_{R_j}\}$ with $i \neq j$
and thus leads also to electric charge
breaking, yields the strongest bound among all
the UFB and CCB constraints.

\subsection{SUGRA predictions for the neutralino-nucleon cross section}

As can be deduced from the above discussion concerning the cross
section in Eq.~(\ref{crosssection}), the values of the
SUSY parameters, such as scalar masses or gaugino masses,
are crucial in the analysis.
Obviously, these values are associated with the mechanism of SUSY
breaking.
A pragmatic attitude to this issue is the addition of
explicit soft SUSY-breaking parameters of the appropriate size
(of order $10^2-10^3$ GeV) in the Lagrangian and with appropriate
flavour symmetries to avoid dangerous flavour-changing neutral
currents (FCNC) transitions.
The problem with this 
pragmatic attitude is that, taken by itself, lacks any theoretical 
explanation. SUGRA theories provide an attractive context
that can justify such a procedure. Indeed, if one considers the 
SUSY standard model and couples it to $N=1$ SUGRA, 
the spontaneous breaking of local SUSY in a hidden sector
generates explicit soft SUSY-breaking terms of the required form 
in the effective low-energy Lagrangian \cite{dilaton}.  
If SUSY is broken at a scale  $\Lambda _S$, the 
soft terms have a scale of order $\Lambda _S^2/M_{Planck}$, and
therefore
one obtains the required size if SUSY is broken at an 
intermediate scale $\Lambda _S \sim  10^{11} $ GeV.

Thus
one usually considers the MSSM in the framework
of SUGRA.
Working in this framework 
the soft parameters
generated once SUSY is broken through
gravitational interactions,
i.e., gaugino masses,
scalar masses,
trilinear couplings and bilinear couplings, 
are denoted at high energy 
by $M_{a}$,
$m_{\alpha}$, $A_{\alpha\beta\gamma}$, and
$B$,
respectively. Although in principle these are free parameters
together with $\mu$,
when electroweak symmetry breaking is imposed $\mu^2$ and $B$ 
are determined.
Finally, 
the
renormalization group equations (RGEs) are
used to derive low-energy SUSY 
parameters\footnote{Let us remark that, given the convention used 
throughout this review for gaugino masses in the Lagrangian, 
${\cal L} =\frac{1}{2}\sum_{a} M_a {\lambda}_a \lambda_a$ + h.c., one has
to use the (one-loop) RGEs obtained e.g.
in Ref.~\refcite{rges}, but with an opposite sign in the
gaugino contributions to the RGE's of the $A$ parameters.}. 
Since the
SUGRA framework 
still 
allows a large
number of free parameters, 
in order to have predictive power, one usually assumes 
that the above soft parameters 
are 
universal at the 
GUT  
scale, $M_{GUT} \approx 2\times 10^{16}$ GeV,
providing also for an understanding of FCNC suppression.
As mentioned in the Introduction, 
this is the mSUGRA scenario, also called in the
literature
the Constrained MSSM (CMSSM). 
Although, apparently, this assumption about universality is arbitrary,
it is worth noticing that  
interesting classes of SUGRA models 
give rise to this kind of universal soft SUSY-breaking terms \cite{dilaton}.
In addition, explicit string constructions
with these universality properties can be found 
in some limits \cite{dilaton}.
We will discuss them in some detail
in Subsections~5.3 and 5.4.

Let us recall that
the full N=1 Supergravity Lagrangian is specified in terms of three functions
which depend on the scalar fields $\phi_M$ of the theory: the
analytic gauge kinetic function $f_a(\phi_M)$,
the real K\"ahler potential $K(\phi_M,\phi^*_M)$ and
the analytic superpotential $W(\phi_M)$.
In particular, 
$f_a$ determines the kinetic terms for the gauge 
fields and is related to the gauge couplings as $Re f_a=1/g_a^2$, where
the
subindex $a$ is associated with the different gauge groups of the 
theory, $K$ determines the kinetic terms for the scalar fields,
and $W$ determines the Yukawa couplings.
Thus, for example, the form of $K$ that leads to canonical kinetic
terms for the observable fields $C_{\alpha}$, namely
$K=\sum_{\alpha} C_{\alpha}C_{\alpha}^*$,
irrespective of the SUSY-breaking mechanism gives rise to
universal soft scalar masses \cite{dilaton}, $m_{\alpha}=m$,
of the type imposed in mSUGRA.


Below 
we will carry out the analysis of mSUGRA concerning dark matter.
We will also discuss how the results are modified when
the above assumptions are relaxed. In particular, we will allow an
intermediate scale instead of the usual GUT one, and also 
non-universal soft scalar and gaugino masses.

\begin{figure}[t]
\begin{center}
\epsfig{file= 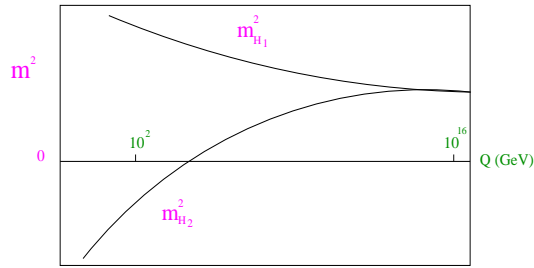,width=7cm,height=3.5cm,angle=0}
\end{center}
\caption{
Running of the soft Higgs masses-squared with energy.}
\label{run}
\end{figure}

\subsubsection{mSUGRA scenario with a GUT scale}

\begin{figure}[t]
\begin{center}
\epsfig{file= 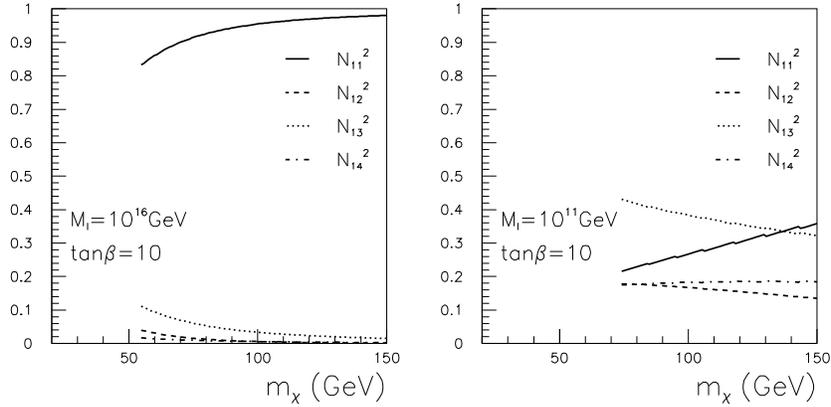, 
height=7cm,angle=0}
\end{center}

\vspace{-0.5cm}

\caption{Gaugino-Higgsino components-squared
of the lightest neutralino as a function of its mass for
the unification scale (left frame), 
$M_I=10^{16}$ GeV, and for the intermediate scale (right frame),
$M_I=10^{11}$ GeV.
}
\label{N_1i}
\end{figure}

As can be deduced from the above discussion, 
in this scenario there are only four free 
parameters\footnote{In fact, scenarios with less free parameters are
also possible. For instance, in the above example with canonical
kinetic terms, if in addition we assume that Yukawa couplings
and $\mu$ are constants, i.e. that they do not depend on hidden sector
fields, one obtains
$B=A-m$. Scenarios with this type of constraints between soft
parameters
have been analyzed in the context of dark matter in 
Ref.~\refcite{softrelation}.}: 
$m$, $M$, $A$, and $\tan \beta$. In addition, the
sign of $\mu$ remains also undetermined (see Eq.~(\ref{electroweak})).  
Let us also remark that in this context the lightest neutralino
is mainly bino \cite{binor,seealso}. 
To understand this result qualitatively, we show
schematically in Fig.~\ref{run} the well known evolution
of $m_{H_u}^2$ (and $m_{H_d}^2$ neglecting
bottom and tau Yukawa couplings) towards large
and negative values with the scale.
Since 
$\mu^2$ given by Eq.~(\ref{electroweak}),
for reasonable values of  $\tan\beta$,
can be approximated as,
\begin{equation}
\mu^2\approx -m_{H_u}^2-\frac{1}{2} M_Z^2\ ,
\label{electroweak2}
\end{equation}
then it becomes also 
large (the effect of the one-loop corrections to the scalar
potential can be minimized by evaluating $\mu$ at the
scale $M_{SUSY}=\sqrt{m_{\tilde t_1} m_{\tilde t_2}}$).
In particular, $|\mu|$ becomes 
much larger than
$M_1$ and $M_2$. Thus, as can be easily understood from 
Eqs.~(\ref{mass}) and (\ref{lneu}),
the lightest neutralino will be mainly gaugino, and in particular
bino, since
at low energy $M_1=\frac{5}{3}\tan^2\theta_W M_2\approx 0.5 M_2$.
We show this fact in the plot on the left frame of Fig.~\ref{N_1i}, 
where for $\tan\beta=10$ the gaugino-Higgsino 
components-squared $N_{1i}^2$ of the lightest neutralino as a function
of its mass $m_{\tilde{\chi}_1^0}$ 
are exhibited. Here we are using an example with $m=150$ GeV and $A=M$. 
Note that
$M$ is essentially fixed for a given
$m_{\tilde{\chi}_1^0}$.
Clearly, 
$N_{11}$ is 
extremely large and therefore $P\gsim 0.9$.

\begin{figure}[t]
\begin{center}
\epsfig{file=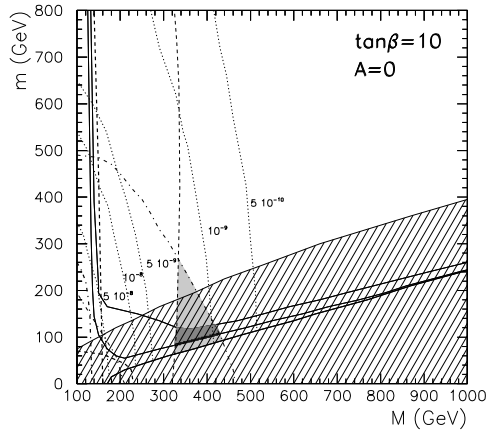,width=7cm}
%


\end{center}

\caption{Scalar neutralino-proton cross section $\sigma_{\tilde{\chi}_1^0-p}$
in the parameter space of the mSUGRA scenario ($m$, $M$) 
for $\tan \beta=10$, $A=0$
and $\mu > 0$.
The dotted curves are contours of $\sigma_{\tilde{\chi}_1^0-p}$.
The region to the left of the
near-vertical dashed line is excluded by the lower bound 
on the Higgs mass $m_h>114.1$ GeV. 
The region to the left of the near-vertical double dashed line
is excluded by the lower bound on the chargino mass
$m_{\tilde\chi_1^{\pm}}>103$ GeV.
The corner in the lower left shown also by a double dashed line
is excluded by the LEP bound on the stau mass
$m_{\tilde{\tau}_1}>87$ GeV.
The region bounded by dot-dashed lines is allowed by $g_{\mu}-2$.
The region to the left of the 
double dot-dashed line is excluded by $b\to s\gamma$.
From bottom to top, the solid lines are the upper bounds of the areas such
as $m_{\tilde{\tau}_1}<m_{\tilde{\chi}_1^0}$ (double solid), 
$\Omega_{\tilde{\chi}_1^0} h^2<0.1$ and $\Omega_{\tilde{\chi}_1^0} h^2<0.3$. 
The light shaded area is favored by all the phenomenological
constraints, 
while the dark one fulfills in addition
$0.1\leq \Omega_{\tilde{\chi}_1^0}h^2\leq 0.3$ (the black region on top of this
indicates the WMAP range $0.094<\Omega_{\tilde{\chi}_1^0}h^2<0.129$).
The ruled region 
is excluded because of
the charge and colour breaking constraint UFB-3.}
\label{a2m}

\end{figure}

Now, using 
Eq.~(\ref{crosssection}) one can compute the
cross section for different values of the 
parameters (for recent results see Refs.~\refcite{Bottino,phases,arna2,masfases,Falk,masfases2,Bott,Ellis2,Arnowitt,focus,Bed,Drees,Gomez,large,Nojiri,tuning,darkcairo,Drees2,EllisOlive,Arnowitt3,focusito,nosopro,Lahanas,Dutta,Elliswmap,Lahanas2,Wnath,Vergados,darkufb,Farrill,softrelation}).
As a consequence of the $\tilde{\chi}_1^0$ being mainly bino,
the predicted $\sigma_{\tilde{\chi}_1^0-p}$  is well below
the accessible experimental regions
for low and moderate values of $\tan\beta$,
since the scattering channels through Higgs exchange shown
in Fig.~\ref{Feynman} are not so important (recall
that the Higgs-neutralino-neutralino couplings are proportional
to $N_{13}$ and $N_{14}$ as shown in Eq.~(\ref{sie})).
In addition,
the (tree-level)
mass of the CP-odd Higgs, $A$,
\begin{equation}
m^2_A=m_{H_d}^2+m_{H_u}^2+2\mu^2\ ,
\label{ma} 
\end{equation}
will be large 
because $\mu^2$ is large.
Since the heaviest CP-even Higgs,
$H$, is almost degenerate in mass with this, 
$m_H$ will also be large
producing a further suppression in the scattering channels.
This fact is shown in 
Fig.~\ref{a2m}, where contours 
of $\sigma_{\tilde{\chi}_1^0-p}$ in the
parameter space ($m$, $M$) for $\tan \beta=10$, $A=0$
and $\mu > 0$ \footnote{Let us remark that
the sign of the dominant contribution to the supersymmetric       
contribution to the $g_{\mu}-2$ is given by $M_2\mu$.
As discussed in Subsection~5.1, we 
are taking this to be positive,
and therefore we will only consider $sign(M)=sign(\mu)$. Now, due to the
symmetry of the RGEs, the results for ($-M,A,-\mu$) are identical to those
for ($M,-A,\mu$), and therefore, one can cover the whole (permitted)
parameter space restricting to positive values for $M$ and $\mu$ and
allowing $A$ to take positive and negative values.} are plotted \cite{darkufb}.
Note that we can deduce the value of the $\tilde{\chi}_1^0$ mass in the
plots from the value of $M$, since 
$m_{{\tilde{\chi}_1^0}}\simeq {M_1}\simeq 0.4\ M$. 
For the gluino mass we can also use the simple relation,
$m_{\tilde g}\simeq 2.5\ M$.

As we can see in the figure,
the experimental bounds discussed in Subsection~5.1
are very important and exclude large regions of the
parameter
space.
This is due to the combination of the Higgs mass bound with
the $g_{\mu}-2$ lower bound\footnote{Recall that we are using only the limit
based on $e^+e^-$ analysis (see footnote $n$), otherwise the allowed region
would extend towards the right hand side of the figure.}. 
One obtains from the Higgs mass 
the lower bound $M\gsim$ 320 GeV, and from $g_{\mu}-2$ the upper bound
$M\lsim$ 440 GeV.
These bounds imply for the neutralino,
$128\lsim m_{\tilde{\chi}_1^0}\lsim 176$ GeV, and 
for the gluino (and squarks) $800\lsim m_{\tilde{g},\tilde{q}}\lsim 1100$ GeV. 
The light shaded area in the figure shows the
region allowed by the experimental bounds. There, the lower 
contour (double solid line)
is obtained including also
the constraint coming from the LSP bound, 
$m_{\tilde{\chi}_1^0}<m_{\tilde{\tau}_1}$.
For this area $\sigma_{\tilde{\chi}_1^0-p}\approx 10^{-9}$ pb.


On the other hand, the astrophysical bounds
$0.1\lsim \Omega_{\tilde{\chi}_1^0}h^2\lsim 0.3$
must be imposed in the computation.
Let us recall that
there are only four regions where the upper bound
$\Omega_{\tilde{\chi}_1^0}h^2 \sim 1/\sigma_{ann}\lsim 0.3$
can be satisfied.
There is the {\it bulk region} at moderate $M$ and 
$m$ \cite{Goldberg,Ellis}.
There, since $\tilde{\chi}_1^0$ is mainly bino,
the annihilation channels into leptons are 
important (see e.g. Fig.~\ref{annihilation}), specially
those from  $\tilde l_R$ exchange because they 
have the largest hypercharge. 
Moreover, the mass of $\tilde l_R$ only receives very small contributions
from gaugino loop diagrams. Here $\Omega_{\tilde{\chi}_1^0}h^2 \lsim 0.3$
requires specific upper bounds on $\tilde{\chi}_1^0$
and  $\tilde l_R$ masses \cite{drno}.
Another region is the {\it coannihilation region}
extending to larger $M$ \cite{Seckel}. There
the tail where the LSP is almost degenerate with the
NLSP, the stau, producing efficient coannihilations,
is rescued \cite{stau}. This is true even for large values of  
$m_{\tilde{\chi}_1^0}$. These two regions can clearly be seen
in Fig.~\ref{a2m}. 
There is also the {\it focus-point region} at 
$m>1$ TeV not shown in this figure, and that will be discussed
at the end of this subsection.
Finally, there is the {\it Higgs pole region}
extending to large $M$ and $m$ because rapid annihilation
through a direct-channel pole \cite{drno,br,large} 
when $2m_{\tilde{\chi}_1^0}\sim m_{A,H}$ (see e.g. Fig.~\ref{annihilation}).
But this region occurs at large $\tan\beta$ and will be discussed
below in Fig.~\ref{a35}.

Given the above bounds, we observe in Fig.~\ref{a2m}
that the allowed (dark shaded) area becomes very small 
(extremely small if the recent WMAP data are taken into account). Only
the beginning of the tail mentioned above 
is rescued.
In addition, 
the restrictions coming from
the UFB-3 constraint
exclude also this area.
In conclusion, the results indicate that the whole parameter
space for $\tan\beta=10$
is excluded
on these grounds. 
This is also true for other values of $A$, as shown explicitly 
in Ref.~\refcite{darkufb}.

\begin{figure}[t]

\epsfig{file=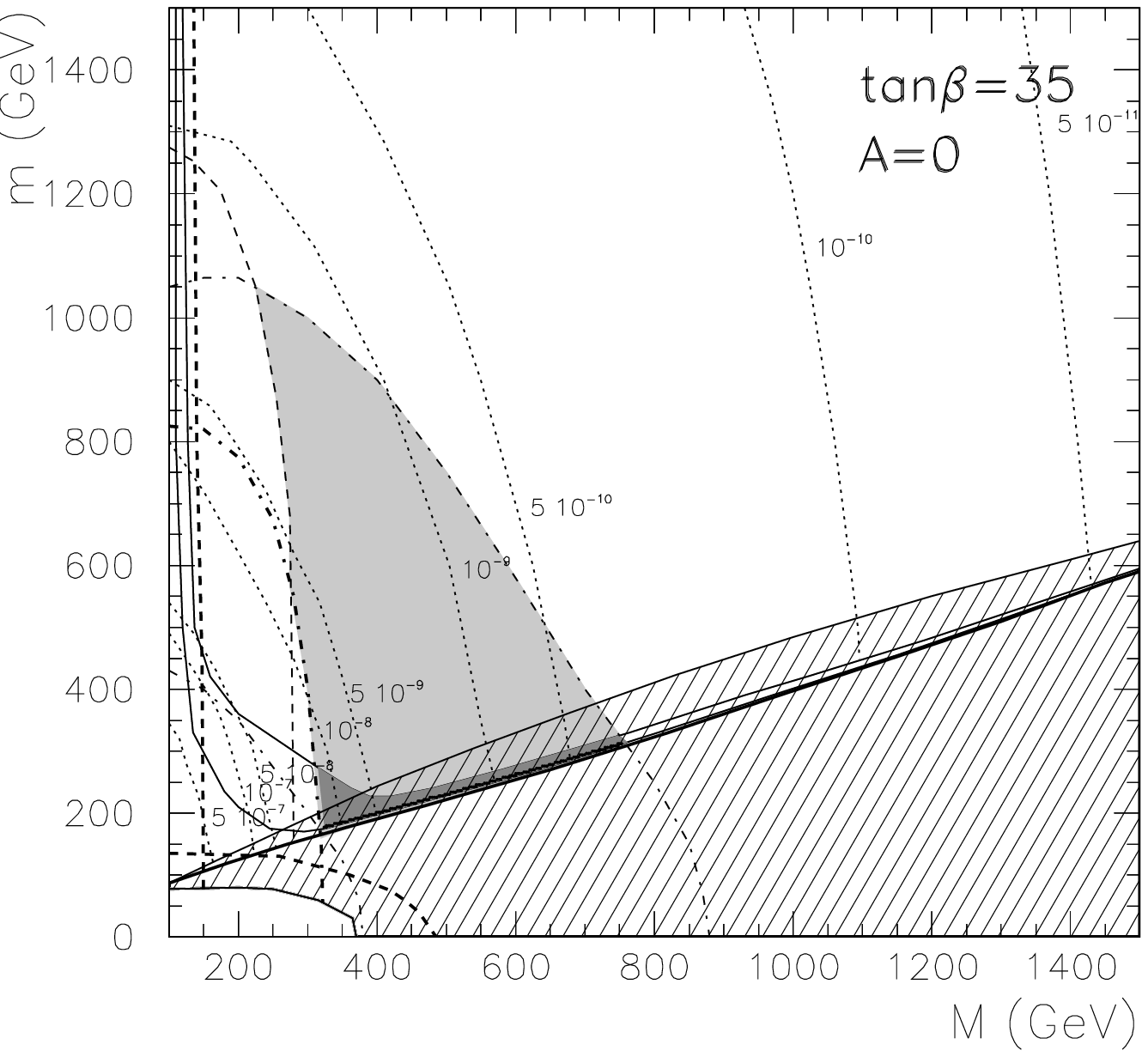,width=6.2cm}
\epsfig{file=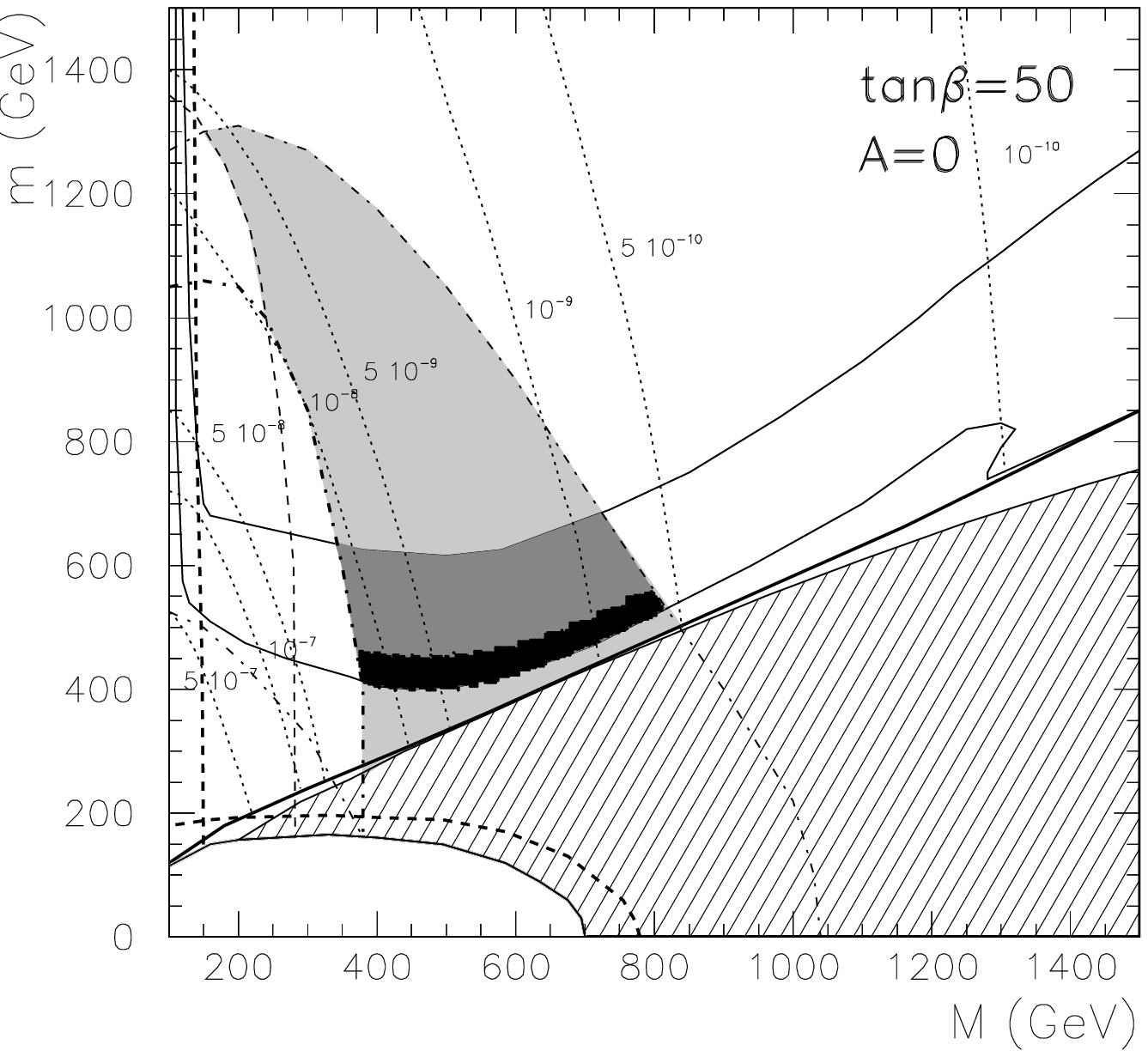,width=6.2cm}


\caption{The same as in Fig.~\ref{a2m} 
but for $\tan\beta=35$ and 50.
The white region at the bottom bounded by a solid line
is excluded because 
$m_{\tilde{\tau}_1}^2$ 
becomes negative.}
\label{a35}

\end{figure}

\begin{figure}[t]
\epsfig{file=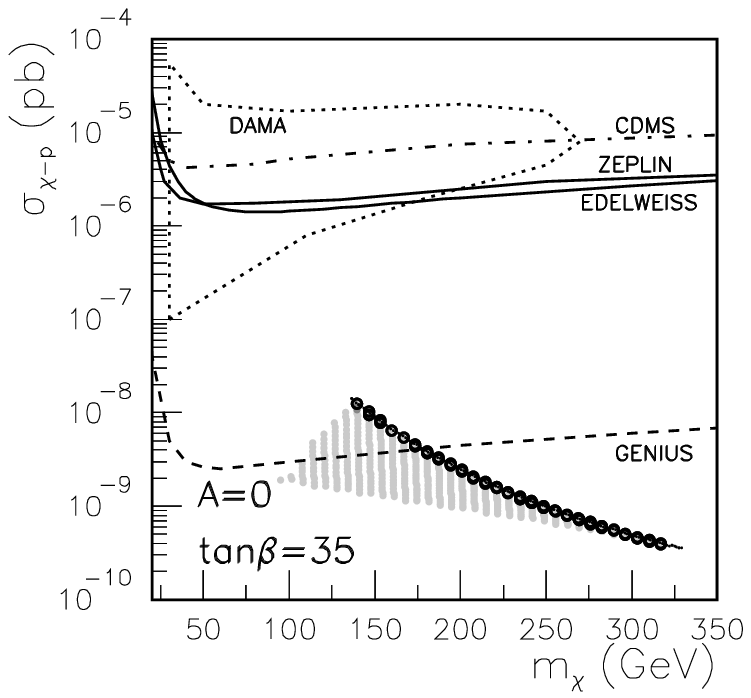,width=6.2cm}
\epsfig{file=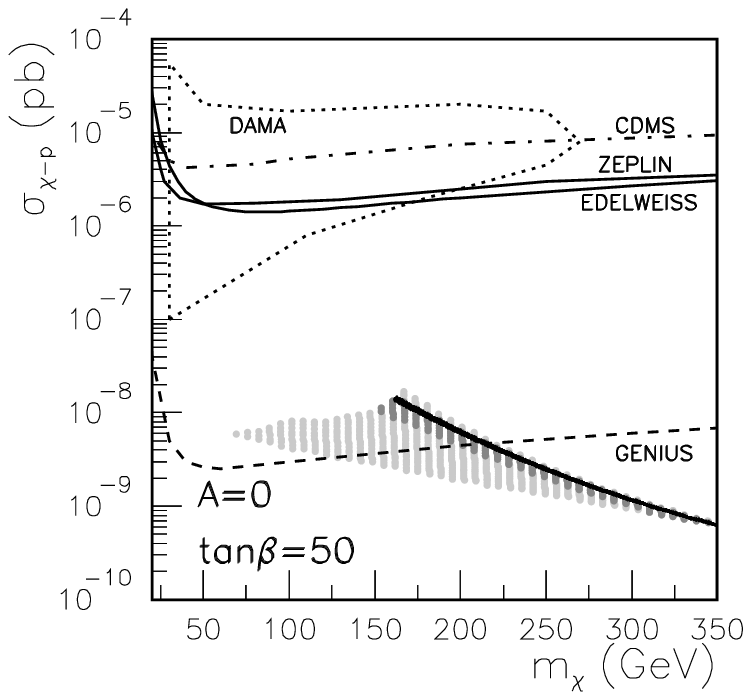,width=6.2cm}
%
\caption{Scatter plot of 
the scalar neutralino-proton cross section
$\sigma_{\tilde{\chi}_1^0-p}$ 
as a function of 
the neutralino mass 
$m_{\tilde{\chi}_1^0}$ in the mSUGRA scenario,
for
$\tan\beta=35$ and 50, $A=0$ and $\mu>0$. 
The light grey dots correspond to points
fulfilling all experimental constraints.
The dark grey dots correspond to points fulfilling in addition
$0.1\leq \Omega_{\tilde{\chi}_1^0}h^2\leq 0.3$
(the black dots on top of these
indicate those fulfilling the WMAP 
range $0.094<\Omega_{\tilde{\chi}_1^0}h^2<0.129$). 
The circles indicate regions 
excluded by the UFB-3 constraint.
The lines corresponding to the different experiments
are as in Fig.~\ref{limits}.
}
\label{cross_scale122}
\end{figure}

The neutralino-proton cross section can be increased when the value
of $\tan\beta$ is increased \cite{Bottino,Arnowitt,Gomez,large}.  
Notice for instance that the contribution of the
down-type quark to the cross section is proportional to
$1/\cos\beta$ (see e.g. Eq.~(\ref{approx})).
In addition, the bottom Yukawa
coupling increases, and as a consequence 
$m_{H_d}^2$ decreases, implying that
$m^2_A$, given by Eq.~(\ref{ma}),
also decreases. Since, as mentioned above, $m_H\approx m_A$,
this will also decrease significantly.
Indeed, scattering channels through Higgs exchange 
are more important now and their contributions to the cross section
will increase it.
Thus, in principle, we can even enter in the DAMA region.
However, 
the present experimental constraints 
exclude this possibility \cite{EllisOlive,Arnowitt3}.
We show this fact for $\tan \beta=35$ and $A=0$ 
in the plot on the left frame 
of 
Fig.~\ref{a35} \cite{darkufb}. 
In principle, 
if we only impose the LEP lower bound $m_{\tilde\chi_1^{\pm}}>103$ GeV,
the cross section can be as large as 
$\sigma_{\tilde{\chi}_1^0-p}\approx 10^{-6}$ pb.
However, at the end of the day,
the 
other 
experimental bounds (Higgs mass, $b\to s\gamma$,
$g_{\mu}-2$ upper bound) 
constrain the cross section  
to be $\sigma_{\tilde{\chi}_1^0-p}\lsim 10^{-8}$ pb.
The region allowed by 
the $g_{\mu}-2$ lower bound is now larger and the cross section 
can be as low as
$\sigma_{\tilde{\chi}_1^0-p}\approx 5\times 10^{-10}$ pb.
These bounds imply 
$260\lsim M\lsim 750$ GeV, and therefore 
$104\lsim m_{\tilde{\chi}_1^0}\lsim 300$ GeV, 
$650\lsim m_{\tilde{g},\tilde{q}}\lsim 1875$ GeV.
Concerning the UFB-3 constraint,
it is worth noticing
that
the larger $\tan\beta$ is,
the larger 
the excluded region becomes. 
However, unlike the case 
$\tan \beta=10$,
this is not sufficient to forbid
the whole dark shaded area allowed also by
the astrophysical bounds.
In fact, one can check that with
$\tan\beta > 20$ one can always find regions 
where all constraints are fulfilled, depending on the
value of $A$. 
For example, for $\tan\beta=35$ 
and $A=M$ essentially the whole dark shaded area is allowed,
whereas for $A=-M,-2M$ this is forbidden.
A
detailed analysis of this issue can be found in Ref.~\refcite{darkufb}. 

The above comments, concerning the cross section, 
can also be applied for very large values of 
$\tan\beta$, as e.g. $\tan \beta=50$\footnote{For this very large values
of $\tan\beta$ results become very sensitive to the precise
values of $m_t$ and $m_b$. This has been discussed e.g. 
in Refs.~\refcite{Gomez,large} and \refcite{Arnowitt3}.}.
We show this in the plot on the right frame 
of Fig.~\ref{a35}. However, note that now, unlike the case $\tan \beta=35$, 
the whole dark shaded area allowed by experimental
and astrophysical bounds is not constrained by the UFB-3.
This is also true for other values of $A$.
Let us finally recall that the region allowed by 
$0.1\leq \Omega_{\tilde{\chi}_1^0}h^2\leq 0.3$
is larger because the 
CP-odd Higgs $A$ becomes lighter as $\tan\beta$ increases, as discussed
above.
This allows the presence of resonances in the Higgs mediated
annihilation channels, resulting in drastic reduction
of the neutralino relic abundance.
In the case of $\tan\beta=50$, the resonant effects in the
annihilation channels are felt in the whole parameter
space displayed in Fig.~\ref{a35}.
We can see as well, that the area of the parameter space where
$\tilde{\chi}_1^0-\tilde{\tau}_1$ 
coannihilations are relevant lead 
to values of 
$\Omega_{\tilde\chi^0_1} h^2<0.1$.

We summarize the above results
for
$\tan\beta=35,50$, and $A=0$, in Fig.~\ref{cross_scale122} \cite{darkufb}. 
There, the values of $\sigma_{\tilde{\chi}_{1}^{0}-p}$
allowed by all experimental constraints 
as a function of the neutralino mass
$m_{\tilde{\chi}_1^0}$ are shown.
Dark grey dots 
correspond to 
those points having a relic neutralino density within
the preferred range $0.1\leq\Omega h^2\leq 0.3$.
Given the narrow range of these points for the case 
$\tan\beta=35$,
they overlap 
in 
the figure with those
excluded by the UFB-3 constraint (shown with circles).
We observe that, generically, the cross section and the neutralino mass
are constrained to be (for any value of $A$)
$5\times 10^{-10}\lsim \sigma_{\tilde{\chi}_1^0-p}\lsim 3\times 10^{-8}$ pb and
$120\lsim m_{\tilde{\chi}_1^0}\lsim 320$ GeV, respectively.

\vspace{0.2cm}

Qualitatively similar results are obtained when dark matter
is analyzed
in the so-called focus-point supersymmetry scenario.
Let us recall that this has been proposed as an alternative
scenario in
order to avoid dangerous SUSY contributions to flavour and
CP violating effects \cite{Feng}. 
The idea is to assume the existence of squarks and sleptons with
masses which can be taken well 
above 1 TeV (of course this scenario rules out SUSY
as an explanation of the
possible deviation in the $g_{\mu}-2$
from the standard model prediction).
It has also been argued that this situation produces no loss of naturalness.
Notice that for $m^2>>M^2$ the electroweak scale given by  
Eq.~(\ref{electroweak2}),
$M_Z^2/2\approx -m_{H_u}^2-\mu^2$,
can easily be obtained
since $m_{H_u}^2$ becomes less negative.

\begin{figure}
\begin{center}
\epsfig{file=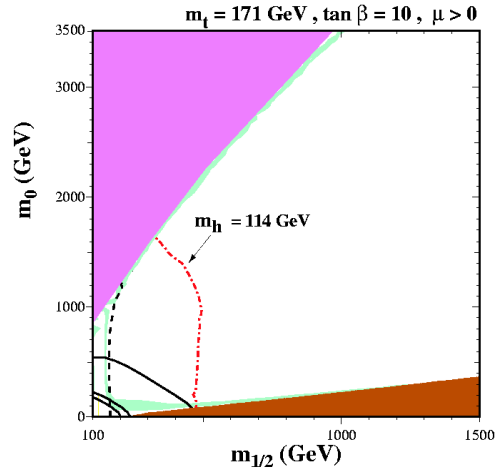,height=
2.5in}
\end{center}
\caption[]{\it An expanded view of the $M - m$ parameter plane
showing the focus-point regions  
at large $m$ for $tan \beta= 10$. 
In the shaded (mauve) region in the 
upper
left corner, there are no solutions with proper electroweak symmetry 
breaking. Note that $m_t = 171$ GeV has been chosen, in 
which case the focus-point region is at lower $m$ than when $m_t = 175$ 
GeV, as assumed in the other figures. The position of this region
is very sensitive to $m_t$. 
}
\label{fig:focus}
\end{figure}

The implications of focus-point supersymmetry for neutralino dark matter
have been considered in Ref.~\refcite{focus}.
In particular, it was pointed out that for 
$m > 1$ TeV  
the lightest neutralino is a gaugino-Higgsino 
mixture over much of parameter space.
This can be understood from Eq.~(\ref{electroweak2}) since
$m_{H_u}^2$ becomes less negative for $m_0 > 1$ TeV
(recall Fig.~\ref{run}), 
and therefore $|\mu|$ decreases.
As a consequence, although as $m$ increases the $t$-channel sfermion exchange
process (see e.g. Fig.~\ref{annihilation})
is more and more suppressed, the LSP gradually acquires a significant
Higgsino component and other diagrams become unsuppressed.
Thus the upper bound in Eq.~(\ref{omega})
is fulfilled. This `focus-point' region, which is adjacent
to the boundary of the region where electroweak symmetry breaking is 
possible, is shown in Fig.~\ref{fig:focus} (from Ref.~\refcite{focusito}).

Concerning the neutralino-proton cross section, 
scattering channels through Higgs exchange will increase it.
However, one still needs large values for 
$\tan \beta$ (of order 50) in order to have 
very large cross sections, and then experimental 
constraints are very important.



\vspace{0.2cm}

Obviously, in the mSUGRA scenario with a GUT scale that we have
reviewed in this Subsection, where
$\sigma_{\tilde{\chi}_1^0-p}\lsim 3\times 10^{-8}$ pb,
more sensitive detectors
producing further data 
are needed.
As discussed in Subsection~4.1.2,
many dark matter detectors are being projected.
Particularly interesting is the case of GENIUS,
where values of the cross section as low as 
$\approx 10^{-9}$ pb will be accessible,
although this might not be sufficient depending on the values
of the parameters (see Fig.~\ref{cross_scale122}).

\subsubsection{mSUGRA scenario with an intermediate scale}

The analysis of the cross section
$\sigma_{\tilde{\chi}_1^0-p}$
carried out above in the context of mSUGRA,
was performed assuming the
unification scale
$M_{GUT} \approx 10^{16}$ GeV.
However, there are several interesting phenomenological 
arguments in favour of SUGRA scenarios
with scales 
$M_I\approx 10^{10-14}$ GeV,
such as to explain neutrino masses, the scale
of axion physics, 
and others \cite{see}.
In addition,
the string scale may be anywhere between the weak and the Planck 
scale, and explicit scenarios with
intermediate scales may arise in the context of 
D-brane constructions from type I strings, 
as we will
discuss in Subsection~5.3.
Inspired by all these scenarios, to use the value of the initial scale 
$M_I$ as a free parameter for the running of the soft terms
is particularly interesting.
In fact, it was  
pointed out in Refs.~\refcite{muas,Bailin,nosotros} 
that 
$\sigma_{\tilde{\chi}_1^0-p}$
is very sensitive to the variation of $M_I$ 
for the running of the soft terms.
For instance, by taking $M_I=10^{10-12}$ GeV rather than 
$M_{GUT}$,
regions in the parameter space of mSUGRA can be found 
where  $\sigma_{\tilde{\chi}_1^0-p}$ is two orders of magnitude
larger than for $M_{GUT}$ \cite{muas,darkcairo,nosopro,darkufb}.

\begin{figure}[t]
\epsfig{file= 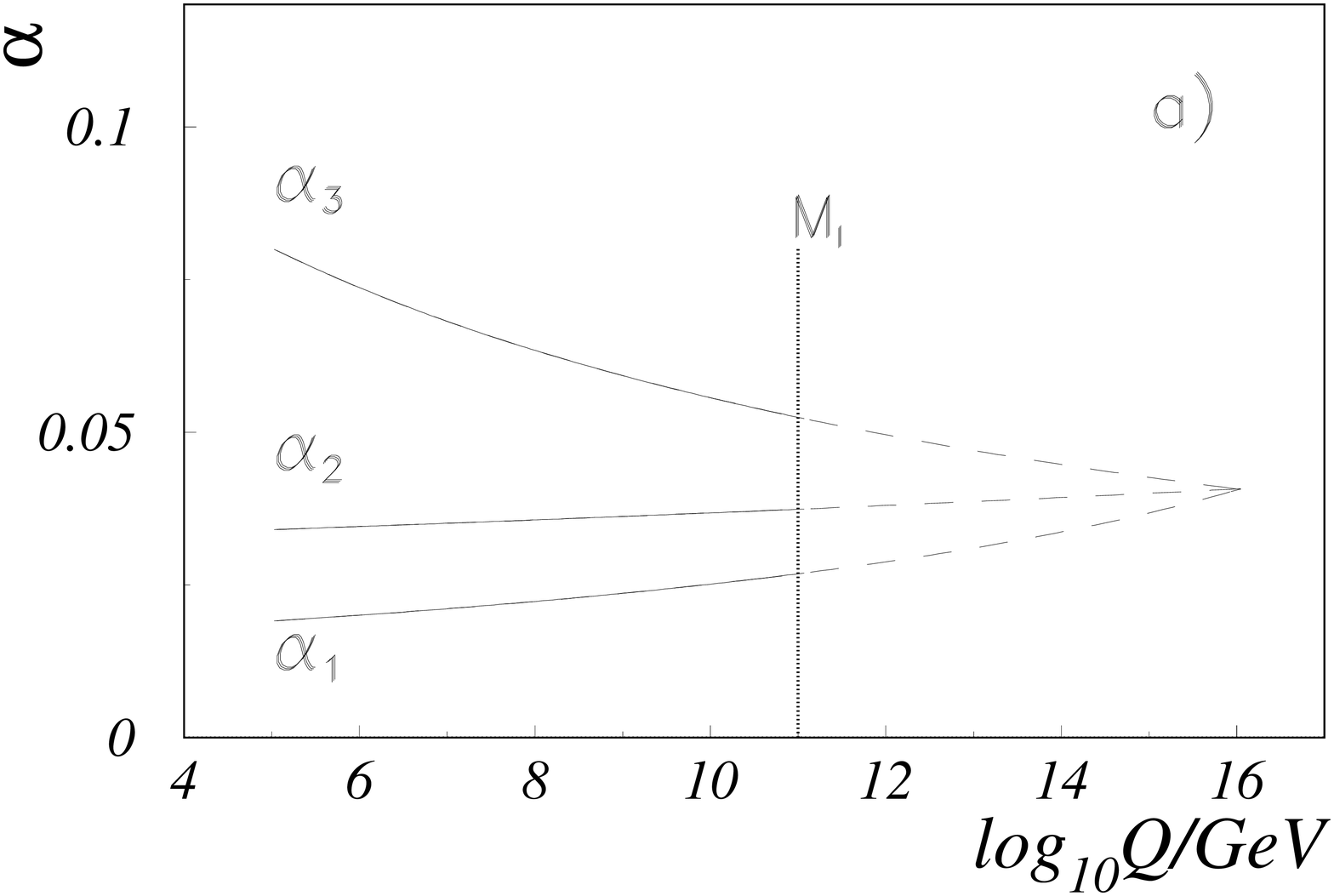,width=5.75cm,height=4.5cm,angle=0}
\hspace{0.5cm}\epsfig{file= 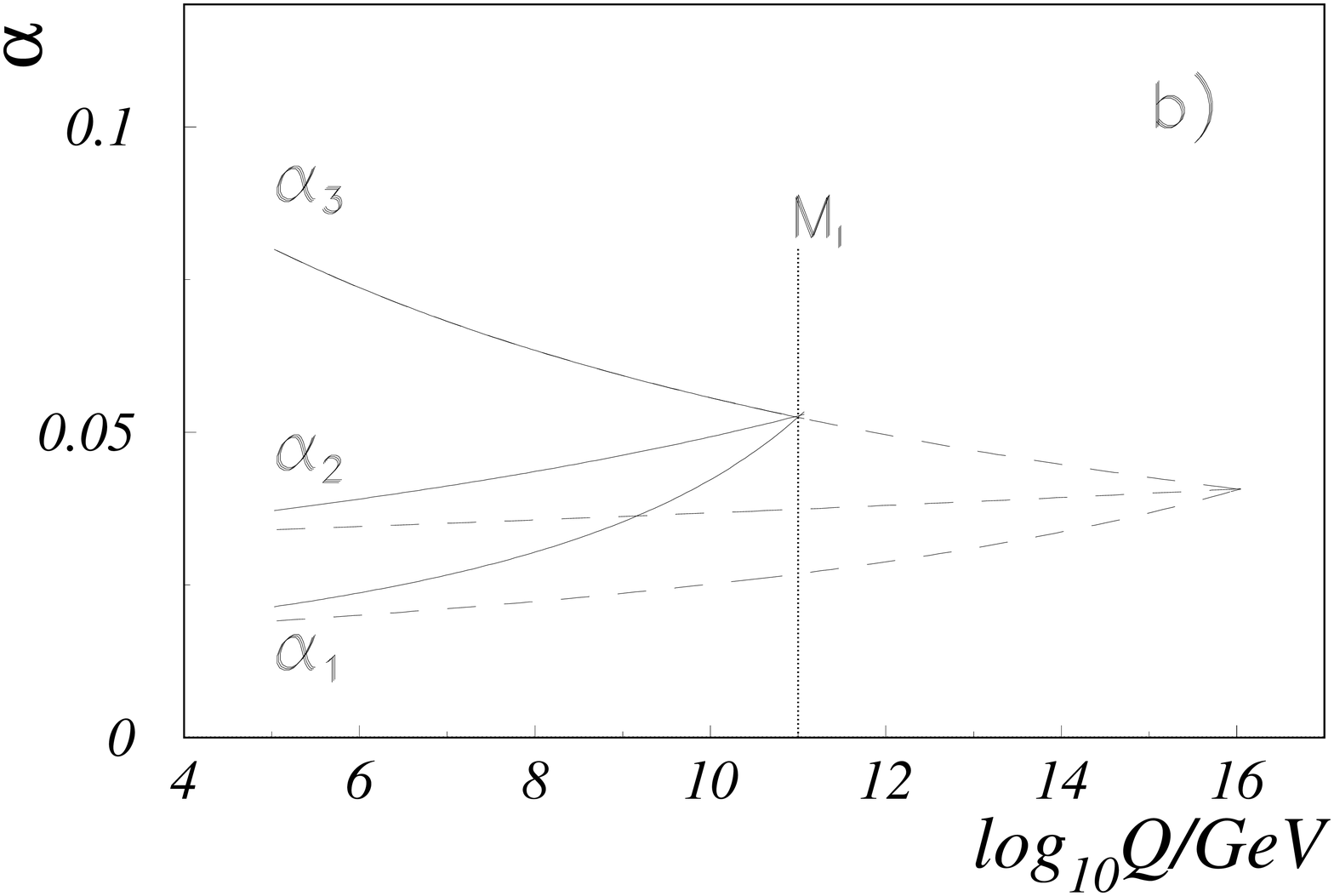, width=5.75cm, height=4.5cm,angle=0}
\caption{Running of the gauge couplings with energy, shown
with solid lines, assuming
(a) non universality and (b) universality of couplings at the
initial scale $M_I$. 
For comparison the usual running of the MSSM couplings
is also shown with dashed lines.}
\label{uni}
\end{figure}

Before trying to understand this result, let us discuss what we
mean by an intermediate unification scale. 
Concerning this point
two possible scenarios
are schematically shown in Fig.~\ref{uni} for the example
$M_I= 10^{11}$ GeV. In scenario {\it (a)} 
the gauge couplings are non universal,
$\alpha_i\neq\alpha$, and their values
depend on the initial scale $M_I$ chosen.
As we will discuss in Subsection~5.3, 
a qualitatively similar scenario may arise
in the context of
type I string constructions  
if the gauge groups of the standard model come from 
different types of D-branes.
Since different D-branes have associated different couplings,
this implies the non universality of the gauge couplings.

On the other hand, scenario {\it (b)} with gauge coupling 
unification at $M_I$, $\alpha_i=\alpha$, can
be obtained with 
the addition of extra fields
in the massless spectrum. For the example of the figure these are
doublets and singlets under the standard model gauge group.
As discussed in Ref.~\refcite{muas},
the values of the gauge coupling constants at the
intermediate
scale are important in the computation of the cross section, and
scenario {\it (a)} gives rise to larger cross sections than scenario
{\it (b)}. Let us then concentrate on the former.

The fact that smaller scales imply a larger 
$\sigma_{\tilde{\chi}_1^0-p}$  
can be explained with the variation in the value of 
$\mu$ with $M_I$.
One observes that, for $\tan\beta$ fixed, the smaller the initial
scale for the running the smaller the numerator in the
first piece of Eq.~(\ref{electroweak}) becomes. 
This can be understood from 
the evolution of $m_{H_u}^2$ with the
scale (see Fig.~\ref{run}).
Clearly, when the value of the initial scale is reduced 
the RGE running is shorter and, as a consequence, 
the negative contribution 
$m_{H_u}^2$ to $\mu^2$ in Eq.~(\ref{electroweak}) becomes less important.
Then, 
$|\mu|$ decreases and therefore
the Higgsino composition of the lightest neutralino increases.
Eventually, $|\mu|$ will be of the order of $M_1$, $M_2$
and $\tilde{\chi}_1^0$ will be a mixed Higgsino-gaugino 
state (see the plot on the frame of Fig.~\ref{N_1i}).
In addition, when $|\mu|$ decreases 
$m^2_A$, given by Eq.~(\ref{ma}),
also decreases. 
As mentioned in the previous Subsection
when talking about increasing $\tan\beta$,
$H$ will decrease and therefore
the scattering channels through Higgs exchange will increase
the cross section.

Let us also remark that, for the same value of the parameters, 
the lightest 
Higgs mass $m_h$ decreases with respect to the GUT scale scenario.
This is because the value of $m_h$ depends on the value of
the gluino mass $M_3$. It increases when $M_3$ increases at low energy.
However, now the running is shorter and therefore $M_3$ at low energy
is smaller than in the GUT scenario.
Although the latter may be welcome in order to obtain larger
cross sections, it may also be dangerous when
confronting with the experimental result concerning 
the Higgs mass.

\begin{figure}
%
\epsfig{file=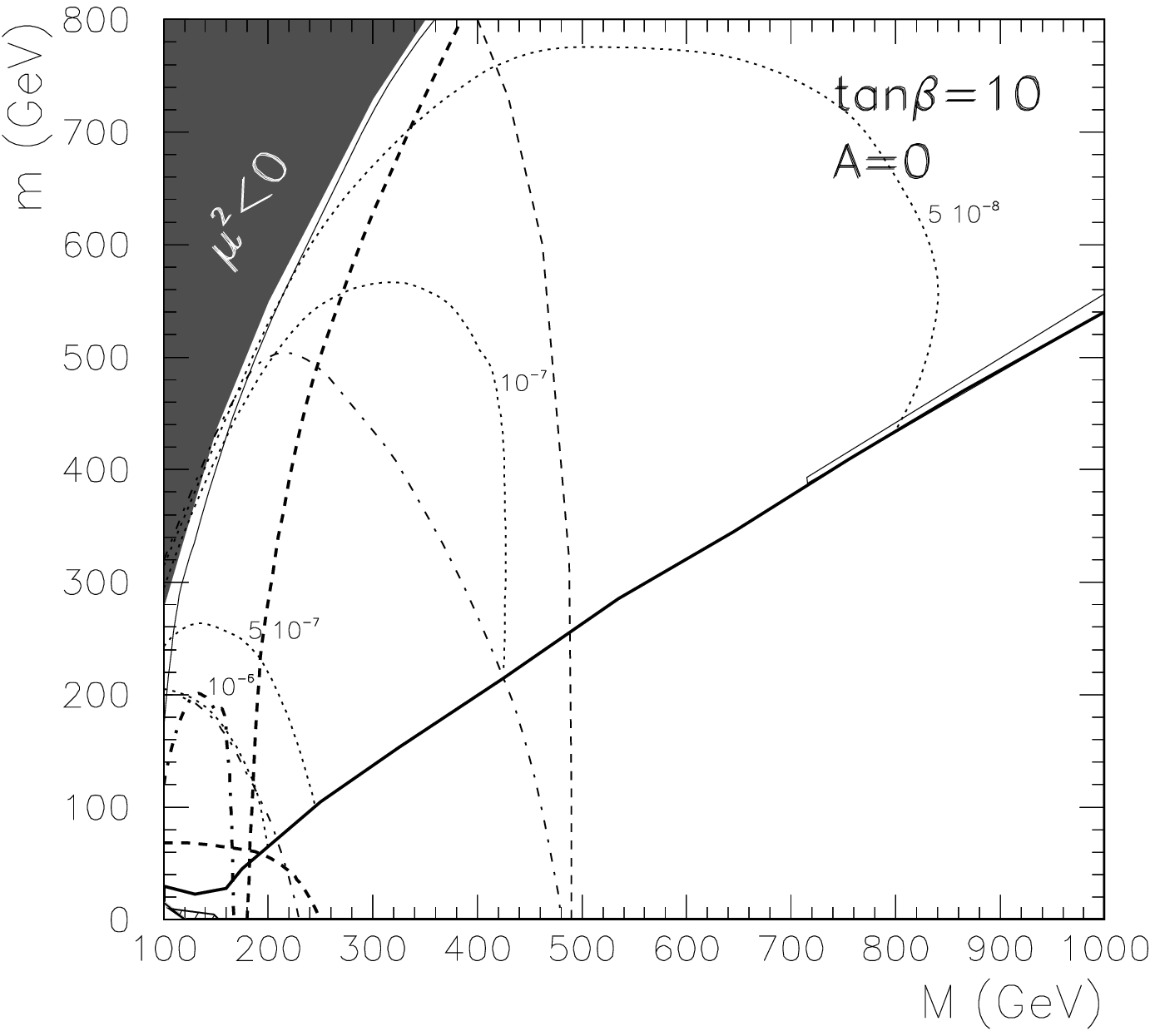,
width=6.2cm}
%
\epsfig{file=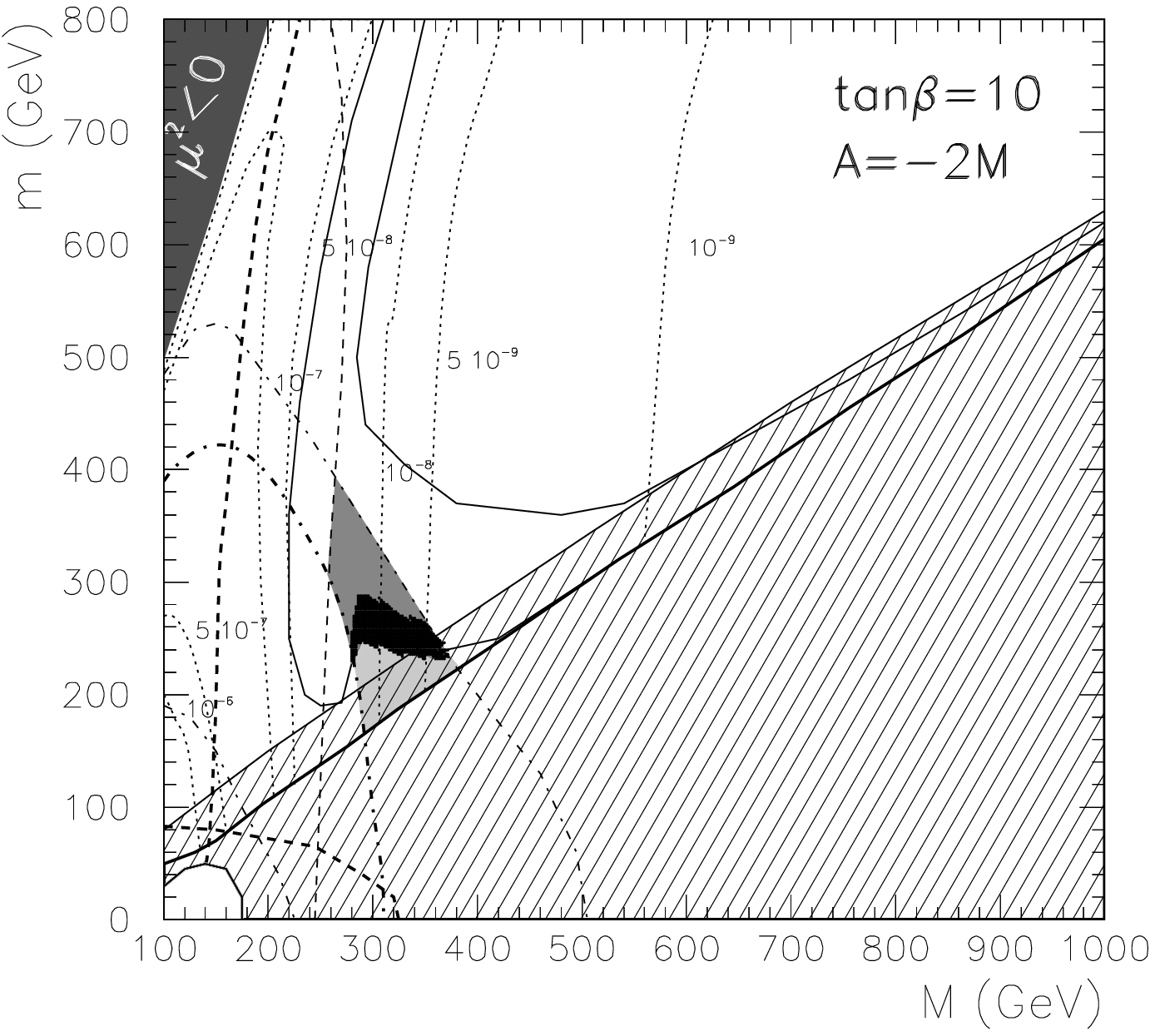,
width=6.2cm}
\caption{The same as in Fig.~\ref{a2m} but for the 
intermediate scale 
$M_I=10^{11}$ GeV, with 
$A=0,-2M$. 
The black area is excluded because
$\mu^2$ becomes negative.
The white region at the bottom bounded by a solid line
is excluded because 
$m_{\tilde{\tau}_1}^2$
becomes negative.}
\label{scale12_10}
\end{figure}

Concerning the value of the relic density,
$\Omega_{\tilde{\chi}_1^0}$ is dramatically reduced
with respect to the $M_{GUT}$ case. 
This is due to a combination of several factors:
1) The Higgsino-gaugino composition of $\tilde{\chi}_1^0$
allows a significant increase of the $\tilde{\chi}_1^0$ annihilation
cross section, due to channels with Higgs and gauge bosons
in the final states; 2) The decrease of the mass of the pseudoscalar
Higgs, along with the value of the $\mu$--term, enables the presence of
resonant annihilation channels even at $\tan\beta=10$;
3) The masses of the lightest chargino and stop are small enough
to allow $\tilde{\chi}_1^0$--$\tilde{\chi}_1^\pm$ \cite{char} 
and $\tilde{\chi}_1^0$--$\tilde{t}_1$ \cite{stop}
coannihilations in some areas of the parameter space. Although 
the later is less relevant, we find some areas at $\tan\beta=50$ and $A<0$.

We show in 
Fig.~\ref{scale12_10} the result for $M_I=10^{11}$ GeV, with
$\tan\beta=10$ and $A=0,-2M$ \cite{darkufb}. 
We choose $A$ proportional to $M$ because this relation is 
particularly interesting, arising naturally in several string 
models \cite{dilaton}. However our conclusions will be independent on 
this assumption. 
For example, if we choose to do the plots for different constant 
values of $A$, a very common procedure in pure SUGRA analyses,
the results will be qualitatively similar.
The case $A=0$
can be compared with the one in
Fig.~\ref{a2m}, where $M_{GUT}$ is used. 
Now the relation $m_{\tilde{\chi}_1^0}\sim 0.4\ M$ 
does not hold, and one has $m_{\tilde{\chi}_1^0}> 0.4M$. In any case, 
$m_{\tilde{\chi}_1^0} <M_1$ since the bino-Higgsino mixing is significant in
this case.
Clearly, for the same values of the parameters,
larger cross sections can be obtained with the intermediate scale.
It is worth noticing that even with this moderate value of 
$\tan\beta$, $\tan\beta=10$,
there are regions where the cross section enters in the DAMA 
area, 
$\sigma_{\tilde{\chi}_1^0-p}\approx 10^{-6}$ pb.
However,
for $A=0$
the whole parameter space is forbidden
due to the combination of the Higgs mass bound with
the $g_{\mu}-2$ lower bound (for $A=M$ the situation is similar).
We have checked explicitly that 114.1 GeV is the correct lower bound to be
used
concerning the Higgs mass, since generically
$\sin^2(\alpha-\beta)\sim 1$ for the intermediate scale.
Notice also that now, for $A=0$ (and also for $A=M$),
$\Omega_{\tilde{\chi}_1^0}h^2$ is smaller than 0.1 in most of the
parameter space. Only tiny regions bounded by solid lines in the 
figure,
and therefore with 
$0.1\leq \Omega_{\tilde{\chi}_1^0}h^2\leq 0.3$,
can be found.


\begin{figure}
\epsfig{file=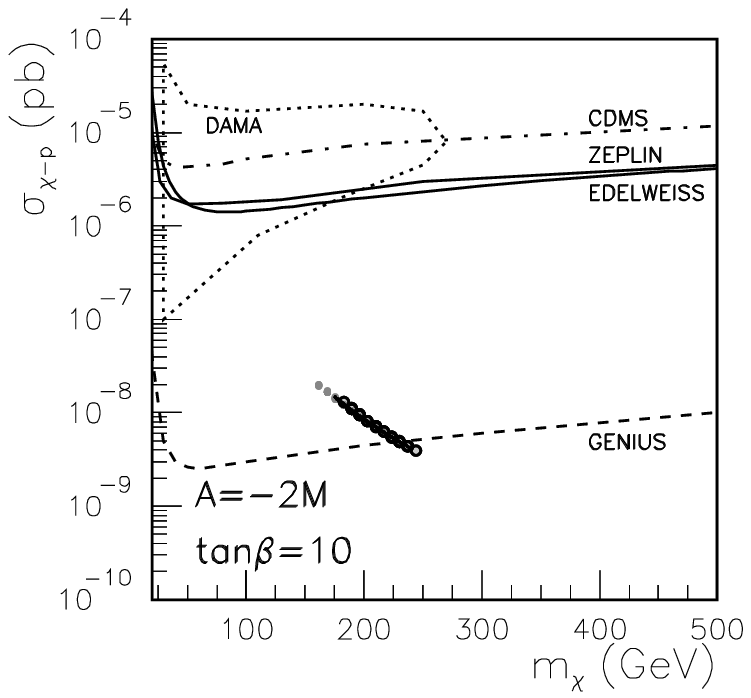,width=6.2cm}
\epsfig{file=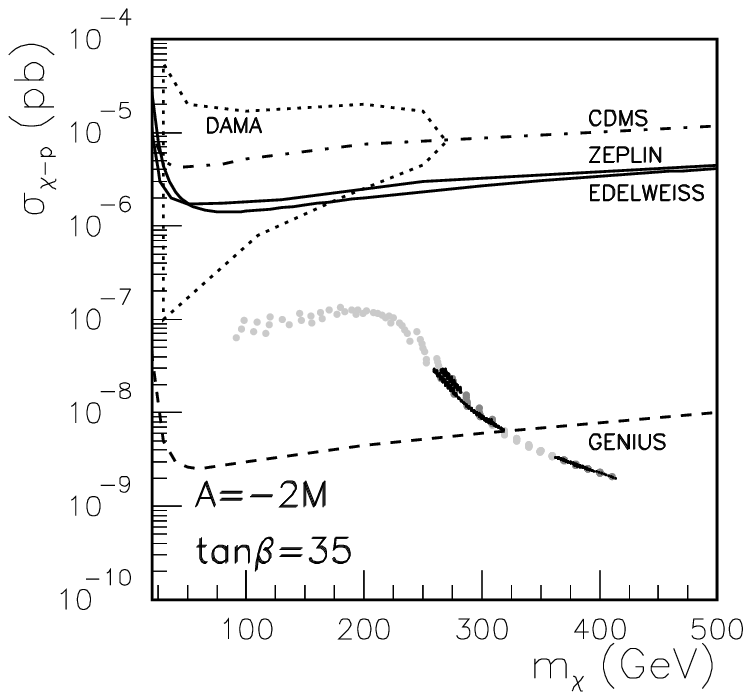,width=6.2cm}
\caption{
The same as in Fig.~\ref{cross_scale122}
but for the intermediate scale
$M_I=10^{11}$ GeV,
with
$\tan\beta=10,35$ and $A=-2M$.
\label{cross_scale12}}
\end{figure}

On the other hand, for
$A=-2M$ (and also for $A=-M$) there are small regions where the
$m_h$ and $g_{\mu}-2$ bounds
are compatible, but finally the constraint $m_h>114.1$ GeV implies
that the allowed cross sections do not enter in the DAMA area.
Although now larger regions with 
$0.1\leq \Omega_{\tilde{\chi}_1^0}h^2\leq 0.3$
are present, for $A=-M$ these are not compatible with the
experimental bounds.





One also finds that the regions excluded by the UFB-3 constraint 
are much smaller than in those cases where the initial 
scale is the GUT one. 
For example for $A=0$ (and $A=M$) no region is excluded
(see however Fig.~\ref{a2m} for the GUT case). For $A=-M$ 
the region excluded by the UFB-3 
is smaller than the one
forbidden by the LSP bound. We have to go to
$A=-2M$ to have it larger.

In the plot on the left frame of Fig.~\ref{cross_scale12} 
we summarize the above results for
$\tan\beta=10$, concerning the
cross section \cite{darkufb}, 
showing the values of $\sigma_{\tilde{\chi}_{1}^{0}-p}$
allowed by all experimental constraints 
as a function of the neutralino mass
$m_{\tilde{\chi}_1^0}$, for $A=-2M$.
Only in this case there are dark grey dots corresponding
to points having a relic neutralino density within the
preferred range
$0.1\leq \Omega_{\tilde{\chi}_1^0}h^2\leq 0.3$.
Given the narrow range of these points,
they overlap in the figure with those
excluded by the UFB-3 constraint.
They correspond to
$150\lsim m_{\tilde{\chi}_1^0}\lsim 250$ GeV, and e.g. the mass
of the lightest stop, $\tilde{t_1}$, is between 250 and 340 GeV. 

Qualitatively, similar results are obtained for larger values of $\tan\beta$.
For example, for $\tan\beta=35$ 
only for $A=-2M$ we obtain
points allowed by all experimental and astrophysical constraints,
and this is shown in the plot on the right frame 
of Fig.~\ref{cross_scale12}
The upper bound in the cross section 
is because of the $b\to s\gamma$ process.
Points within the preferred astrophysical range correspond to 
$\sigma_{\tilde{\chi}_{1}^{0}-p}\lsim 10^{-8}$ pb.
Now, there are two allowed regions with 
$275\lsim m_{\tilde{\chi}_1^0}\lsim 325$ GeV and
$370\lsim m_{\tilde{\chi}_1^0}\lsim 420$ GeV. For these, 
$570\lsim m_{\tilde{t_1}}\lsim 720$ GeV.
Let us finally mention
that in the case
of $\tan\beta=50$,
for $A=-M$ there are points allowed by all experimental and
astrophysical
constraints, and 
$\sigma_{\tilde{\chi}_{1}^{0}-p}\lsim 10^{-7}$ pb. 

It is worth noticing  
that in this analysis 
gaugino mass universality has been assumed
at the high-energy scale, 
although in this scenario gauge couplings do not unify. 
This situation is in principle
possible in generic SUSY models, however it is not so natural in
SUSY models from SUGRA 
where gaugino masses and gauge couplings
are related through the gauge kinetic function. 
Since an explicit string construction with nonuniversal
gauge couplings and gaugino masses will be analyzed in detail 
in Subsection~5.3,
we have chosen to simplify the discussion here
assuming gaugino mass universality.

\vspace{0.2cm}

Summarizing,
when an intermediate scale is considered in mSUGRA,
although 
the cross section increases significantly
the experimental bounds
impose
$\sigma_{\tilde{\chi}_1^0-p}\lsim 4\times 10^{-7}$ pb.
And, in fact, at the end of the day, the preferred astrophysical range
for the relic neutralino density, 
$0.1\leq\Omega_{\tilde{\chi}_1^0}h^2\leq 0.3$,
imposes 
$\sigma_{\tilde{\chi}_1^0-p}\lsim  10^{-7}$ pb.
Clearly, present experiments are not still sufficient,
and more sensitive detectors
producing further data 
are needed, as in the case of a GUT scale.

\subsubsection{SUGRA scenario with non-universal soft terms}

The general situation for 
the soft parameters in SUGRA is to have a
non-universal structure \cite{dilaton}. 
For the case of the observable scalar masses
this is due to the non-universal couplings
in the K\"ahler potential
between the hidden sector fields breaking SUSY and the
observable sector fields.
For example, 
$K=\sum_{\alpha}{\tilde K}_\alpha(h_m,h_m^*) 
\ C_{\alpha} C_{\alpha}^*$, with ${\tilde K}_\alpha$ a function of
the hidden-sector
fields $h_m$, will produce non-universal scalar masses
$m_{\alpha}\neq m_{\beta}$
if ${\tilde K}_\alpha\neq {\tilde K}_\beta$.
For the case of the gaugino masses this is due to the
non-universality of the gauge kinetic functions associated with the
different gauge groups $f_a(h_m)$.
For example, non-universal gaugino masses are obtained if 
the $f_a$ have a different dependence on the hidden sector fields.
We will see in 
Subsection~5.3 
that general string constructions whose low-energy limit is SUGRA,
exhibit these properties \cite{dilaton}.

It was shown in the literature
that the non-universality of the soft parameters allows to 
increase 
the neutralino-proton cross section with respect to the universal
case. 
This can be carried out with non-universal 
scalar masses 
and/or gaugino masses 
We will concentrate on this possibility here.

\vspace{0.5cm}

\noindent {\it (i) Non-universal scalar masses}

\vspace{0.5cm}

Let us analyse a SUGRA scenario with GUT scale
and non-universal soft scalar masses. In fact, 
non-universality in the Higgs sector, concerning dark matter, was first 
studied in 
Refs.~\refcite{Ole,Fornengo,Bottino}. Subsequently, non-universality in
the sfermion sector was added in the analysis \cite{arna,arna2,Arnowitt}.
Analyses of the dark matter cross section using 
generic non-universal soft masses
were carried out in 
Refs.~\refcite{Bottino,arna2,Arnowitt,Santoso,darkcairo,nosopro,Dutta,darkufb},
whereas
in Refs.~\refcite{Drees,Nojiri,Arnowitt3,Rosz,Farrill,Profumo}
$SU(5)$ or 
$SO(10)$ GUT relations
were used. 
In the light of the recent experimental results,
one important consequence of the non-universality is that
the cross 
section can be increased in some regions of the parameter 
space.

Let us then parameterized
this non-universality in the Higgs sector as
follows:
\begin{eqnarray}
m_{H_{d}}^2=m^{2}(1+\delta_{1})\ , \quad m_{H_{u}}^{\ 2}=m^{2}
(1+ \delta_{2})\ .
\label{Higgsespara}
\end{eqnarray}
Concerning squarks and sleptons, in order to avoid potential problems
with FCNC, one can assume that the first
two generations have a common scalar mass $m$ at $M_{GUT}$, and
that non-universalities are allowed only for the third generation:
\begin{eqnarray}
m_{Q_{L}}^2&=&m^{2}(1+\delta_{3})\ , \quad m_{u_{R}}^{\ 2}=m^{2}
(1+\delta_{4})\ , 
\nonumber\\
m_{e_{R}}^2&=&m^{2}(1+\delta_{5})\ ,  \quad m_{d_{R}}^{\ 2}=m^{2}
(1+\delta_{6})\ , \nonumber\\
m_{L_{L}}^2&=&m^{2}(1+\delta_{7})\ ,     
\label{Higgsespara2}
\end{eqnarray}
where  
$Q_{L}=(\tilde{t}_{L}, \tilde{b}_{L})$, $L_{L}=(\tilde{\nu}_{L},
\tilde{\tau}_{L})$, $u_{R}=\tilde{t}_{R}$ and $e_{R}=\tilde{\tau}_{R}$.
Note that whereas $\delta_{i} \geq -1 $, $i=3,...,7$, in order to avoid
an UFB direction breaking charge and colour, 
$\delta_{1,2} \leq -1$
is possible as long as the conditions 
$m_1^2=m_{H_{d}}^2+\mu^2>0$, $m_2^2=m_{H_{u}}^2+\mu^2>0$  
are fulfilled. 

As discussed for intermediate scales in Subsection~5.2.2,
an important factor in order to increase the cross section 
consists in reducing the value of $|\mu|$.
This value is determined by condition (\ref{electroweak2}) and can
be significantly reduced for some choices of the $\delta$'s. We can have
a qualitative
understanding of the effects of the $\delta$'s on $\mu$ from
the following.
First, when $m_{H_u}^2$ at $M_{GUT}$ increases
its negative low-energy contribution to Eq.~(\ref{electroweak2})
becomes less important. Second, when $m_{Q_{L}}^2$ and $m_{u_{R}}^2$
at $M_{GUT}$ decrease, due to their contribution proportional
to the top Yukawa coupling in the RGE of $m_{H_u}^2$, the negative
contribution of the latter to $\mu^2$ is again less important.
Thus one can deduce that 
$\mu^2$ will be reduced (and hence $\sigma_{\tilde{\chi}_1^0-p}$ increased)
by choosing $\delta_{3,4} < 0$ and $\delta_2 >0$.
In fact non-universalities in the Higgs sector give the most important
effect, and including the one in the sfermion sector the cross
section only increases slightly. Thus in what follows we will take
$\delta_{i}=0$, $i=3,...,7$.

\begin{figure}
\epsfig{file=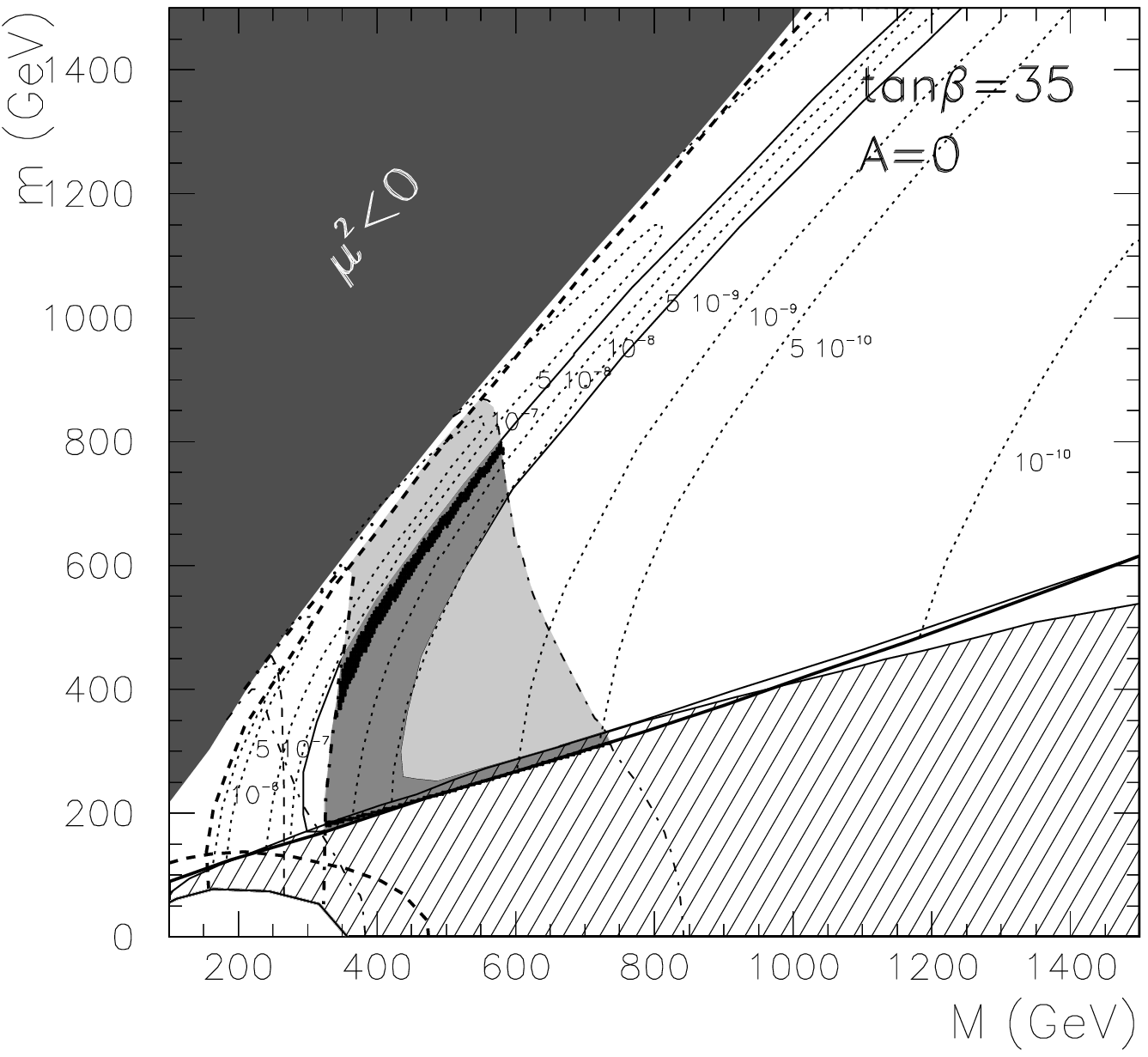,width=6.5cm}
%
\hspace*{-0.51cm}
\epsfig{file=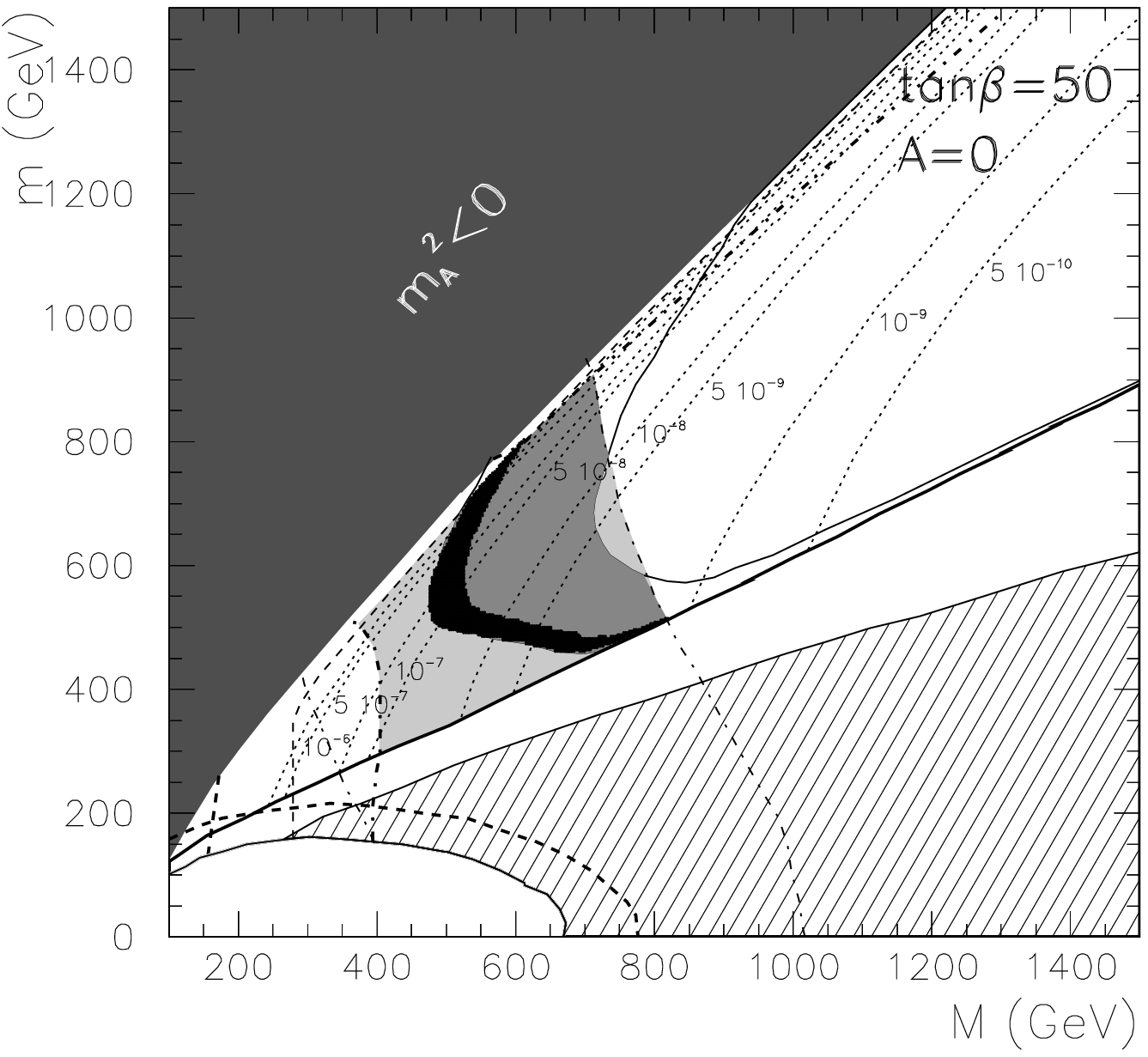,width=6.5cm} 

\caption{The same as in 
Fig.~\ref{a35} 
but for the non-universal case {\it a)} $\delta_1=0$, $\delta_2=1$, 
discussed in Eq.~(\ref{3cases}), with $\tan\beta=35,50$ and $A=0$.}
\label{anu}
\end{figure}

Concerning the value of the relic density,
$\Omega_{\tilde{\chi}_1^0}$ is affected due to the increase
of the Higgsino  components of $\tilde{\chi}_1^0$ with respect
to
the dominant bino component of the universal case. The change in $\mu$ also
determines the presence of the Higgs mediated resonant channels. 
In contrast
to Subsection~5.2.2, the most relevant coannihilation
scenarios are ${\tilde{\chi}_1^0}$--${\tilde{\tau_1}}$, in particular
$\tilde{\chi}_1^0$--$\tilde{\chi}_1^\pm$ coannihilations are only sizeable
for $\tan\beta=35$ in Fig.~\ref{anu} (see the discussion below), 
when the $\mu$--parameter becomes small.
However, even in this case the area inside the WMAP bounds corresponds to
neutralino annihilations, which are enhanced due to enlargement of its
Higgsino components.

On the other hand, there is another relevant 
way of increasing the cross section using the non-universalities
of the Higgs sector. Note that 
decreasing 
$m_{H_d}^2$, i.e. choosing $\delta_1 < 0$,
leads to a decrease in 
$m^2_A$ given by Eq.~(\ref{ma}),
and therefore in the mass of the heaviest 
Higgs $H$ (on the contrary, the lightest
Higgs mass, $m_h$, is almost unaltered, it only decreases less than 1\%).
This produces an increase in the
cross section\footnote{This effect might also be important when
non-universal gaugino masses are taken into account.
The contribution of $M_3$ proportional
to the bottom Yukawa coupling in the RGE of $m_{H_d}^2$ will
do this smaller if $M_3$ is large \cite{prepa}.}. 

\begin{figure}
\epsfig{file=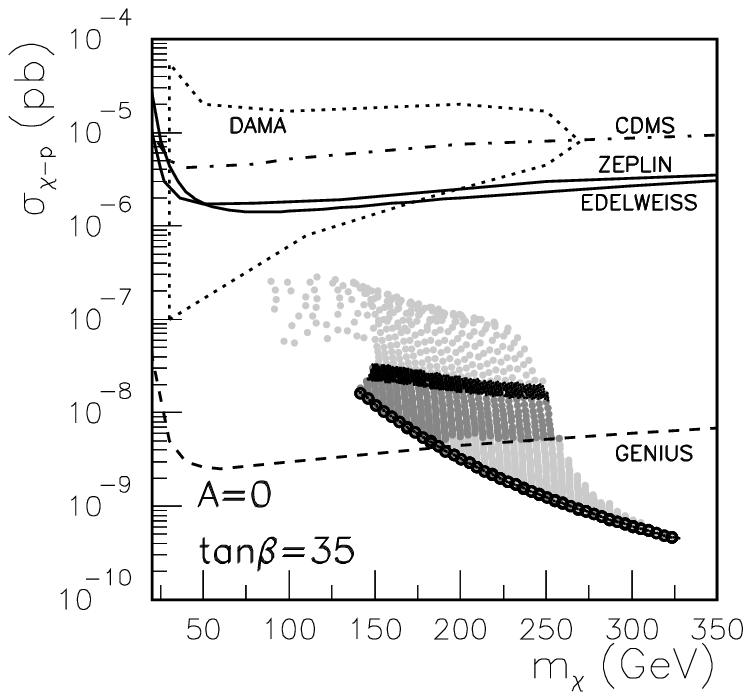,width=6.5cm}
%
\hspace*{-0.51cm}\epsfig{file=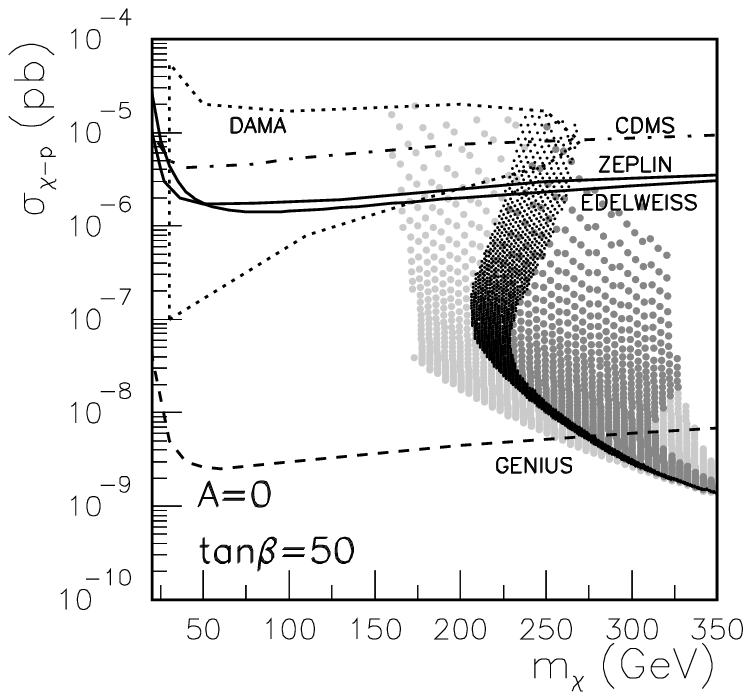,width=6.5cm}
\caption{The same as in Fig.~\ref{cross_scale122} but for 
the non-universal case {\it a)} $\delta_1=0$, $\delta_2=1$, 
discussed in Eq.~(\ref{3cases}), with $\tan\beta=35,50$ and $A=0$.}
\label{anusec}
\end{figure}

Thus we will see that,
unlike the universal scenario in Subsection~5.2.1., 
with non-universalities is possible
to obtain large
values of the cross section, and even some points
enter in the DAMA area fulfilling
all constraints. 
Let us analyze three representative cases \cite{darkufb} with
\begin{eqnarray}
a)\,\, \delta_{1}&=&0\ \,\,\,\,\,\,\,\,,\,\,\,\, \delta_2\ =\ 1\ ,
\nonumber\\
b)\,\, \delta_{1}&=&-1\ \,\,\,\, ,\,\,\,\, \delta_2\ =\ 0\ ,
\nonumber\\
c)\,\, \delta_{1}&=&-1\ \,\,\,\, ,\,\,\, \delta_2\ =\ 1 
\ .
\label{3cases}
\end{eqnarray}

\begin{figure}
\epsfig{file=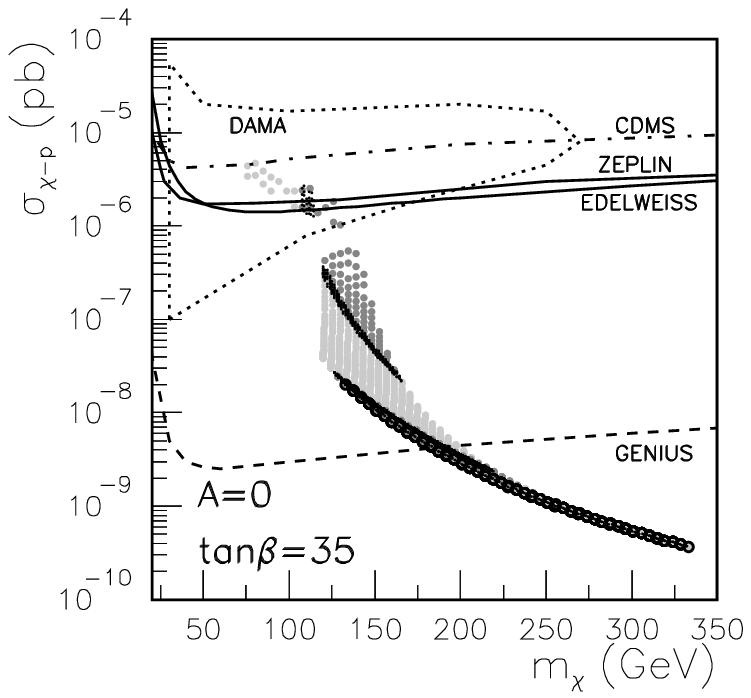,width=6.5cm}
%
\hspace*{-0.51cm}\epsfig{file=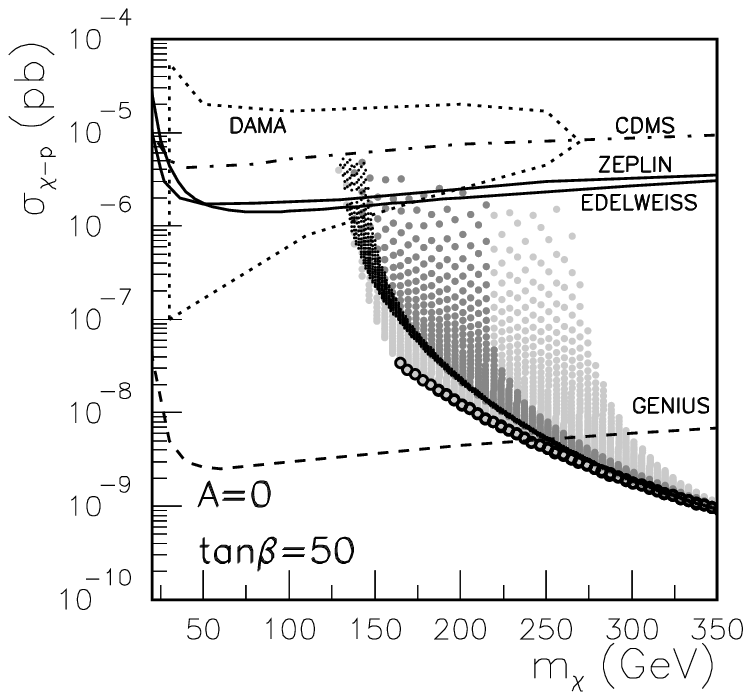,width=6.5cm}


\caption{The same as in Fig.~\ref{cross_scale122}
but for the non-universal case {\it b)} $\delta_1=-1$, $\delta_2=0$, 
discussed in Eq.~(\ref{3cases}), with $\tan\beta=35,50$ and $A=0$.}
\label{anusec2}
\end{figure}





\begin{figure}
\epsfig{file=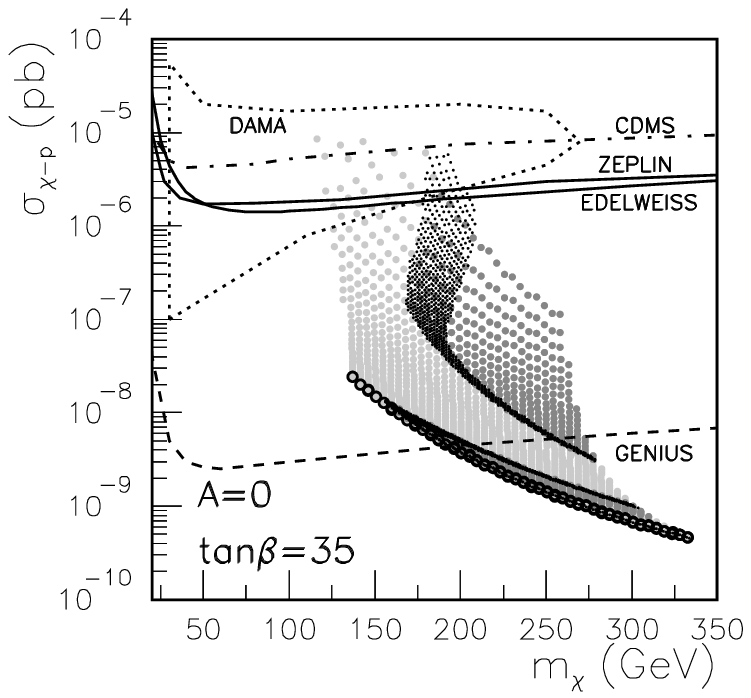,width=6.5cm}
%
\hspace*{-0.51cm}\epsfig{file=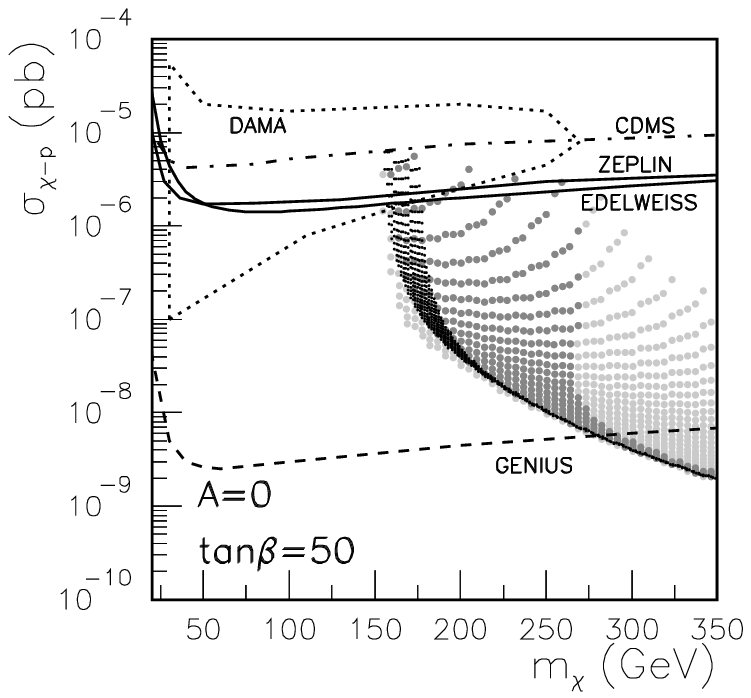,width=6.5cm}


\caption{The same as in Figs.~\ref{cross_scale122}
but for the non-universal case {\it c)} $\delta_1=-1$, $\delta_2=1$, 
discussed in Eq.~(\ref{3cases}), with $\tan\beta=35,50$ and $A=0$.}
\label{anusec3}
\end{figure}

Clearly, the above discussion about decreasing $\mu^2$ applies
well to case {\it a)}, where the variation in
$m_{H_u}^2$ through $\delta_2$ is relevant.
This is shown in Fig.~\ref{anu} for $\tan\beta=35,50$ and $A=0$, which
can be compared with 
Fig.~\ref{a35}.
Note that now, for $\tan\beta=35$, 
there is an important area in the upper left
where $\mu^2$ becomes negative due to the increasing in
$\delta_2$ with respect to the universal case.
A larger area is forbidden for large values of $\tan\beta$, as e.g.
$\tan\beta=50$, 
but now because $m_A^2$ becomes negative. This is similar to what occurs
in the universal scenario when $\tan\beta\gsim 60$,
as discussed in Subsection~5.1.
Notice that from Eq.~(\ref{electroweak2}) 
one can write $m^2_A$ in Eq.~(\ref{ma}) 
as $m^2_A\approx m_{H_d}^2-m_{H_u}^2-M_Z^2$.
Since
$m_{H_u}^2$ at $M_{GUT}$ increases
its negative low-energy contribution 
becomes less important. In addition,  
the bottom Yukawa coupling is large and the $m_{H_d}^2$
becomes negative. 
As a consequence $m_A^2<0$.

For $\tan\beta=35$,
although the cross section increases with respect to the universal
case, and is generically above the GENIUS lower limit,
the present experimental constraints exclude points
entering in the DAMA area. 
This can be seen more clearly comparing Figs.~\ref{anusec} 
and \ref{cross_scale122}.
Notice also that the astrophysical bounds
$0.1\lsim\Omega_{\tilde{\chi}_1^0}h^2\lsim 0.3$ imply
$\sigma_{\tilde{\chi}_1^0-p}\approx 10^{-8}$ pb.
On the contrary, for $\tan\beta=50$ there are points
entering in the DAMA area, and even part of them            
fulfil the astrophysical bounds.
We have checked that for $A=M$ the figures are similar, although
no points enter in the DAMA area, even for $\tan\beta=50$.
On the other hand, the region forbidden by
the LSP bound is larger than the one forbidden by the UFB-3 constraint.

We have also checked that larger values of $\delta_2$,
as e.g. $\delta_2=1.5$, give rise to similar figures.
For small values, $\delta_2\gsim 0.2$ is sufficient to enter in DAMA
fulfilling the
experimental bounds
with $\tan\beta=50$. In fact, e.g., for $\delta_2=0.5, 0.75$ one also gets
many points entering in DAMA as for $\delta_2=1$, however, they do not
fulfil the astrophysical bounds. For the latter one needs
$\delta_2>0.85$.

Let us finally remark that although $\sin^2(\alpha-\beta)$ is close to 1 in
most of the points, some of them can have smaller values
when $\tan\beta=50$. As discussed in Subsection~5.1, these must be points 
with small values for $m_A$, and in fact in this case are those
close to the region with
$m_A^2<0$.
The same
situation occurs for the other cases studied below. 
Thus, according to our discussion
in Subsection~5.1, we use for these points the appropriate bound on the Higgs
mass \cite{barate}. In particular, in Fig.~\ref{anusec}, the light grey dots
above the CDMS line correspond to these points.





For case {\it b)} the cross section increases also substantially
with respect to the universal case.
Now $\delta_2$ is taken vanishing and therefore the value 
of $\mu$ is essentially not modified with respect to the universal
case.
However, taking $\delta_1=-1$
produces an increase in the
cross section through 
the decrease in $m_A^2$,
as discussed previously. 
As
shown explicitly in Fig.~\ref{anusec2}, for
$\tan\beta=35$ and $A=0$,
there are points in the DAMA region. 
All of them correspond to $\sin^2(\alpha-\beta)$ not close to 1, and therefore
with an experimental bound on the Higgs mass smaller than 114.1 GeV.
All points with $\sin^2(\alpha-\beta)\sim 1$ have $\sigma_{\tilde{\chi}_1^0-p}\lsim 6\times 10^{-7}$ pb.
Note that points
with large values of the cross section
fulfil in this case the astrophysical bound
$\Omega_{\tilde{\chi}_1^0}h^2\gsim 0.1$.
Large values of $m$ reduce the resonant effects produced
by the Higgs $A$, and are sufficient to place the relic
abundance inside the bounds.
For $\tan\beta=50$, similarly to case {\it a)}, there are points
entering in the DAMA 
area, and part of them
fulfil the astrophysical bounds. Those above the ZEPLIN line correspond
to $\sin^2(\alpha-\beta)$ not close to 1.
For $A=M$ the figures as similar.

We have checked that smaller values of $\delta_1$,
as e.g. $\delta_1=-1.5,-2$, give also rise to similar figures.
For larger values, $\delta_1\lsim -0.4$ is sufficient to enter in DAMA
fulfilling the
experimental and astrophysical bounds
with $\tan\beta=50$.

Finally, given the above situation concerning the enhancement
of the neutralino-proton cross section for {\it a)} and 
{\it b)}, it is clear that the combination of both cases
might be interesting.
This is carried out
in case {\it c)} where we take $\delta_1=-1$ and 
$\delta_2=1$. As shown in Fig.~\ref{anusec3}, cross sections 
as large as $\sigma_{\tilde{\chi}_1^0-p}\gsim 10^{-6}$ pb, entering
in DAMA
and fulfilling all experimental and astrophysical bounds, 
can be obtained for $\tan\beta=35,50$ and $A=0$. 
Those above the ZEPLIN line correspond
to $\sin^2(\alpha-\beta)$ not close to 1.
On the other hand, for $A=M$ and $\tan\beta=35$ no points with the
correct relic density enter in DAMA.
For other cases the results are similar.
For example, if we consider 
$\delta_1=-0.5$ and $\delta_2=1$, one obtains for
$\tan\beta=35$ points entering in DAMA but with
$\Omega_{\tilde{\chi}_1^0}h^2< 0.01$. For 
$\tan\beta=50$ a similar plot to the one in
Fig.~\ref{anusec3} is obtained.




Concerning the restrictions coming from the UFB-3 constraint, 
we can see in Fig.~\ref{anu}
that these are slightly less important than in the
universal scenario (see Fig.~\ref{a35}).
Of course, this is not a general result, and different choices of
the $\delta$'s can modify the situation.
For example, 
for the same case as in
Fig.~\ref{anu}, with $\tan\beta=35$, 
but using the opposite choice for the sign of the
$\delta$ parameters,
not only the cross section is smaller,
$\sigma_{\tilde{\chi}_1^0-p}< 10^{-8}$ pb,
but also the UFB-3 constraint is very
restrictive, forbidding all points which are allowed by
the experimental and astrophysical constraints.

\vspace{0.2cm}

In summary,
when
non-universal scalars are allowed in SUGRA, for 
some special choices of the non-universality,  
the cross section can 
be increased a lot with respect to the 
universal scenario.
It is even possible, for some particular values of the
parameters, to find points allowed by all experimental and
astrophysical 
constraints with
$\sigma_{\tilde{\chi}_1^0-p}\approx 10^{-6}$ pb, and therefore
inside the DAMA area. 
Note however that these points would be basically excluded by
the other underground experiments.
In any case, the interesting result is that
large regions accessible for future experiments are present.

\vspace{0.6cm}

\noindent {\it (ii) Non-universal gaugino masses}

\vspace{0.5cm}

The effects of the non-universality of gaugino masses on the dark matter
relic density in SUGRA scenarios were studied
in detail in
Refs.~\refcite{Griro,Yamaguchi,Nelson,jeong}.
Analyses of the neutralino-proton cross section using $SU(5)$ GUT relations
for gaugino masses
were carried out in Refs.~\refcite{Nath2,Orloff,Roy2}, whereas
in Refs.~\refcite{darkcairo,nosopro,Dutta,Birkedal,darkufb} 
generic non-universal soft masses were used. 
It was also realized
that the non-universality in the gaugino masses can increase the cross 
section. 



\begin{figure}
\epsfig{file=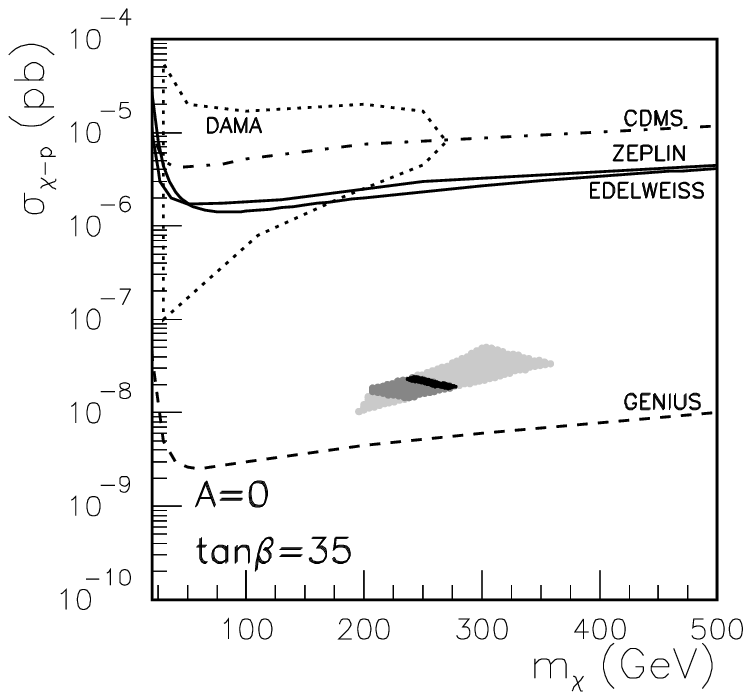,width=6.5cm}
\hspace*{-0.51cm}\epsfig{file=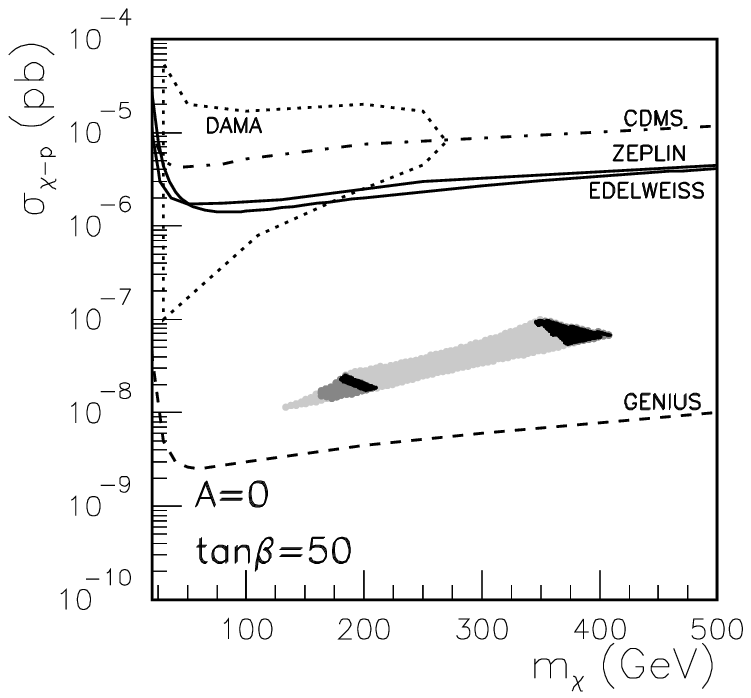,width=6.5cm}


\caption{The same as in Fig.~\ref{cross_scale122}
but for 
the case discussed in Eq.~(\ref{gauginospara}) with
non-universal soft gaugino masses, 
$\delta'_{1,2}=0, \delta'_3=-0.5$.
}
\label{nuniversal2}
\end{figure}





Let us parameterize this non-universality at $M_{GUT}$
as follows:
\begin{eqnarray}
M_1=M(1+\delta'_{1})\ , \quad M_2=M(1+ \delta'_{2})\ ,
\quad M_3=M(1+ \delta'_{3})
\ ,
\label{gauginospara}
\end{eqnarray}
where $M_{1,2,3}$ are the bino, wino and gluino masses, respectively.
Let us discuss now which values of the parameters are interesting in
order to increase the cross section with respect to the universal
case $\delta'_i=0$. In this sense,
it is worth noticing that
$M_3$ appears in the RGEs of squark masses, so e.g.
their contribution proportional
to the top Yukawa coupling in the RGE of $m_{H_u}^2$ will
do this less negative if $M_3$ is small.
Although this effect increases the cross section, it is also worth
noticing that
small values of $M_3$ also lead to
an important decrease in the Higgs 
mass (we have checked that generically the
value of $\sin^2(\alpha-\beta)$ is very close to 1, and therefore
we are using the lower bound $m_h=114.1$ GeV as in the mSUGRA scenario).
In addition,
$b \to s \gamma$ and $g_{\mu}-2$ constraints are also relevant.

Summarizing, 
although the cross section increases with respect to the universal
case, the present experimental constraints exclude points
entering in the DAMA region.
This is shown in Fig.~\ref{nuniversal2} \cite{darkufb} 
for $\tan\beta=35,50$ and $A=0$, using
$\delta'_{1,2}=0, \delta'_3=-0.5$,
where one can see that
there are 
points allowed by all experimental and astrophysical constraints, but
they
correspond to 
$\sigma_{\tilde{\chi}_1^0-p}\lsim 10^{-7}$ pb.


Finally, as in the previous case with non-universal scalars,
increasing the cross section through values at low energy
of $m_{H_u}^2$ less negatives implies
less important UFB constraints.
Now these are not very relevant, and in fact they correspond
to the UFB-1 ones.

\subsubsection{Effective MSSM scenario}

In Refs.~\refcite{Gondolo,muchos,Bed,Bottinoeff,Austri} 
the authors considered that
the uncertainties involved in SUGRA scenarios
(such as e.g. the values of the soft parameters, i.e. universality versus
non-universality, or
the choice of the scale, i.e. $M_{GUT}$ versus
$M_I$)
were problematic enough as
to work better with a phenomenological 
SUSY model whose parameters are defined directly at the electroweak 
scale. 
This effective scheme of the MSSM was denoted by effMSSM in
Ref.~\refcite{Bottinoeff}. For example, in this work
it was imposed for simplicity a set of
assumptions at the electroweak scale: 
a) all trilinear parameters are set to zero except those of the third family, 
which are unified to a common value $A$;
b) all squark soft-mass parameters are taken  
degenerate: $m_{\tilde q_i} \equiv m_{\tilde q}$; 
c) all slepton  soft--mass parameters are taken  
degenerate: $m_{\tilde l_i} \equiv m_{\tilde l}$; 
d) the gaugino masses, $M_1$ and $M_2$, are 
assumed to be linked by the GUT relation 
$M_1= \frac{5}{3} \tan^2 \theta_W M_2$ (this assumption was relaxed
in the fifth and fourth works of Ref.~\refcite{Bottinoeff}
obtaining relic neutralinos significantly lighter 
than those commonly considered, 
$m_{\tilde{\chi}^0_1} \gsim 50$ GeV).
As a consequence, the SUSY
parameter space consists of seven independent parameters, chosen them
to be: 
$M_2, \mu, \tan\beta, m_A, m_{\tilde q}, m_{\tilde l}, A$.

Obviously, the scenarios studied in the previous Subsections, where
some boundary conditions at the high scale are imposed and
then the RGEs are used in order to obtain the
low-energy spectrum, should be a subset of the general effMSSM 
scenario (i.e. without any assumption).
In the effMSSM with the above assumptions, after imposing the
experimental and astrophysical constraints,
points with cross sections entering in the DAMA area
can be obtained. This is similar to the result obtained when
non-universal soft scalar masses were considered in the previous Subsection.
Note in this sense that varying $\mu$ and $m_A$
arbitrarily corresponds to vary 
$m_{H_u}$ and $m_{H_d}$.

\subsection{Superstring predictions for the 
neutralino-nucleon cross section}

Although the standard model provides
a correct description of the observable world, there exist, however,
strong indications that it is just an effective theory at low energy
of some fundamental one. The only candidate for such a theory is,
nowadays, the string theory, which have the potential to unify
the strong and electroweak interactions with gravitation in a
consistent
way.

In the late eighties, working in the context of the (perturbative) 
$E_8\times E_8$ heterotic superstring, 
a number of interesting four-dimensional vacua
with particle content not far from that of the supersymmetric 
standard model were found \cite{viejos}. 
Until recently, 
it was thought that this was the only way in order to construct
realistic superstring models. However, in the late nineties, it was
discovered that explicit models with realistic properties 
can also be constructed using D-brane configurations from 
type I string vacua \cite{nnosotros}.
We will review below these two superstring constructions concerning
dark matter detection.

Such constructions
have a natural hidden sector built-in: the
complex dilaton field $S$ and the complex moduli fields $T_i$. 
These gauge singlet fields are generically present in four-dimensional
string models: the dilaton arises from the gravitational sector of the theory
and the moduli parameterize the size and shape of the compactified
variety.
Therefore 
the auxiliary fields of those multiplets can be the seed of SUSY
breaking,
solving the arbitrariness of SUGRA where the hidden sector is not
constrained.
In addition, in superstrings, $K$ and $f$ can be computed explicitly
leading to 
interesting predictions for the soft parameters, 
and therefore for the value of the neutralino-proton
cross section.

\subsubsection{$E_8\times E_8$ heterotic superstring constructions}

For any four-dimensional construction coming from the perturbative 
10-dimensional heterotic
superstring,
the tree-level gauge kinetic function is independent of the moduli sector
and is simply given by
\begin{eqnarray}
{f_a} &=& k_a S\ , 
\label{kahler3}
\end{eqnarray}
where usually 
one takes $k_3=k_2=\frac{3}{5}k_1=1$. In any case, the values
$k_a$ are
irrelevant for the tree-level computation since they do not contribute to
the soft parameters. 

On the other hand, the K\"ahler potential 
has been computed for several compactification schemes. This is 
for example the case of 6-dimensional Abelian orbifolds, 
where
three moduli $T_i$ are generically present.
For this  class of models the K\"ahler potential has the 
form 
\begin{eqnarray}
K &=& -\log(S+S^*) - \sum _i \log(T_i+T_i^*) 
+ \sum _{\alpha } |C^{\alpha }|^2 \Pi_i(T_i+T_i^*)^{n_{\alpha }^i} \ .
\label{orbi}
\end{eqnarray}
Here $n_{\alpha }^i$ are (zero or negative) fractional numbers usually 
called ``modular weights" of the matter fields $C^{\alpha }$. 

It is important to know what fields, either $S$ 
or $T_i$, play the predominant
role in the process of SUSY breaking. This will have relevant consequences
in determining the pattern of soft parameters, and therefore the spectrum
of physical particles. That is why it is very useful to introduce the
following parameterization 
for the VEVs 
of dilaton and moduli auxiliary fields \cite{dilaton}
\begin{eqnarray}
&& F^S= \sqrt{3} (S+S^*) m_{3/2} \sin \theta\;, \nonumber \\
&& F^i= \sqrt{3} (T_i+T^*_i) m_{3/2} \cos \theta\; \Theta_i\;,
\label{parameterize}
\end{eqnarray}
where
$i=1,2,3$ labels the three complex compact dimensions,
$m_{3/2}$ is the gravitino mass,
and
the angle $\theta$ and the $\Theta_i$ with $\sum_{i} |\Theta_i|^2=1$,
just parameterize the direction of the goldstino in the $S$, $T_i$ field
space.

Using this parameterization and eqs.~(\ref{kahler3}) and (\ref{orbi})
one obtains
the following results for the soft parameters \cite{dilaton}:
%
\begin{eqnarray}
M_a &=& \sqrt{3}m_{3/2} \sin\theta\ ,
\nonumber\\
m_{\alpha }^2 &=& m_{3/2}^2\left(1 + 3\cos^2\theta\ 
\sum_i n^i_{\alpha } {\Theta }_i^2
\right)
\ ,
\nonumber\\
A_{\alpha \beta \gamma } &=& -\sqrt{3} m_{3/2} \left( \sin\theta
+ \cos\theta \sum_{i} {\Theta }_i 
\left[1 +\
n^i_{\alpha }+n^i_{\beta
}+n^i_{\gamma
}-
(T_i+T_i^*) \partial_i \log Y_{\alpha \beta \gamma}\! \right]\right)\ . 
\nonumber\\
\label{masorbi}
\end{eqnarray}
Although in the case of the $A$-parameter 
an explicit $T_i$-dependence may appear in
the term proportional to $\partial_i \log Y_{\alpha \beta \gamma }$, 
where $Y_{\alpha \beta \gamma}(T_i)$ are the Yukawa couplings, 
it disappears in 
several interesting 
cases, and we will only consider this possibility here.
Using the above information, one can analyze the structure of
soft parameters available in Abelian orbifolds. 

In the dilaton-dominated case ($\sin\theta =1$) the 
soft parameters are universal, and fulfil
\begin{eqnarray}
& m
\ =\ m_{3/2}\ , \ M
\ =\ \sqrt{3} m_{3/2}\ ,\
A
\ =\ -M
\ . &
\label{cuatro}
\end{eqnarray}
Of course, they are a subset of the parameter space of mSUGRA,
and therefore one should expect small dark matter cross sections,
as discussed in Subsection~5.2.1.
However, in general, the soft parameters 
show a lack of universality due to the
modular weight dependence (see scalar masses and trilinear
parameters in Eq.~(\ref{masorbi})).
For example, assuming an overall modulus, i.e.  $T=T_i$ 
and $\Theta_i=1/\sqrt 3$, one obtains
\begin{eqnarray}
m_{\alpha }^2 &=& m_{3/2}^2\left(1 + n_{\alpha }\cos^2\theta 
\right)
\ ,
\nonumber\\
A_{\alpha \beta \gamma } &=& -\sqrt{3} m_{3/2} \sin\theta
- m_{3/2} \cos\theta 
\left(
3 +
n_{\alpha }+n_{\beta
}+n_{\gamma
}
\right)\ ,
\label{masorbi2}
\end{eqnarray}
where we have defined the overall modular weights 
$n_{\alpha}=\sum_i n^i_{\alpha }$, and in the
case of $Z_n$ Abelian orbifolds they can take
the values -1,-2,-3,-4,-5.
Fields belonging to the untwisted sector of the orbifold have $n_{\alpha}=-1$.
Fields in the twisted sector but without oscillators have usually
modular weight -2 and those with oscillators have 
$n_{\alpha}\leq -3$.
Note that e.g. for scalars in the untwisted sector
\begin{eqnarray}
& m_{\alpha}^2+m_{\beta}^2+m_{\gamma}^2
\ =\ M^2
\ ,\;\;\;\;\;\;\;\;\;\;\;\;\;\;
A
\ =\ -M
\ . &
\label{cuatrom}
\end{eqnarray}
These type of relations between fields in the
same Yukawa coupling, for instance
$m_{Q_L}^2+m_{u_R}^2+m_{H_u}^2=M^2$,
were applied in Ref.~\refcite{Drees}
to the analysis of dark matter detection, with the
final result that the neutralino-proton cross section 
is very similar to the one of the mSUGRA scenario.

On the other hand, the apparent success of the joining of gauge
coupling constants at $M_{GUT} \approx 2\times 10^{16}$ GeV in the MSSM 
is not automatic
in the heterotic superstring, where the
natural unification scale is 
$M_{H}=\sqrt{\frac{\alpha}{8}} M_{Planck}$,
with $\alpha\approx 1/24$ the gauge coupling.
Thus unification takes place at energies around a factor
$\approx 12$ smaller than expected in the heterotic superstring.
This problem might be solved with the presence of large
string threshold corrections which explain the mismatch between
$M_{GUT}$ and $M_H$. In a sense, what would happen is that
the gauge coupling constants will cross at $M_{GUT}$ and
diverge towards different values at $M_H$. These different
values appear due to large one-loop stringy threshold corrections.
It was found in Ref.~\refcite{Ibanez} that these corrections
can be obtained for restricted values of the modular weights of the
fields. In fact, assuming flavour independence, 
one finds that the simplest possibility corresponds to taking the
following values for the standard model fields:
\begin{eqnarray}
n_{Q_L} = n_{d_R} = -1\ ,\;\; n_{u_R} = -2\ ,\;\;
n_{L_L} = n_{e_R} = -3\ ,\;\;
n_{H_u} + n_{H_d} = -5,-4\ ,
\label{scalars1}
\end{eqnarray}
where e.g. $u_R$ denotes the three family squarks $\tilde{u}_R$, 
$\tilde{c}_R$, $\tilde{t}_R$.

The above scenario was studied in Ref.~\refcite{Shafi}
for $n_{H_u}=-3, n_{H_d} = -2$,
paying special attention to the calculation of the relic neutralino 
density.
Subsequently, in Ref.~\refcite{Shafi2}, the authors
estimate the direct detection rate for the same scenario.
The conclusion was that this is small and large scale detectors
are needed to discover the neutralino.

Although, apparently, gaugino masses are always universal in
this superstring construction (see Eq.~(\ref{masorbi})),
this is not in fact completely true.
It is true that they are universal at tree level because in this
case the gauge kinetic function in Eq.~(\ref{kahler3}) is 
always proportional to the dilaton. However, threshold corrections
turn out to be crucial in the $\sin\theta\to 0$ limit,
when SUSY breaking is not dominated by the dilaton.
In this case one can observe that gaugino masses turn out to be
non-universal, and a different phenomenology is obtained.
The relic neutralino density was analyzed in this scenario
in Ref.~\refcite{Dreess}.
More recently, the relic density was also analyzed in the general
context when the soft Lagrangian is dominated by loop 
contributions in Ref.~\refcite{Mambrini}.
For another recent work studying also direct and indirect detection in
these scenarios see Ref.~\refcite{Mambrini2}.

Let us finally remark that an analysis of the dark matter using a 
specific mechanism for the breaking of SUSY has also been carried out
in the literature.
In particular, it was pointed out in Ref.~\refcite{nilles} 
that,
since scalar masses are much larger than gaugino masses when
SUSY is spontaneously broken by non-perturbative 
hidden-sector gaugino condensates, the relic abundance of the
neutralinos is too large and incompatible with the astrophysical
observations in most of the parameter space.

Summarizing,
in the specific compactification models of the heterotic superstring
studied here, 
although the soft terms can be non-universal, the final 
cross section is generically small and 
similar to the one of the mSUGRA scenario.

\subsubsection{D-brane constructions}

D-brane constructions are explicit scenarios where two of the interesting
situations studied in Section~5.2,
non-universality and intermediate scales, may occur.
Concerning the latter, it was recently 
realized that 
the string scale may be anywhere between the weak and the Plank 
scale \cite{nnosotros}.
For instance, embedding the SM inside D3-branes in type I
strings, the string scale is given by
\begin{equation}
M_I^4= \frac{\alpha M_{Planck}}{\sqrt 2} M_c^3\ ,
\end{equation}
where $\alpha$ is the gauge coupling and $M_c$ is the compactification scale. 
Thus one gets $M_I\approx 10^{10-12}$ GeV with $M_c\approx 10^{8-10}$ GeV.

As mentioned in Section~5.2.2, these intermediate scales are
interesting from the phenomenological viewpoint \cite{see}.
They are also interesting from the theoretical viewpoint. For example,
in supergravity models supersymmetry can be spontaneously broken in a 
hidden sector of the theory and the gravitino mass,
which sets the overall scale of the soft terms, is given by:
\bea
m_{3/2}\approx \frac{F}{M_{Planck}}\ ,
\label{gravitino}
\eea
where $F$ is the auxiliary field whose vacuum expectation value
breaks supersymmetry. 
Since in supergravity one would expect $F\approx M_{Planck}^2$, one  
obtains
$m_{3/2}\approx M_{Planck}$ and therefore 
the hierarchy problem solved in principle by supersymmetry
would be re-introduced,
unless non-perturbative effects such as gaugino condensation
produce $F\approx M_W M_{Planck}$. However, if the scale
of the fundamental theory is $M_I\approx 10^{11}$ GeV instead of
$M_{Planck}$,
then $F\approx M_I^2$
and one gets $m_{3/2}\approx M_W$ in a natural way, without invoking any
hierarchically suppressed non-perturbative effect.
In the above example with a D3-brane, with a modest
input hierarchy \footnote{It is worth noticing, however, that 
those values would imply
${\rm Re}(S)=1/\alpha\approx 24$ and
${\rm Re}(T)=\frac{1}{\alpha}\left(\frac{M_I}{M_c}\right)^4 \approx
10^9$,
i.e. one has again a hierarchy problem but now for the VEVs of the 
fields that one has to determine dynamically.}
between string and compactification scales,
$M_I\approx 10^{11}$ GeV and $M_c\approx 10^9$ GeV,  
one obtains the desired hierarchy 
$M_W/M_{Planck}\approx 10^{-16}$.

The first attempts to study dark matter within these constructions
were carried out in scenarios with the unification scale  
$M_{GUT} \approx 10^{16}$ GeV
as the initial scale \cite{Khalil,Nath2,Arnowitt2}
and dilaton-dominated SUSY-breaking scenarios 
with an intermediate scale as the initial scale \cite{Bailin}.
However,
the important issue of the D-brane origin of the $U(1)_Y$ gauge group
as a combination of other $U(1)$'s 
and its influence on the matter distribution in these scenarios
was not included in those analyses.
When this is taken into account, interesting results are 
obtained \cite{nosotros}. In particular, 
scenarios with the gauge group and particle content of the
SUSY standard model lead naturally to intermediate values for the
string
scale, in order to reproduce the value of gauge couplings
deduced from experiments. In addition, the soft terms 
turn out to be generically non universal.
Due to these results, in principle, 
large cross sections
can be obtained.

Let us first recall that there are
two possible avenues to construct the supersymmetric standard model
with D-branes:
{\it (a)} The $SU(3)_c$, $SU(2)_L$ and $U(1)_Y$ groups
of the standard model come from different sets of D$p$-branes. 
{\it (b)} They come from the same set of D$p$-branes. 
In the first scenario, it is worth remarking
the difficulty of  
obtaining three copies of quarks and leptons
if the gauge groups are attached to different
sets of D$p$-branes
Thus
whether or not this scenario
may arise from different sets of D$p$-branes 
in explicit string constructions
is an important issue which is worth attacking in the future.
Concerning the other scenario {\it (b)}, models 
with the gauge group of the standard model and
three families of particles have been explicitly built.
Let us concentrate first on scenario {\it (a)}.

\begin{figure}[t]
\begin{center}
\begin{tabular}{c}
\epsfig{file= 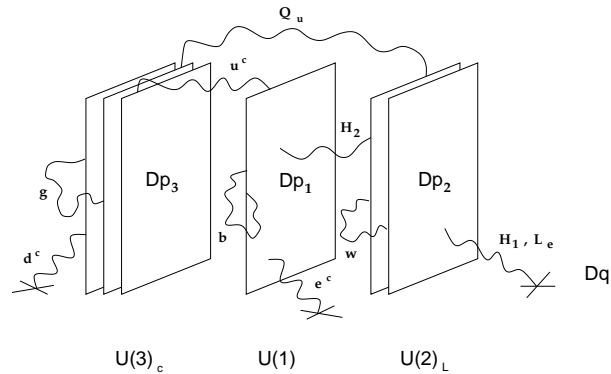, width=8cm}\\
\end{tabular}
\end{center} 
\caption{A generic D-brane scenario giving rise to the gauge bosons
and matter of the standard model. It contains three D$p_3$-branes,
two D$p_2$-branes and one D$p_1$-brane, where $p_N$ may be
either 9 and $5_i$ or 3 and $7_i$. The presence of extra D-branes,
say D$q$-branes, is also
necessary.
For each set the D$p_N$-branes
are in fact on the top of each other.
} 
\label{dbranesf}
\end{figure}

Consider for example a type I string scenario \cite{nosotros}
where the gauge group 
$U(3)\times U(2)\times U(1)$, giving rise to
$SU(3)\times SU(2)\times U(1)^3$, arises
from three different types of D-branes as shown schematically
in Fig.~\ref{dbranesf}, where open strings starting and
ending on the same sets of D$p_N$-branes give rise to the gauge bosons
of the standard model. For the sake of visualization each set is
depicted at parallel locations, but in fact they are intersecting each
other.
Let us recall that
the presence of extra D-branes,
say D$q$-branes, is also
necessary as explained in 
Ref.~\refcite{nosotros}

\begin{figure}[t]
\begin{center}
\begin{tabular}{c}
\epsfig{file= 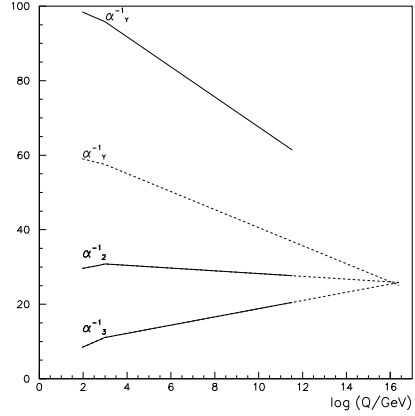, width=6.2cm}\\
\end{tabular}
\end{center} 
\caption{Running of the gauge couplings of the MSSM with energy $Q$ 
embedding the gauge groups within different sets of D$p$-branes (solid lines).
For comparison
the running of the MSSM couplings with
the usual normalization factor for the hypercharge, $3/5$,
is also shown with dashed lines. An overall SUSY scale
of 1 TeV is used.} 
\label{rugauge}
\end{figure}

Since the standard model gauge group arises from different types of 
D-branes, the gauge kinetic functions associated with them are
different in general, and therefore 
the gauge couplings are
non-universal. On the other hand,
$U(1)_Y$ is a linear combination
of the three $U(1)$ gauge groups arising from $U(3)$, $U(2)$ and
$U(1)$ within the three different D-branes (with
the extra $U(1)$'s anomalous and with the associated gauge
bosons with masses of the order of the string scale $M_I$).
This implies
\begin{equation}
\frac{1}{\alpha_Y(M_I)} =     
\frac{2}{\alpha_1(M_I)} + \frac{1}{\alpha_2(M_I)} 
+ \frac{2}{3 \alpha_3(M_I)}\ ,
\label{couplings}
\end{equation}
where $\alpha_k$ correspond to the gauge couplings of the $U(k)$ branes.
As shown in Ref.~\refcite{nosotros}, in order
to reproduce the low-energy value of the gauge couplings 
deduced from experiments,   
the above equation  
leads to solutions with
the string scale $M_I \approx 10^{10-12}$ GeV. 
This scenario is shown in Fig.~\ref{rugauge} 
for $M_I = 10^{12}$ GeV. Notice that it is similar to
the one shown in Fig.~\ref{uni}a, when discussing intermediate
scales in mSUGRA, but with the qualitatively difference
that the running of the 
$U(1)_Y$ gauge coupling must fulfil
relation (\ref{couplings}).

%

The analysis of the soft terms has been done under
the assumption that only the
dilaton ($S$) and moduli ($T_i$) fields contribute to SUSY
breaking and it has been found  that these soft terms  are generically
non-universal. Considering the assignment of gauge bosons and
matter of Fig.~\ref{dbranesf}, and
using the standard parameterization of Eq.~(\ref{parameterize}), 
one is able to obtain the following soft terms \cite{nosotros}.
The gaugino masses associated with the three gauge groups of the
standard model are given by
%
\bea
M_3 & = & \sqrt{3} m_{3/2} \sin \theta \ , \nn\\
M_{2} & = & \sqrt{3}  m_{3/2}\ \Theta_1 \cos \theta  \ , \nn\\
M_{Y} & = &  \sqrt{3}  m_{3/2}\ \alpha_Y (M_I)
\left(\frac{2\ \Theta_3 \cos \theta}{\alpha_1 (M_I)}
+\frac{\Theta_1 \cos \theta}{\alpha_2 (M_I)}
+\frac{2\ \sin \theta}{3 \alpha_3 (M_I)}
\right)\ .
\label{gaugino1}
\eea
The soft scalar masses of the three families are given by
\begin{eqnarray}
m^2_{Q_L} & = & m_{3/2}^2\left[1 -
\frac{3}{2}  \left(1 - \Theta_{1}^2 \right)
\cos^2 \theta \right] \ , \nn \\
m^2_{d_R} & = & m_{3/2}^2\left[1 -
\frac{3}{2}  \left(1 - \Theta_{2}^2 \right)   
\cos^2 \theta \right] \ , \nn \\
m^2_{u_R} & = & m_{3/2}^2\left[1 -
\frac{3}{2}  \left(1 - \Theta_{3}^2 \right)
\cos^2 \theta \right] \ , \nn \\
m^2_{e_R} & = & m_{3/2}^2\left[1- \frac{3}{2}
\left(\sin^2\theta + \Theta_{1}^2 \cos^2\theta  \right)\right] \ , \nn \\
m^2_{L_L} & = & m_{3/2}^2\left[1- \frac{3}{2}
\left(\sin^2\theta + \Theta_{3}^2 \cos^2\theta  \right)\right] \ , \nn \\
m^2_{H_u} & = & m_{3/2}^2\left[1- \frac{3}{2}
\left(\sin^2\theta + \Theta_{2}^2 \cos^2\theta  \right)\right] \ , \nn \\
m^2_{H_d} & = & m^2_{L_L} \;,     
\label{scalars11}
\end{eqnarray}
where e.g. $u_R$ denotes the three family squarks $\tilde{u}_R$, 
$\tilde{c}_R$, $\tilde{t}_R$.
Finally the trilinear parameters of the three families are
\begin{eqnarray}
A_{u} & = &  \frac{\sqrt 3}{2}m_{3/2}
   \left[\left(\Theta_{2} - \Theta_1
 - \Theta _{3}  \right) \cos\theta
- \sin\theta \; \right] \ ,
\nn \\
A_{d} & = &  \frac{\sqrt 3}{2}m_{3/2}
   \left[\left(\Theta_{3} - \Theta_1
  - \Theta _{2}  \right) \cos\theta
- \sin\theta \; \right] \ ,
\nn \\
A_{e} & = &  0\; .
\label{trilin11}
\end{eqnarray}

Although these formulas for
the soft terms imply that one has in principle
five free parameters, $m_{3/2}$, $\theta$ and  $\Theta_i$ with $i=1,2,3$,
due to relation $\sum_i |\Theta_i|^2=1$ only four of them are
independent.
In the analysis the parameters $\theta$ and $\Theta_i$
are varied in the whole allowed range, $0\leq \theta\leq 2\pi$,
$-1\leq\Theta_i\leq 1$. For 
the gravitino mass, 
$m_{3/2}\leq 300$ GeV is taken.
Concerning Yukawa couplings, their values are fixed imposing 
the correct fermion mass spectrum at low energies, i.e.,
one is assuming that Yukawa structures of D-brane scenarios
give rise to those values.

\begin{figure}[t]
\hspace*{-0.51cm}\epsfig{file= 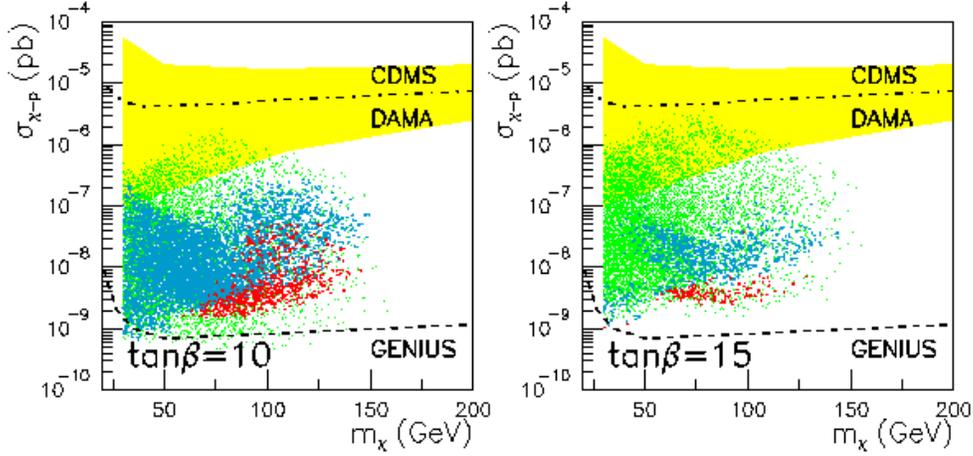, 
height=6.2cm
}
\vspace{-0.5cm}
\caption{
Scatter plot of the scalar neutralino-proton cross section
$\sigma_{\tilde{\chi}_1^0-p}$ 
as a function of 
the neutralino mass $m_{\tilde{\chi}_1^0}$ in the D-brane scenario
with 
the string scale $M_I=10^{12}$ GeV
discussed
in the text, and for $\tan\beta= 10$ and 15.
Only the big (red and blue) dots 
fulfil $b \to s \gamma$ and $g_{\mu}-2$ constraints.
The red ones 
correspond to points 
with $m_h\geq 114$ GeV whereas the blue ones correspond
to points
with $91 <m_h < 114$ GeV. 
DAMA and CDMS current 
experimental limits and projected GENIUS limits are also shown.
}
\label{dbrane}
\end{figure}
%





Fig.~\ref{dbrane} displays a scatter plot of 
$\sigma_{\tilde\chi_1^0-p}$ as a function of the
neutralino mass $m_{\tilde\chi_1^0}$ for a scanning of the parameter space
discussed above \cite{nosopro}. 
Two different values of $\tan\beta$, 10 and 15, are shown.
Although the astrophysical and UFB 
constraints have not been taken into account 
in the
analysis, and 
$\sin^2(\alpha-\beta)$ has not been computed explicitly,
several conclusions can already be drawn.
Regions of the parameter
space consistent with DAMA limits exist, but the 
$b \to s \gamma$ and $g_{\mu}-2$ constraints forbid 
most of them. The latter are shown with small (green) points,
and they have $91 <m_h < 114$ GeV. 
In Fig.~\ref{dbrane} only regions with big (red and blue) dots 
fulfil the above mentioned constraints.
The red ones
correspond to points 
with $m_h\geq 114$ GeV whereas the blue ones correspond
to points
with $91 <m_h < 114$ GeV. 
It is worth noticing that the larger $\tan\beta$ is, 
the smaller the regions allowed by the experimental constraints become.
For example, increasing
$\tan\beta$ the value of 
$g_{\mu}-2$ turns out to be larger and may exceed the 
experimental bounds\footnote{However, for larger values of the
gravitino mass, in some special regions of the parameter space the
allowed points may even enter in DAMA \cite{prepa}.}.

Let us finally mention that 
other examples with the standard model gauge group embedded in D-branes in 
a different way, or with larger values of the string scale, 
can also be found \cite{nosotros}.
The dark matter cross section is qualitatively similar.

Concerning scenario {\it (b)}, where all gauge groups of the 
standard model are
embedded
within the same set of D-branes, and therefore with gauge coupling
unification, it can be analyzed similarly to scenario {\it (a)}.
However, in this case
the value of the cross section is generically smaller.
As mentioned above, in the context of mSUGRA with an intermediate
scale, this is due to the
different values of $\alpha$'s at the
string scale in both types of scenarios.

\vspace{0.2cm}

Summarizing,
in D-brane configurations from type I string
non-universal soft terms and 
intermediate scales arise naturally.
Although we saw in Subsection~5.2
that for some values of the parameters
this may be interesting in order to increase the value of the cross
section, in our case these values are in such a way that
the cross section turns out to be essentially below DAMA.
However, it is worth noticing that regions accessible for future
experiments are present.

\subsection{M-Theory predictions for the neutralino-nucleon cross section}

\begin{figure}[t]
\begin{center}
\epsfig{file=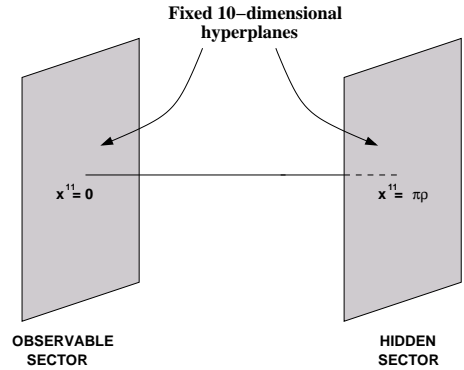,width=6cm}
\end{center}
\caption{Schematic of heterotic M-Theory.
\label{mtheory}}
\end{figure}

The proposal of $11$-dimensional M-Theory 
as a fundamental theory which 
contains the five $10$-dimensional 
superstring theories, as well as $11$-dimensional 
SUGRA, as different vacua of its 
moduli space 
has motivated a large amount of 
phenomenological analyses \cite{mioreview}. 
The cornerstone of most of these works is the 
construction due to Ho\v{r}ava and Witten, who 
showed that the low-energy limit of M-Theory, 
compactified on a $S^1/Z_2$ orbifold 
with $E_8$ gauge multiplets on each of the $10$-dimensional 
orbifold fixed planes, was indeed the strong 
coupling limit of the $E_8\times E_8$ heterotic string 
theory. This is shown schematically in Fig.~\ref{mtheory}.

A resulting $4$-dimensional SUGRA can be obtained if six of the
remaining 
dimensions are compactified in a Calabi-Yau manifold. 
A certain number of virtues were noticed in this theory. 
The most relevant one was the possibility of 
tuning the $11$-dimensional Planck scale and the 
orbifold radius so that the GUT-scale, 
$M_{GUT}\approx 2\times10^{16}$ GeV, which is here 
identified with the inverse of the Calabi-Yau volume, 
and the Planck scale, $M_{Planck}=1.2\times10^{19}$ GeV, 
were recovered.

On the other hand, the structure of the soft SUSY-breaking terms has been
determined 
with the result \cite{cerde2}
\begin{eqnarray}
M &=& \frac{\sqrt{3}m_{3/2}}{1+\epsilon_O}\left(\sin\theta + 
\frac{1}{\sqrt{3}}\epsilon_O\cos\theta \right)
\ ,
\nonumber \\
m^2 &=& m_{3/2}^2 
- 
\frac{3m_{3/2}^2}{\left(3+\epsilon_O\right)^2}\left[
\epsilon_O\left(6+\epsilon_O\right)sin^2\theta
+ 
\left(3+2\epsilon_O\right)\cos^2\theta
- 2\sqrt{3}\epsilon_O\sin\theta\cos\theta \right]\ , 
\nonumber \\
A &=&
-\frac{\sqrt{3}m_{3/2}}{3+\epsilon_O}\left[\left(3-2\epsilon_O\right)
\sin\theta 
+ \sqrt{3}\epsilon_O\cos\theta \right] \ ,
\label{softterms}
\end{eqnarray}
where $-1<\epsilon_O<1$, and only one
modulus $T$ (also valid in the overall modulus case),
with the standard 
parameterization \cite{dilaton} 
$F^S= \sqrt{3} (S+S^*) m_{3/2} \sin \theta$,
$F^T= (T+T^*) m_{3/2} \cos \theta$,
has been assumed. 
Let us remark that this assumption about the Calabi-Yau
compactification 
leads to interesting phenomenological virtues \cite{mioreview}.
In particular,
the soft SUSY-breaking terms are automatically universal,
and therefore the presence of dangerous FCNC 
is avoided. Examples of such compactifications exist, as e.g.
the quintic hypersurface $CP^4$. Although these spaces were also
known in the context of the weakly-coupled heterotic string,
the novel fact in heterotic M-theory is that model building 
is relatively simple, and the construction of
three generation models might be considerably easy.

Using the above results for the soft terms, 
one can analyze how compatible is the parameter space 
of heterotic M-theory with the sensitivity of current dark matter detectors. 
Several analyses of dark matter in M-theory were 
performed \cite{bailinlove,nilles,bailinlove2,pallis}, 
paying special attention to the calculation of the relic density. 
Since in this scenario the soft terms 
(\ref{softterms}) are universal, and 
it 
is very unnatural to obtain low scales \cite{cerde},
all examples concerning dark matter
can be considered as a subset into
the 
parameter space of mSUGRA with a GUT scale.
In this sense, one should not expect to obtain high values for 
$\sigma_{\tilde\chi_1^0-p}$. Let us review this result.

\begin{figure}[!t]
\hspace*{-0.51cm}\epsfig{file=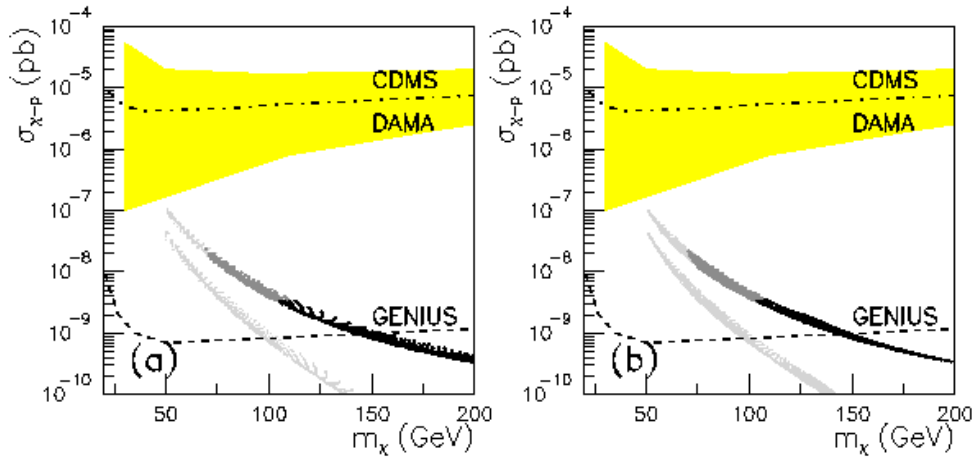,height=6.2cm}
\caption{a) Neutralino-nucleon cross-section versus 
neutralino mass for
$-0.6\leq\epsilon_O\leq-0.1$. b) The same for $0.1\leq\epsilon_O<1$. 
In both cases only the big (black and dark gray) dots fulfill 
both $b \to s \gamma$ and $g_{\mu}-2$ constraints. The black ones 
correspond to points with $m_h\geq114$ GeV, 
whereas the dark gray ones correspond to points 
with $91\leq m_h\leq114$ GeV.
Current DAMA and CDMS limits
and the projected GENIUS limit are shown.}
\label{nofbcross}
\end{figure}

The soft terms are expressed in terms of three free parameters: 
the gravitino mass $m_{3/2}$, 
the Goldstino angle, $\theta$, and the parameter $\epsilon_O$. 
The values of the $\epsilon_O$ parameter will be chosen 
in order to guarantee the correct GUT scale
(i.e.,  
$-0.6\leq\epsilon_O\leq-0.1$ and $0.1\leq\epsilon_O<1$ 
for non-standard and standard embeddings, respectively). 
We will take $m_{3/2}=300$ GeV, and $\tan\beta=10$, 
performing a variation of the goldstino angle, $\theta$, in $[0,2\pi)$.
Both cases are depicted in Fig.~\ref{nofbcross} \cite{cerde}.
Although the astrophysical and UFB 
constraints have not been taken into account 
in this
analysis,
several conclusions can already be drawn.
The experimental constraints, 
$b \to s \gamma$ and $g_{\mu}-2$,
put severe bounds, but
still neutralinos as light as $\sim100$ GeV can be 
obtained.
Once the lower limit on the Higgs mass is applied, 
the cross-section is as small 
as $\sigma_{\tilde\chi_1^0-p}\sim3\times10^{-9}$ pb, 
far beyond the reach of present detectors, 
and close to the lower limit of the projected GENIUS.
Let us recall that
all the points represented satisfy the experimental constraints 
on the lower masses of the supersymmetric particles and 
satisfy $m_h\geq91$ GeV. Small (light gray) dots represent 
points not fulfilling the  $b \to s \gamma$ constraint. Large dots 
do satisfy that constraint, and among these, dark gray points 
have $91$ GeV$\leq m_h\leq114$ GeV, while black dots satisfy 
the stronger lower bound for the Higgs mass $m_h>114$ GeV. 
Of course, as discussed above, this constraint on the Higgs mass holds 
in general for the cases with universal soft terms 
for $\tan\beta\lsim50$ and therefore it is the one 
we should consider here. However, due to the strong 
restrictions it imposes it is shown explicitly.

Although the predicted values for the cross-section increase, 
in principle, when larger values of $\tan\beta$ are 
taken into account, the experimental bounds become much more 
important in these cases (especially those 
corresponding to $b \to s \gamma$ and $g_{\mu}-2$), excluding larger 
regions in the parameter space and 
thus forbidding large values of $\sigma_{\tilde\chi_1^0-p}$.

Finally, it is worth noticing that
non-perturbative objects of M-Theory, 
such as M$5$-branes, 
can be shown to survive 
the orbifold projection 
of Ho\v{r}ava-Witten 
construction under certain circumstances, 
permitting much more 
freedom to play with gauge groups and 
with the matter fields 
that appear. In this context,
dark matter implications of vacua 
with five-branes were investigated in the 
limit where the five-brane modulus, $Z$, is the 
only one responsible for the breaking of supersymmetry \cite{krani}.
However, the authors used previous 
soft-terms computed in the literature, 
where some corrections were not included \cite{cerde2}. 
The soft terms are now more involved than those in
Eq.~(\ref{softterms}) and include the $F$ terms associated
with $S$, $T$ and $Z$.
A similar analysis as above, concerning dark matter detection, 
can be carried out in this case \cite{cerde}.
The highest value of the cross section,
$\sigma_{\tilde\chi_1^0-p}\sim 10^{-8}$ pb,
can be obtained in the special case in which the five-brane
modulus is the only one responsible for SUSY breaking.

\vspace{0.2cm}

Summarizing, as it could be expected from the universality of the
soft terms, 
the predictions of heterotic M-theory with one modulus 
for the neutralino-nucleon cross section 
are too low to be probed by the present dark 
matter detectors. 
Only future experiments, as e.g. GENIUS, would be able 
to explore such low values. 


\section{Conclusions}

Nowadays there is overwhelming evidence that most of the mass in the universe
(90\% and probably more) 
is some (unknown) non-luminous `dark matter'.
At galactic and cosmological scales it 
only manifests through 
its gravitational interactions with 
ordinary matter.
However, at microscopical scales it might manifest through weak interactions,
and this raises the hope 
that it may be detected in low-energy particle physics
experiments.

Plausible candidates for dark matter 
are Weakly Interacting Massive Particles, the so-called WIMPs.
They are very interesting because they can be present in the
right amount to explain the observed matter density of the
Universe 
$0.1\lsim \Omega h^2\lsim 0.3$.
The leading candidate for WIMP is 
the so-called neutralino, a particle predicted by the
supersymmetric extension of the standard model.
These neutralinos are stable and therefore may be left over from the
Big
Bang.
Thus they 
will cluster gravitationally with ordinary stars in the galactic halos, 
and 
in particular they will be present in our own galaxy, the Milky Way.
As a consequence there will be a flux of these dark matter particles
on the Earth. 

Many underground 
experiments have been carried out around the world in order to detect this
flux. One of them, the DAMA collaboration, 
even claims to have detected it.
They obtain that 
the preferred range of the WIMP-nucleon cross section
is $\approx$
$10^{-6}-10^{-5}$ pb 
for a WIMP mass between 30 and 270 GeV. 
Unfortunately, this result is controversial because of the negative
search result obtained by other experiments like
CDMS, EDELWEISS AND ZEPLIN in the same range of
parameters.
Thus we will have to wait for the next generation of experiments,
which are already starting operations or in project, to 
obtain more information about
whether or not neutralinos, or generically WIMPs, are 
the evasive dark matter filling the whole Universe.
The most sensitive 
detector, GENIUS,
will be able to test a WIMP-nucleon cross section
as low as $\approx 10^{-9}$ pb. 
Indeed such a sensitivity covers a large range
of the parameter space of SUSY models with
neutralinos as dark matter.

Concerning this point, we have reviewed the known SUSY scenarios, 
and in particular
how big the cross section for the direct detection of neutralinos 
can be.
This analysis is crucial in order to know the
possibility of detecting dark matter 
in the experiments.
In particular, 
we have concentrated on SUGRA and
superstring and M-Theory scenarios.

Let us recall that the analysis has been carried out imposing
the most recent experimental and astrophysical constraints
on the parameter space. 
Concerning the former, 
the lower bound on the Higgs mass,
the $b\to s\gamma$ branching ratio, and the
muon $g-2$ have been considered.
The astrophysical bounds on the matter density
mentioned above have also been
imposed on the 
theoretical computation of the relic neutralino density,
assuming thermal production.
In addition, 
the constraints that the absence of dangerous charge
and colour breaking minima imposes on the parameter space
has also been taken into account.

In the usual mSUGRA scenario, where the soft terms
are assumed to be universal, and the GUT scale is considered, 
the cross section is
constrained to be 
$\sigma_{\tilde{\chi}_1^0-p}\lsim 3\times 10^{-8}$ pb.
Obviously, in this case, present experiments are not sufficient and
more sensitive detectors
producing further data 
are needed.
A similar conclusion is obtained when an intermediate scale is considered.
Although 
the cross section increases significantly,
the experimental bounds
impose
$\sigma_{\tilde{\chi}_1^0-p}\lsim 4\times 10^{-7}$ pb.
And, in fact, at the end of the day, the preferred astrophysical range
for the relic neutralino density, 
$0.1\leq\Omega_{\tilde{\chi}_1^0}h^2\leq 0.3$,
imposes 
$\sigma_{\tilde{\chi}_1^0-p}\lsim  10^{-7}$ pb.
Still present experiments are not sufficient.

When
non-universal scalars are allowed in SUGRA, for 
some special choices of the non-universality,  
the cross section can 
be increased a lot with respect to the 
universal scenario.
It is even possible, for some particular values of the
parameters, to find points allowed by all experimental and
astrophysical 
constraints with
$\sigma_{\tilde{\chi}_1^0-p}\approx 10^{-6}$ pb, and therefore
inside the DAMA area. 
This is similar to what occurs in the so-called
effMSSM scenario.
For non-universal gauginos,
although the cross section increases, the experimental bounds
exclude this possibility.

On the other hand, 
the low-energy limit of superstring theory and M-theory is 
SUGRA, and therefore the neutralino will also be a candidate for
dark matter in these scenarios.
In the context of superstring theory we have reviewed two interesting
constructions, the perturbative heterotic superstring and
D-branes configurations from type I string, where the soft terms
can be computed explicitly in some models.
In the former, although the soft terms can be non-universal, the final 
cross section is similar to the one of the mSUGRA scenario. 
In the latter, in addition to non-universality,
intermediate scales arise naturally, but again the cross section
is in general small (although some regions may even be compatible 
with DAMA).

Finally, 
due to the universality of 
the soft SUSY-breaking 
terms in the heterotic M-Theory scenario
analyzed with only one modulus,
and the fact that the most 
natural value for the initial scale 
is of order $10^{16}$ GeV, the parameter 
space can be considered as a subset of mSUGRA.
Therefore, the predicted cross-section is very low,
$\sigma_{\tilde{\chi}_1^0-p}\lsim 10^{-8}$ pb, 
far beyond the reach of the present dark matter experiments. 

\vspace{0.2cm}

In summary, underground physics as the one discussed here is crucial 
in order
to detect dark matter. Even if neutralinos are discovered first
at future particle accelerators such as LHC, only their direct detection
due to their presence in our galactic halo will confirm that they
are the sought-after dark matter of the Universe.
For that, many new underground experiments are already starting 
operations or in project,
and they will be able to cover an important range of the parameter space
of SUSY models.
\\

\noindent {\bf Acknowledgments}

\noindent
We gratefully acknowledge D.G. Cerde\~no for his valuable help,
and interesting comments.
This work was supported in part by the Spanish
DGI of the MCYT under contracts
BFM2003-01266 and FPA2003-04597, and
under Acci\'on Integrada Hispano-Alemana HA2002-0117;
and the European Union under contract 
HPRN-CT-2000-00148. 




\end{document}